%% file: state-operator.tex
\documentclass{SciPost}
\input{preamble.tex}
\begin{document}

\begin{center}
	{\Large\textsc{	Generalised symmetries and  state-operator correspondence  for nonlocal operators }}
\end{center}

\begin{center}
	Diego M. Hofman and
	Stathis Vitouladitis
\end{center}

\begin{center}
	\footnotesize
	Institute for Theoretical Physics, University of Amsterdam, \\ 1090 GL
	Amsterdam, the Netherlands	\\[\baselineskip]
	\footnotesize{
		\href{mailto:d.m.hofman@uva.nl}{\sf d.m.hofman@uva.nl} \qquad \href{mailto:e.vitouladitis@uva.nl}{\sf e.vitouladitis@uva.nl}
    }
\end{center}

\vspace{3em}

\begin{abstract}
	\noindent
	We provide a one-to-one correspondence between line operators and states in four-dimensional CFTs with continuous 1-form symmetries. In analogy with $0$-form symmetries in two dimensions, such CFTs have a free photon realisation and enjoy an infinite-dimensional current algebra that generalises the familiar Kac--Moody algebras. We construct the representation theory of this current algebra, which allows for a full description of the space of states on an arbitrary closed spatial slice. On \(\S^2\times\S^1\), we rederive the spectrum by performing a path integral on \(\B^3\times\S^1\) with insertions of line operators. This leads to a direct and explicit correspondence between the line operators of the theory and the states on \(\S^2\times\S^1\). Interestingly, we find that the vacuum state is not prepared by the empty path integral but by a squeezing operator. Additionally, we generalise some of our results in two directions. Firstly, we construct current algebras in \((2p+2)\)-dimensional CFTs, that are universal whenever the theory has a \(p\)-form symmetry, and secondly we provide a non-invertible generalisation of those higher-dimensional current algebras.
\end{abstract}

\pagebreak
    \vspace{10pt}
    \tableofcontents
\pagebreak

\section{Introduction}

Conformal Field Theory (CFT) is a fundamental framework in theoretical physics. It plays a crucial role in various distinct areas of theoretical physics. In statistical mechanics, it characterises the universality classes of different systems at criticality \cite{Cardy:1984bb,Cardy:2008jc}. Additionally, it describes the long-range behaviour of many quantum field theories (QFTs), and the short-range behaviour of ultraviolet-complete (UV-complete) QFTs via the renormalisation group \cite{Wilson:1971bg,Wilson:1971dh,Wilson:1974mb}. Finally, it serves as a gateway to quantum gravity, most notably, through string theory and the AdS/CFT correspondence \cite{Maldacena:1997re,Witten:1998qj,Aharony:1999ti}. These points alone highlight the fundamental importance of a deep, non-perturbative understanding of CFT. In flat space a lot is known. In any given CFT, the complete set of scaling dimensions of local operators and operator product expansion (OPE) coefficients suffices to reconstruct arbitrary correlation functions, effectively solving the theory. Significant progress has been achieved using traditional methods like conformal perturbation theory, as well as non-perturbative approaches such as the conformal bootstrap \cite{Poland:2018epd,Poland:2022qrs,Hartman:2022zik}.

However, this emphasis on local objects is not well suited for some physical situations of interest. For instance, condensed matter theory has seen renewed interest in nonlocal excitations, such as anyons \cite{Burnell_2018} and fractons \cite{Nandkishore:2018sel}. In quantum field theory, the question of quark confinement is connected to the physics of line operators \cite{Polyakov:1979gp,Makeenko:1980vm,Polyakov:1980ca,Gaiotto:2014kfa,Gaiotto:2017yup}. Finally, quantum gravity is inherently linked with nonlocal operators, based on the simple argument that diffeomorphism invariance forbids local operators.\footnote{Topological local operators, corresponding to \((d-1)\)-form symmetries, are an exception. However, arguments about the absence of (higher-form) global symmetries suggest their nonexistence \cite{Harlow:2018tng}.} The physics of nonlocal operators is most effectively probed by placing the corresponding physical system on a space with interesting topology.

Reconciling this with the theme of the first paragraph, necessitates an understanding of conformal field theory on topologically non-trivial spacetimes. This understanding is effectively complete for two-dimensional CFTs, where Moore and Seiberg \cite{Moore:1988qv} showed that in the rational case, modular covariance of torus one-point functions ensures that the CFT can be defined and solved on arbitrary Riemann surfaces. Central to this are
\begin{enumerate*}[label=(\roman*)]
	\item  the geometry of Riemann surfaces \cite{Friedan:1986ua}, and
	\item  the state-operator correspondence.
\end{enumerate*}
The idea can be summarised as follows. Any Riemann surface can be viewed as a collection of discs and pair-of-pants geometries sewn together along circles. Alternatively, one can insert a resolution of the identity operator along a given circle, expressed in terms of states on that circle. Using the state-operator correspondence these states are mapped to local operators at the centre of a disc bounding that circle. Thus, the computation of any arbitrary correlation function is reduced to the two- and three-point functions on the sphere and one-point functions on the torus.

In higher dimensions, significantly less is known.\footnote{See, however, \cite{Shaghoulian:2015kta,Belin:2016yll,Shaghoulian:2016gol,Horowitz:2017ifu,Belin:2018jtf,Luo:2022tqy,Allameh:2024qqp,lei2024modularity} for some of what is known.} While the main ideas are the same, technical difficulties arise. More explicitly, locality suggests that QFT observables should be reconstructable from basic building blocks via cutting and sewing. However, this process is much more involved than its two-dimensional counterpart. Therefore, while local operators still completely determine the spectrum of states on a spatial \((d-1)\)-sphere, this is insufficient for reconstructing the entire CFT. A natural question to ask is, then: how feasible is a state-operator correspondence that relates nonlocal operators to states on other spatial manifolds? Belin, de Boer, and Kruthoff \cite{Belin:2018jtf} attempt to answer this question for the three-torus and argue that such a correspondence is not straightforward for a generic CFT.

What is the situation when additional symmetries are available? Recent years have seen an incredible surge of interest in the study of symmetries in quantum field theory. Starting with higher-form symmetries \cite{Gaiotto:2014kfa}, and generalising onwards to higher-group, non-invertible symmetries and other generalisations,\footnote{See, e.g. \cite{Cordova:2022ruw} for a more complete list of references} these notions of symmetry provide powerful organising principles for quantum field theories. Thus, one may reformulate the question as follows: Is there a state-operator correspondence for nonlocal operators, in higher-dimensional CFTs with generalised global symmetries? In this paper, we provide an affirmative answer to this question.

More precisely we consider unitary CFTs in \(d=2p+2\) dimensions with continuous \(p\)-form symmetries (invertible or non-invertible). It turns out that this is a very powerful combination. The photonisation argument \cite{Hofman:2018lfz} relates them to circle or orbifold branches of theories of free \(p\)-forms. This further implies an infinite collection of codimension-one topological operators, labelled by chiral and anti-chiral \(p\)-forms. This is in complete analogy with two-dimensional CFTs, where holomorphic conserved currents can be dressed with arbitrary holomorphic functions, remaining conserved. This analogy goes further: the spectrum of states of such theories is organised by current algebras, generalising the Kac--Moody algebras of two-dimensional CFTs. We focus on the four-dimensional case, enjoying one-form symmetries, and realised by free Maxwell theory. We explicitly construct the representation theory of these algebras on generic closed spatial slices, \(\Sigma\). A major result is the complete characterisation of the space of states on a generic spatial topology and geometry. This Hilbert space consists of Kac--Moody descendants built on top of primary states, charged under the one-form symmetries. As a non-trivial check, the extended character of those representations matches exactly the partition function of Maxwell theory on \(\S^1_\beta\times\Sigma\), as obtained via path integral methods, and is given by strikingly simple formulas that are very reminiscent of two-dimensional rational CFTs:
\begin{equation}\label{eq:parti-Maxwell-intro}
	\parti_\t{Maxwell}\qty[\S^1_\beta\times\Sigma,\tt] = \frac{\Theta_\Sigma(q,\tt)}{\upeta_\Sigma(q)^2}, \qquad q=\ex{-\beta}.
\end{equation}
In the above, \(\Theta_\Sigma(q)\), defined in \cref{eq:theta-def}, generalises the Siegel--Narain Theta function and \(\upeta_\Sigma(q)\), defined in \cref{eq:eta-def}, generalises the Dedekind eta function, while \(\tt\) is the complexified coupling constant of Maxwell theory.\footnote{See also \cite{Fliss:2023uiv} for a related story in generic dimensions.} On \(\Sigma = \S^2\times\S^1\), we match this spectrum to states prepared by path integrals with insertions of line operators. This has two important consequences. Firstly, a classification of line operators in four-dimensional CFTs with global symmetries. There are \emph{primary}  operators, carrying charge under the one-form symmetry and preparing the highest-weight states of our current algebra. In the case of Maxwell theory these operators are given by Wilson--'t Hooft lines. Similarly to local operators, these primary line operators have a definite scaling weight, as defined in \cite{Kapustin:2005py} and discussed in  \cite{Verlinde:1995mz}. Then, there are \emph{descendant} line operators, obtained by dressing the Wilson--'t Hooft lines with photon modes. Relatedly, the second consequence, and our main result is a state-operator correspondence stating the following:
\vspace{-0.5em}
\begin{center}
	In four-dimensional CFTs with a continuous one-form symmetry, \\
	states on \(\S^{2}\times\S^1\) are in one-to-one correspondence with line operators on \(\R^3\times\S^1\).
\end{center}
\vspace{-0.5em}
Interestingly, we find that the lack of Weyl transformation from \(\R^3\times\S^1\) to the Lorentzian cylinder, \(\R\times\S^2\times\S^1\), coupled with the two polarisations of the photon, implies that radial evolution is equivalent to a \emph{squeezing transformation} of the photon states. This implies, in turn, that the vacuum state is not prepared by a path integral with no operator insertions, but rather, by a path integral with insertions of photon modes of all frequencies.

An organising summary is as follows. In \cref{sec:photonisation-higherKM} we explain the photonisation argument, and its implications. We show that invertible continuous \(p\)-form symmetries in \((2p+2)\)-dimensional CFTs lead to a higher-dimensional generalisation of Kac--Moody algebras. We supplement this with a discussion on non-invertible continuous \(p\)-form symmetry, where we construct a non-invertible generalisation of these current algebras. In \cref{sec:local state-op} we focus on the case (\(p=0\)) of two-dimensional CFTs, where we review the well-known state-operator correspondence for the free compact scalar, in terms of its organising Kac--Moody algebra. We then jump, in \cref{sec:4d}, to the case of \(p=1\) and free Maxwell theory, detailing the path integral on generic closed manfiolds. We then quantise the theory using our current algebras, matching it to the path integral expressions. This allows us to reach \cref{sec:nonlocal state-op}, where we set up our state-operator correspondence on \(\S^2\times\S^1\), by performing a path integral on \(\B^3\times\S^1\) with line operator insertions. To that end, we first classify the line operators of the theory in terms of our current algebra. We then perform the radial evolution on \(\B^3\times\S^1\), and explain how it leads to squeezing the photon states, to finally land directly onto the nonlocal state-operator correspondence. We finish, in \cref{sec:discussion} with a discussion on our results and interesting future directions. In \cref{app:Laplacian,app:KM-simple-spectrum,app:matrices} we collect details regarding the spectral analysis of the Hodge Laplacian, the current algebra on a generic manifold, and the radial evolution, respectively.

\section{Photonisation and higher-dimensional current algebras}\label{sec:photonisation-higherKM}

In this section we explore the photonisation argument of \cite{Hofman:2018lfz} and its various incarnations. We first show that a unitary conformal field theory in \(2p+2\) dimensions with a continuous \(p\)-form symmetry (when \(p=0\), we restrict to abelian \(0\)-form symmetries) has a realisation as a theory of free \(p\)-forms. We go on to show that this gives rise to a  current algebra, akin to the two-dimensional Kac--Moody algebras. We comment on non-abelian versions of the arguments for the \(p=0\) case. Finally, we derive non-invertible current algebras, for theories enjoying continuous non-invertible symmetries and comment on their applications.

\subsection{Photonisation}\label{ssec:photonisation}

Our starting point is a unitary conformal field theory in \(d=2p+2\) dimensions, with a continuous, invertible \(p\)-form symmetry. We focus on the case of a single \(\gf{\U(1)}{p}\) symmetry.\footnote{We denote a \(p\)-form symmetry group as \(\gf{\U(1)}{p}\).} Higher-form symmetries are abelian, so for \(p\geq 1\), this is all we can have.\footnote{We can also have a product of decoupled \(\gf{\U(1)}{p}\)s; the generalisation is trivial.} For \(p=0\) we can also have non-abelian symmetries; we comment on those in \cref{ssec:nonabelian}. The continuous symmetry induces a conserved \((p+1)\)-form current, \(\f{J}{p+1} \coloneqq J_{\mu_1 \cdots \mu_{p+1}}(x) \dd{x^{\mu_1}}\w\cdots\w\dd{x^{\mu_{p+1}}}\), satisfying
\begin{equation}\label{eq:conservation}
	\dd\star \f{J}{p+1} = 0.
\end{equation}
This current leads to a conserved charge, supported on a codimension-\((p+1)\) manifold
\begin{equation}
	Q[\Sigma_{d-p-1}] \coloneqq \int_{\Sigma_{d-p-1}} \star\f{J}{p+1},
\end{equation}
or equivalently a codimension-\((p+1)\) topological operator
\begin{equation}
	U_{\Sigma_{d-p-1}}(\alpha) \coloneqq \exp(\ii \alpha \int_{\Sigma_{d-p-1}} \star\f{J}{p+1}).
\end{equation}
Since this is a compact \(\U(1)\) symmetry --- as opposed to non-compact \(\U(1)\), i.e. \(\R\) ---, the parameter \(\alpha\) is circle-valued: \(\alpha\sim \alpha+2\pi\).

Let us now see the implications of conformality, reviewing the argument of \cite{Hofman:2018lfz}. We start by discussing the theory on Euclidean flat space, \(\R^d\), with metric \(g_{\mu\nu}=\delta_{\mu\nu}\). The \((p+1)\)-form currents are primary operators of scaling dimension \(p+1\), so their two-point function\footnote{Here and in the following we write the correlation functions of products of currents in differential form languange to suppress indices. This is not a wedge product of forms, it should rather be seen as a form-valued form, or in other words a section of \(\bigwedge^{\blob}\ctb \R^d \otimes \bigwedge^{\blob}\ctb \R^d\).} is completely fixed by conformal symmetry, up to a constant, \(\k\) \cite{Costa:2014rya}:
\begin{equation}
	\ev{\f{J}{p+1}\qty(x_1)\ \f{J}{p+1}\qty(x_2)} = \frac{\k}{\norm{x_{12}}^d}\;\mathrm{G}\qty(x_{12}),
\end{equation}
where \(x_{12}\coloneqq x_1-x_2\) and \(\mathrm{G}\qty(x_{12})\) is a uniquely determined tensor structure:
\begin{equation}
	\mathrm{G}^{\mu_1\cdots\mu_{p+1}\nu_1\cdots\nu_{p+1}}(x_{12}) = \delta^{\mu_1}_{[\rho_1}\cdots \delta^{\mu_{p+1}}_{\rho_{p+1}]}\delta^{\nu_1}_{[\sigma_1}\cdots \delta^{\nu_{p+1}}_{\sigma_{p+1}]} \prod_{\ell=1}^{p+1}\qty(g^{\rho_\ell \sigma_\ell}-2 \frac{\qty(x_{12})^{\rho_\ell}\qty(x_{12})^{\sigma_\ell}}{\norm{x_{12}}^2}),
\end{equation}
where \(g_{\rho\sigma}\) is the flat Euclidean metric and \(g^{\rho\sigma}\) its inverse. From here on, it is easy to show that
\begin{equation}
	\ev{\dd{\f{J}{p+1}(x_1)}\ \f{J}{p+1}(x_2)} = 0,
\end{equation}
which immediately implies
\begin{equation}
	\ev{\dd{\f{J}{p+1}(x_1)}\ \dd{\f{J}{p+1}(x_2)}} = 0.
\end{equation}
Using the standard state-operator correspondence, this represents the norm of a state on \(\cH_{\S^{d-1}}\), created by the local operator \(\dd{\f{J}{p+1}}=0\). Since this is a unitary CFT, this norm can be zero only if the operator \(\dd{\f{J}{p+1}}\) is zero itself. Said differently, conformal invariance and unitarity implies that there is a dual conserved current, \(\star \f{J}{p+1}\), i.e.:
\begin{equation}
	\dd\f{J}{p+1}=0.
\end{equation}

An equivalent, implicit argument, follows the reasoning of \cite{leeCommentsCompatibilityConformal2021}. All the states in a conformal field theory are organised in representations of the Cartan subalgebra of the Euclidean conformal algebra, \(\mathfrak{so}(d+1,1)\). This results in each state, \(\ket{\cO}\in\cH_{\S^{d-1}}\), being labelled by its scaling dimension, \(\Delta_\cO\), and a set of highest weights, \(\set{h_{1},h_{2},\ldots,h_{\lfloor d/2 \rfloor}}_\cO\), of the \(\mathfrak{so}(d)\) irreducible representation. Let us focus on the primary states; the descendants can be obtained by acting with the ladder operators of the conformal algebra. Unitarity, i.e. \(\braket{\cO}{\cO}\geq 0\) for all primary states \(\cO\), imposes that \(\Delta_\cO \geq f\qty(\set{h_i}_\cO)\), for some function \(f\), that depends on the dimension, \(d\), and the \(\mathfrak{so}(d)\) representation \cite{Minwalla:1997ka}.

Now, consider the state \(\ket*{\f{J}{p+1}}\), corresponding to a conserved current. Its scaling dimension is \(\Delta_J=d-p-1\). Moreover, the conservation equation \(\dd\star\f{J}{p+1}=0\), implies that this state belongs in a short conformal multiplet, since its first descendant is null. As such, the above unitarity bound, \(\Delta_J \geq f\qty(\set{h_i}_J)\), must be saturated on this state. In dimensions \(d=2p+2\), the conserved currents lie in reducible representations of \(\mathfrak{so}(d)\): they can be decomposed into self-dual and anti-self-dual currents, \(\f{J}{p+1}^\pm \coloneqq \half\qty(\f{J}{p+1}\pm\ii^{p-1}\star \f{J}{p+1})\), each of which is in an irreducible representation. To saturate the unitarity bound for \(\ket*{\f{J}{p+1}}\), the unitarity bound for both \(\ket*{\f{J}{p+1}^\pm}\) must be saturated. Therefore both \(\ket*{\f{J}{p+1}^\pm}\) lie in short multiplets, which in turn implies that \(\dd\star\f{J}{p+1}^\pm=0\), or in other words, one has \(\dd\f{J}{p+1}=0\), on top of \(\dd\star\f{J}{p+1}=0\).

To summarise, what both of the above arguments show is that a unitary \((2p+2)\)-dimensional CFT with a continuous \(p\)-form symmetry, must also have a dual \(p\)-form symmetry. In other words, it has:
\begin{equation}
	\dd\star\f{J}{p+1} = 0 \qq{and} \dd\f{J}{p+1} = 0.
\end{equation}
This is nothing but the equations of motion of a \(p\)-form abelian gauge field, \(\f{a}{p}\),  upon identifying \(\f{J}{p+1}\) with the field strength, \(\f{f}{p+1}\), of \(\f{a}{p}\). In other words, unitary \((2p+1)\)-dimensional CFTs with a \(\gf{\U(1)}{p}\) \(p\)-form symmetry can always be realised by \(p\)-form Maxwell theories. In the words of \cite{Hofman:2018lfz} we say that the CFT \emph{photonises}.

\subsection{Abelian current CFTs}\label{ssec:abelian current CFTs}

We will now show that the symmetry is actually enhanced even more, into a current algebra --- a higher-dimensional version of the two-dimensional Kac--Moody algebras. This algebra will turn out to be spectrum-generating (up to decoupled neutral dressing) and it will eventually be key to reaching the main result of this paper in \cref{sec:nonlocal state-op}. We will call CFTs with such a current algebra \emph{current CFTs}. We will mainly analyse abelian current CFTs, but we will comment on non-abelian and non-invertible generalisations in \cref{ssec:nonabelian,ssec:noninvertible}.

We go temporarily to a more general setup and relax conformality. Consider a \((d=2p+2)\)-dimensional Euclidean quantum field theory, with two, dual, \(\gf{\U(1)}{p}\) \(p\)-form symmetries, i.e. two conserved currents \(\f{J}{p+1}\), \(\f{\star J}{p+1}\):
\begin{equation}\label{eq:J *J conservation}
	\dd\star\f{J}{p+1} = 0 \qq{and} \dd\f{J}{p+1} = 0.
\end{equation}
The space of \((p+1)\)-forms, \(\Omega^{p+1}(X)\), admits a \(\Z_2\) grading,  \(\Omega^{p+1}(X)=\Omega^{p+1}_+(X)\oplus\Omega^{p+1}_-(X)\), graded by the Hodge-star operator, or more precisely, by the operator \(\ii^{1-p}\star\). Namely, there exist projectors\footnote{Recall that we are in Euclidean signature; in Lorentzian signature there would be an extra \(\ii\).}
\begin{equation}\label{eq:proj}
	\bbP_\pm \coloneqq \half \qty(\id\pm \ii^{1-p}\,\star),
\end{equation}
that allow us to write any \((p+1)\)-form, \(\f{\omega}{p+1}\), as \(\bbP_+ \f{\omega}{p+1}+\bbP_- \f{\omega}{p+1}\), where \(\bbP_\pm \f{\omega}{p+1}\) satisfy
\begin{equation}\label{eq:SDeq}
	\ii^{1-p}\star \bbP_\pm \f{\omega}{p+1} = \pm \bbP_\pm \f{\omega}{p+1}.
\end{equation}
We will call equation \cref{eq:SDeq}, self-duality equation and will refer to \(\bbP_\pm \f{\omega}{p+1}\) as (anti-)self-dual forms, in all dimensions, even though, technically, they are (anti-)self-dual only when \(p\) is odd, or equivalently in dimensions \(d=4k\). In \(d=4k+2\), corresponding to even \(p\), they are just the eigenforms of the operator \(\ii\star\).

An important property of the projectors \cref{eq:proj} is that the can pass through wedge products of \((p+1)\)-forms:
\begin{equation}
	\qty(\bbP_\pm \f{\omega}{p+1})\w \f{\eta}{p+1} = \f{\omega}{p+1}\w\qty(\bbP_\pm \f{\eta}{p+1}).
\end{equation}
To see that, it suffices to note that \(\star\f{\omega}{p+1}\w\f{\eta}{p+1}\) is a \((2p+2)\)-form and hence is proportional to the volume form. The simultaneous conservation, \cref{eq:J *J conservation}, of \(\f{J}{p+1}\) and \(\f{\star J}{p+1}\) is equivalent to the conservation of
\begin{equation}\label{eq:self-dual-current}
	\f{J}{p+1}^\pm \coloneqq \bbP_\pm \f{J}{p+1},
\end{equation}
i.e.
\begin{equation}
	\dd\f{J}{p+1}^\pm = 0.
\end{equation}

Once we have these two conserved currents, we can construct an infinite number of conserved charges. In particular, note that the family of currents
\begin{equation}
	\cJ_\Lambda^\pm \coloneqq -\star\qty(\f{J}{p+1}^\pm \w \f{\Lambda}{p}^{\mp})
\end{equation}
is conserved for any (anti)chiral \(p\)-form, i.e. for any \(\f{\Lambda}{p}^{\mp}\) such that
\begin{equation}\label{eq:chiral-forms}
	\bbP_\pm \dd \f{\Lambda}{p}^{\mp} = 0.
\end{equation}
These can be thought of as \(p\)-form analogues of (anti)holomorphic functions.\footnote{Equation \cref{eq:chiral-forms} is a local definition, like the defining equations of (anti)holomorphic functions.} Indeed,
\begin{equation}
	\dd\star \cJ_\Lambda^\pm = \f{J}{p+1}^\pm \w \dd\f{\Lambda}{p}^{\mp} = \f{J}{p+1} \w \bbP_\pm\dd\f{\Lambda}{p}^{\mp} = 0.
\end{equation}
Integrating on a codimension-one closed manifold, \(\Sigma_{d-1}=\Sigma_{2p+1}\),  gives us two families of conserved charges:
\begin{equation}
	Q_\Lambda^\pm\qty[\Sigma_{d-1}] \coloneqq \int_{\Sigma_{d-1}} \f{J}{p+1}^\pm \w \f{\Lambda}{p}^{\mp},
\end{equation}
or equivalently topological operators,
\begin{equation}\label{eq:ULambda}
	U_{\Sigma_{d-1}}^\pm(\Lambda) \coloneqq \exp(\ii \int_{\Sigma_{d-1}} \f{J}{p+1}^\pm \w \f{\Lambda}{p}^{\mp}).
\end{equation}
Here we mention two important points. First, note that these are zero-form symmetries, regardless of \(p\). As such, they can a priori, have non-trivial commutators. They stem, however from an abelian symmetry, therefore their commutation relations, if non-trivial, indicate a central extension. We will indeed find such a central extention in the next paragraphs. Second, note that there is a gauge redundancy in defining the conserved charges. Namely, we must identify \(\f{\Lambda}{p}^\mp \sim \f{\Lambda}{p}^\mp + \f{\lambda}{p}^\mp\), where \(\dd\f{\lambda}{p}^\mp=0\), since the shift by a closed \(p\)-form leaves the charges invariant.

Now we reinstate conformality. In \cref{ssec:photonisation} we saw that a CFT with these two conserved currents obeys the dynamics of \(p\)-form Maxwell theory. We can therefore, locally, realise the currents as the curvature of a \(p\)-form gauge field \(\f{J}{p+1} = \dd{\f{a}{p}}\), whose equation of motion is a free wave equation \(\cdd\dd\f{a}{p}=0\). We then recognise \(\f{a}{p}\) and \(\star\f{J}{p+1}\), restricted to a codimension-one slice, as conjugate phase-space variables, giving rise to a (pre-)symplectic form:
\begin{equation}
	\Omega_{\Sigma_{d-1}} = \int_{\Sigma_{d-1}} \var{\cmom}\w \var{\cpos},
\end{equation}
identifying
\begin{equation}\label{eq:ccoords}
	\cpos=\eval{\f{a}{p}}_{\Sigma_{d-1}} \qq{and} \cmom= 2\k \eval{\qty(\star \f{J}{p+1}^\pm)}_{\Sigma_{d-1}}.
\end{equation}
In the above, the factor \(2\k\) is conventional, and is related to the strength of the interaction, i.e. the electric charge of the \(p\)-form Maxwell theory realising the CFT. In fact, we should identify as the canonical position, \(\cpos\), a conjugacy class, or a specific representative thereof, of the equivalence relation \(\f{a}{p}\sim\f{a}{p}+\f{\lambda}{p}\), for \(\f{\lambda}{p}\) flat. This is what takes care of the degeneracy of the symplectic form, corresponding to gauge transformations. We will proceed with \cref{eq:ccoords}, keeping this issue in mind.

We can then compute the commutation relations between the charges, simply by plugging the vector fields they generate on the phase space in the definition of the symplectic form. First we compute
\begin{equation}
	\begin{aligned}
		\fdv{Q_\Lambda^\pm}{\cpos} = -\Half\dd{\f{\Lambda}{p}^\mp} \qq{and}
		\fdv{Q_\Lambda^\pm}{\cmom} = \pm\ii^{1-p}\k\, \f{\Lambda}{p}^\mp.
	\end{aligned}
\end{equation}
With this, we immediately get the algebra:
\begin{empheq}[box=\obox]{equation}\label{eq:p-KM}
	\comm{Q_{\Lambda_1}^\sigma}{Q_{\Lambda_2}^{\sigma'}}_{\Sigma_{d-1}} = \frac{\ii^{1-p}\k}{2}\qty(\sigma+(-1)^p \sigma')\int_{\Sigma_{d-1}} {\Lambda_1^{-\sigma}}\w\dd{\Lambda_2^{-\sigma'}},
\end{empheq}
where \(\sigma\) and \(\sigma'\) are signs, \(\sigma,\sigma'\in\set{+,-}\). Here and in the following, we drop the subscript \([p]\) indicating the form-degree of \(\Lambda\) to declutter the notation. This algebra is a higher-dimensional generalistion of the familiar two-dimensional Kac--Moody algebra, and it will be one of the protagonists of the story that follows. To better illustrate \cref{eq:p-KM} let us elaborate further on two specific values of \(p\).

\paragraph{\(p=0\).} This case corresponds to the case of a two-dimensinal CFT, with a (zero-form) \(\U(1)\) symmetry. There, \(\Lambda_{i}^\pm\) are just scalar functions and the above commutators become,
\begin{equation}
	\begin{aligned}
		\comm{Q_{\Lambda_1}^\pm}{Q_{\Lambda_2}^\pm}_{\Sigma_{1}} & = \pm\ii \k \int_{\Sigma_1} \Lambda_1^{\mp}\ \dd{\Lambda_2^{\mp}} \\
		\comm{Q_{\Lambda_1}^+}{Q_{\Lambda_2}^-}_{\Sigma_{1}} & = 0.
	\end{aligned}
\end{equation}
This is just the familiar \(\widehat{\fu}(1)\times\widehat{\fu}(1)\) Kac--Moody algebra. Tracing back the logical tower that led us here, and running the arguments again, in language familiar from two-dimensional CFTs, we recover the following familiar statement. A conserved current, in a unitary two-dimensional CFT, can always be split into a holomorphic and an antiholomorphic piece, which are separately conserved. These can, in turn, be dressed with arbitrary holomorphic and antiholomorphic functions to give rise to a holomorphic and an antiholomorphic current algebra. Quantising on the theory on a spatial circle and expanding the currents in Fourier modes, gives rise to the usual abelian affine Kac--Moody algebra. Indeed, taking \(\Sigma_1\) to be \(\S^1\) and expanding in Fourier modes, \(\ex{\pm \ii \sfn \theta}\), gives the familiar form of the \(\widehat{\fu}(1)\times\widehat{\fu}(1)\) Kac--Moody algebra:
\begin{equation}\label{eq:2d-KM}
	\begin{aligned}
		\comm{J^\pm_\sfn}{J^\pm_\sfm} & = \k\,\sfn\, \delta_{\sfn+\sfm,0},      \\
		\comm{J^+_\sfn}{J^-_\sfm} & = 0, \qquad\qquad\qquad \sfn,\sfm\in\Z.
	\end{aligned}
\end{equation}
Finally, recall that all these CFTs have a free-field realisation as a free compact scalar, \(\phi\sim \phi+2\pi R\). The level, \(\k\), of the algebra is related to the radius of the scalar.

\paragraph{\(p=1\).} This is the case of a four-dimensional CFT with two, dual, one-form symmetries. As we have argued before, this is realised by regular Maxwell theory. Here \(\Lambda_i^\pm\) are one-forms, and the charges are defined on a three-dimensional manifold. The commutation relations become, in this case,
\begin{equation}
	\begin{aligned}
		\comm{Q_{\Lambda_1}^+}{Q_{\Lambda_2}^+}_{\Sigma_{3}} & = 0  = \comm{Q_{\Lambda_1}^-}{Q_{\Lambda_2}^-}_{\Sigma_{3}}                           \\
		\comm{Q_{\Lambda_1}^+}{Q_{\Lambda_2}^-}_{\Sigma_{3}} & = \k \int_{\Sigma_3} \Lambda_1^{-}\w \dd{\Lambda_2^{+}}. \label{eq:1-KM}
	\end{aligned}
\end{equation}
We recognise again the structure of a \(\widehat{\fu}(1)\times\widehat{\fu}(1)\) current algebra, with the important difference that here the central extension mixes the two factors, instead of acting on each of them separately. Observing the pattern set by \cref{eq:p-KM} it is clear that in \(4n+2\) dimensions the chiral and antichiral components will go their own, centrally extended, way separately, whereas in \(4n\) dimesions, they mix. This has to do with the existence of real self- and anti-self-dual forms, as noted earlier. Let us note that, just like the two-dimensional case, we can expand the currents in \textquote{modes,} to obtain an algebra of the individual modes. We will do so in \cref{ssec:current-Maxwell}. Moreover, the level, \(\k\), of the algebra is now related to the coupling of the Maxwell theory that realises the CFT, i.e. the electric charge. Finally, in this case, the algebra \cref{eq:1-KM} was constructed in \cite{Hofman:2018lfz}, where it was also arrived at through a twistor formalism, without appealing to the phase space structure of Maxwell theory.

Going back to the general case, let us mention yet another presentation of the current algebra, \cref{eq:1-KM}, that will be useful to obtain a non-invertible version thereof in \cref{ssec:noninvertible}. This presentation is in terms of the topological operators, that act on the Hilbert space. A generic topological operator in a current CFT takes the form
\begin{equation}\label{eq:ULambda-general}
	U\qty(\Lambda_1^-,\Lambda_2^+) \coloneqq \exp(\ii \int_{\Sigma_{d-1}} \f{J}{p+1}^+\w \Lambda_1^{-} + \ii \int_{\Sigma_{d-1}} \f{J}{p+1}^-\w \Lambda_2^{+}),
\end{equation}
where we supperss the (topological) dependence on \(\Sigma_{d-1}\). The fusion of two such operators is
\begin{equation}\label{eq:p-KM-exp}
	U\qty(\Lambda_1^-,\Lambda_2^+)\otimes U\qty(\Lambda_3^-,\Lambda_4^+) = f_p^{(\k)}\qty(\Lambda_1^-,\Lambda_2^+,\Lambda_3^-,\Lambda_4^+)\ U\qty(\Lambda_1^-+\Lambda_3^-,\Lambda_2^++\Lambda_4^+),
\end{equation}
where
\begin{align}\label{eq:central-function}
	f_p^{(\k)}\qty(\Lambda_1^-,\Lambda_2^+,\Lambda_3^-,\Lambda_4^+) \coloneqq \exp\!\Bigg( &\frac{\ii^{p+1}\, \k}{2}\int_{\Sigma_{d-1}} \left\{\,\delta_{p,\t{even}}\;\qty(\Lambda_1^- \w \dd\Lambda_3^-+\Lambda_2^+ \w \dd\Lambda_4^+)\right. \nonumber \\                                                                                   &~-\left. \delta_{p,\t{odd}}\; \qty(\Lambda_1^- \w \dd\Lambda_4^++\Lambda_2^+ \w\dd\Lambda_3^-)\right\}\Bigg).
\end{align}
This is a simple abelian fusion rule, with a central extension, proportional to \(\k\),  specified by the function \(f_p^{(\k)}\qty(\Lambda_1^-,\Lambda_2^+,\Lambda_3^-,\Lambda_4^+) \). As before, we will illustrate \cref{eq:p-KM-exp} in the cases \(p=0\) and \(p=1\).

At \(\boldsymbol{p=0}\), there is a central term when fusing topological operators with the same chirality:
\begin{equation}
	U\qty(\Lambda_1^-,0)\otimes  U\qty(\Lambda_3^-,0) = \exp(\frac{\ii\, \k}{2}\int_{\Sigma_1}\Lambda_1^- \w \dd\Lambda_3^-)\ U\qty(\Lambda_1^-+\Lambda_3^-,0)
\end{equation}
and similarly for \(\Lambda_2^+\) and \(\Lambda_4^+\). Operators with opposite chirality do not see each other:
\begin{equation}
	U\qty(\Lambda_1^-,0)\otimes  U\qty(0,\Lambda_4^+) = U\qty(\Lambda_1^-,\Lambda_4^+)
\end{equation}
and similarly for \(\Lambda_2^+\) and \(\Lambda_3^-\). This is, of course, a reflection of the fact we mentioned above, that in two dimensional CFTs the hoolomorphic and antiholomorphic sectors do not see each other in the central extension.

Coming to \(\boldsymbol{p=1}\), we see the central extension in the fusion of operators of opposite chirality:
\begin{equation}
	U\qty(\Lambda_1^-,0)\otimes  U\qty(0,\Lambda_4^+) = \exp(\frac{\k}{2} \int_{\Sigma_3} \Lambda_1^-\w\dd{\Lambda_4^+})U\qty(\Lambda_1^-,\Lambda_4^+),
\end{equation}
while those with the same chirality have a simple abelian fusion:
\begin{equation}
	U\qty(\Lambda_1^-,0)\otimes  U\qty(\Lambda_3^-,0) = U\qty(\Lambda_1^-+\Lambda_3^-,0),
\end{equation}
and similar equations involving the rest of the operators.

Let us also comment on the relation of \cref{eq:p-KM}, with other appearances of higher-dimensional generalisations of Kac--Moody algebras in the literature. As explained above, \cref{eq:p-KM} is the natural generalisation of the four-dimensional Kac--Moody algebra of \cite{Hofman:2018lfz}. Moreover, it complements (and coincides with, in some cases) the current algebras that organise the edge-modes of topological field theories \cite{Fliss:2023uiv}. However, it is not the same as Mickelsson--Faddeev algebra \cite{Mickelsson:1982zv,Mickelsson:1983xi,Faddeev:1984jp}, and by extension also not the same as the higher-dimensional loop algebras of \cite{Cederwall:1993de}. Moreover it is also different from the higher Kac--Moody algebras of \cite{faonteHigherKacMoodyAlgebras2019,Gwilliam:2018lpo}.

\subsection{Non-abelian current CFTs}\label{ssec:nonabelian}

We only comment very briefly on non-abelian current CFTs, as we have nothing to add besides what is already well-known. This case can only occur at \(p=0\), since \((p\geq 1)\)-form symmetries are abelian. So, consider a two-dimensional CFT with a conserved one-form current \(\f{J}{1}\), valued in the Lie algebra, \(\fg\), of a semi-simple Lie group \(G\). The argument of \cref{ssec:photonisation} goes through, and shows that \(\f{\star J}{1}\) must also be conserved. The consequence is that this time instead of photonisation, it corresponds to non-abelian bosonisation \cite{Witten:1983ar}. Namely, the conservation of the currents corresponds to the equations of motion of a Wess--Zumino--Witten (WZW) model. We can locally write the current \(\f{J}{1}\) as \(\inv{g}\dd g\), for \(g\) a \(G\)-valued scalar and the conservation of the current is
\begin{equation}
	\dd\star\qty(\inv{g}\dd g) = 0.
\end{equation}
From here on, it is a standard exercise in two-dimensional CFT (see e.g. \cite{DiFrancesco:1997nk}) to derive the associated current algebra, which will naturally be the the affine Kac--Moody algebra \(\widehat{\fg}_\k\otimes\widehat{\fg}_\k\).

\subsection{Non-invertible current CFTs}\label{ssec:noninvertible}

A different option is to consider current CFTs whose underlying symmetry is non-invertible. The simplest way to obtain a continuous non-invertible symmetry is to begin with the setup of \cref{ssec:photonisation} and impose the equivalence relation \(\f{J}{p+1}\sim \f{-J}{p+1}\). From the point of view of the photonised theory, this corresponds to gauging the \(\Z_2\) charge-conjugation symmetry \(\f{A}{p}\mapsto \f{-A}{p}\). This corresponds to the orbifold branch of \(p\)-form Maxwell theory, or equivalently, an \(\O(2)=\U(1)\rtimes\Z_2\) \(p\)-form gauge theory. Variants of this theory (mostly for \(p=0\) and \(p=1\)), and its (non-invertible) symmetries have been subject of intense study in recent years \cite{Chang:2020imq,Thorngren:2021yso,Heidenreich:2021xpr,Nguyen:2021yld,Bhardwaj:2022yxj,Antinucci:2022eat,Damia:2023gtc}.

Let us first review the non-invertible symmetries of \(\O(2)\) \(p\)-form gauge theory. It is clear that the operator
\begin{equation}
	U(\alpha)=\exp(\ii \alpha\int_{\Sigma_{d-p-1}} \star\f{J}{p+1}),
\end{equation}
albeit still topological, is no longer gauge-invariant, for generic values of \(\alpha\). Note that \(\alpha\) is no longer circle-valued, but rather it is valued in the segment \(\alpha\sim -\alpha \sim \alpha+2\pi\). In what follows, it will be implictly assumed that we take a representative of the equivalence classes defined by the above equivalence relations in \(\closed{0}{\pi}\).  The operator \(U(0) = \mathbf{1}_{d-p-1}\), is the identity \((d-p-1)\)-dimensional operator and is, naturally, gauge-invariant. Moreover,
\begin{equation}
	U(\pi) = \exp(\ii \pi \int_{\Sigma_{d-p-1}} \star\f{J}{p+1}) \eqqcolon (-\mathbf{1})_{d-p-1}
\end{equation}
is also gauge-invariant and gnerates a \(\Z_2^{[p]}\) \(p\)-form symmetry. For the rest we can construct topological and gauge-invariant operators by taking a direct sum of the original topological operators. Explicitly, the operators
\begin{equation}
	D(\alpha) \coloneqq U(\alpha)\oplus \inv{U(\alpha)},
\end{equation}
for \(\alpha\in\open{0}{\pi}\), pass all the tests of being a symmetry of the theory.\footnote{Provided that there are states on which they act non-trivially. In this case there are: e.g. gauge invariant sums of the original \(\U(1)\) Wilson operators \cite{Heidenreich:2021xpr}.} What was sacrificed, is obviously invertibility. This is reflected on the fusion rules:
\begin{equation}\label{eq:noninv-fusion}
	D(\alpha)\otimes D(\beta) = D(\alpha+\beta)\oplus D(\alpha-\beta),
\end{equation}
for \(\alpha\neq \beta\neq \pi-\beta \in\open{0}{\pi}\). Moreover, these operators have quantum dimension two, namely, \(\ev{D(\alpha)}_{\S^{p+1}}=2\), in contrast to the invertible topological operators, which have quantum dimension one.

Furthermore, as was explained in \cite{Heidenreich:2021xpr}, since charge-conjugation is a zero-form symmetry, after gauging there is a Pontryagin-dual (or quantum) \(\Z_2^{[2p]}\)  \(2p\)-form symmetry (\(d-0-2=2p\)), generated by the \(\Z_2\) charge-conjugation Wilson lines: \(\mathbf{1}_1\) and \((-\mathbf{1})_1\). While their fusion rules are simply \(\Z_2^{[2p]}\) fusion rules:
\begin{equation}
	(-\mathbf{1})_1\otimes(-\mathbf{1})_1 = \mathbf{1}_1,
\end{equation}
they can also dress the \((d-p-1)\)-dimensional operators we discussed above, and appear in their fusion rules. The operator \((-\mathbf{1})_1\) is known as a \textquote{determinant line}.

Therefore, the full set of codimension-\((p-1)\) topological operators in \(\O(2)\) \(p\)-form gauge theory is
\begin{equation}
	\set{\mathbf{1}_{d-p-1},\; (-\mathbf{1})_{d-p-1},\; \mathbf{1}_{d-p-1}^{(-\mathbf{1})_1},\; (-\mathbf{1})_{d-p-1}^{(-\mathbf{1})_1},\; D(\alpha),\; D(\alpha)^{(-\mathbf{1})_1} \suchthat \alpha\in\open{0}{\pi}},
\end{equation}
where the superscript \((-\mathbf{1})_1\) indicates dressing with the determinant line. The rest of the fusion rules follow by reconciling \cref{eq:noninv-fusion} with the allowed dressings \cite{Heidenreich:2021xpr,Bhardwaj:2022yxj}. These are:
\begin{equation}\label{eq:noninv-fusion-rest}
	\begin{aligned}
		(-\mathbf{1})_{d-p-1}\otimes (-\mathbf{1})_{d-p-1} & = \mathbf{1}_{d-p-1},                                                                                                                                                \\[0.5em]
		D(\alpha)\otimes (-\mathbf{1})_{d-p-1}             & = D(\pi-\alpha),                                                                                                                                                     \\[0.5em]
		D(\alpha)\otimes D(\alpha)                         & = \mathbf{1}_{d-p-1} \oplus \mathbf{1}_{d-p-1}^{(-\mathbf{1})_1} \oplus D(2\alpha),   \qquad                                              & \alpha\neq \frac{\pi}{2} \\[0.5em]
		D(\alpha)\otimes D(\pi-\alpha)                     & = (-\mathbf{1})_{d-p-1} \oplus (-\mathbf{1})^{(-\mathbf{1})_1}_{d-p-1} \oplus D(2\alpha-\pi),        \qquad                               & \alpha\neq \frac{\pi}{2} \\[0.5em]
		D(\pi/2)\otimes D(\pi/2)                           & = \mathbf{1}_{d-p-1}\oplus \mathbf{1}_{d-p-1}^{(-\mathbf{1})_1}\oplus(-\mathbf{1})_{d-p-1}\oplus (-\mathbf{1})_{d-p-1}^{(-\mathbf{1})_1}.
	\end{aligned}
\end{equation}

Now, seeing this from the lens of conformal field theory, we can easily repeat the argument of \cref{ssec:abelian current CFTs}. If the theory before gauging was a CFT, we have that \(\dd\f{J}{p+1}^\pm=0\). After gauging, the operators
\begin{align}
	D^\pm(\alpha) & \coloneqq U^\pm(\alpha)\oplus \inv{U^\pm(\alpha)}, \qq{with}                                       \\[0.8em]
	U^\pm(\alpha) & \coloneqq \exp(\ii \alpha\int_{\Sigma_{d-p-1}} \f{J^\pm}{p+1}), \qquad \alpha\in\open{0}{\pi}
\end{align}
are topological operators of the \(\O(2)\) \(p\)-form gauge theory. Their fusion rules follow from \cref{eq:noninv-fusion,eq:noninv-fusion-rest}. Moreover, there are non-invertible analogues of \cref{eq:ULambda,eq:ULambda-general}. These are
\begin{equation}
	D\qty(\Lambda_1^-, \Lambda_2^+) \coloneqq U\qty(\Lambda_1^-, \Lambda_2^+)\oplus U\qty(\Lambda_1^-, \Lambda_2^+)^{-1},
\end{equation}
with \(\Lambda_{1,2}^\pm\) (anti-)chiral \(p\)-forms, taking values in the open interval \(\open{0}{\pi}\). Their fusion rules define a non-invertible current algebra:
\begin{empheq}[box=\obox]{equation}\label{eq:noninv-KM}
	\begin{aligned}
	D\qty(\Lambda_1^-, \Lambda_2^+)\otimes D\qty(\Lambda_3^-, \Lambda_4^+) &= f_p^{(\k)}\ D\qty(\Lambda_1^-+\Lambda_3^-, \Lambda_2^++\Lambda_4^+) \\
	&\oplus\inv{\qty(f_p^{(\k)})}\ D\qty(\Lambda_1^--\Lambda_3^-, \Lambda_2^+-\Lambda_4^+),
	\end{aligned}
\end{empheq}
where \(f_p^{(\k)}=f_p^{(\k)}\qty(\Lambda_1^-,\Lambda_2^+,\Lambda_3^-,\Lambda_4^+)\), given by \cref{eq:central-function}. Note that this fusion rule is valid on generic chiral forms, \(\Lambda_i^\pm\). At special points, i.e. when the \(\Lambda_i^\pm\) coincide at a point, or differ by \(\pi\), \cref{eq:noninv-KM} should be understood with the operators dressed appropriately, as in \cref{eq:noninv-fusion-rest}. Let us now illustrate and interpret this non-invertible current algebra in the cases \(p=0\) and \(p=1\).

\paragraph*{\(\boldsymbol{p=0.}\)} At \(p=0\) we find ourselves on the orbifold branch of the two-dimensional compact boson CFT. The fusion ring \cref{eq:noninv-KM} splits, again, into holomorphic and antiholomorphic and we have, for example:
\begin{align}\label{eq:noninv-KM-2d}
	D\qty(\Lambda_1^-, 0)\otimes D\qty(\Lambda_3^-,0) &=
	\exp(\frac{\ii\, \k}{2}\int_{\Sigma_1}\Lambda_1^- \w \dd\Lambda^-_3) D\qty(\Lambda_1^-+\Lambda_3^-, 0) \nn 
	&\oplus \exp(-\frac{\ii\, \k}{2}\int_{\Sigma_1}\Lambda_1^- \w \dd\Lambda^-_3) D\qty(\Lambda_1^--\Lambda_3^-, 0),
\end{align}
and similarly for \(\Lambda_2^+\), \(\Lambda_4^+\). In complete analogy with the circle branch, where the \(\widehat{\fu}(1)\) Kac--Moody algebra contains all the information to construct the full spectrum of the theory, here too, \cref{eq:noninv-KM-2d} is, in principle, sufficient to reconstruct the full orbifold branch.\footnote{Together with its special points, coming from \cref{eq:noninv-fusion-rest}. The special points will be important to obtain the twisted sectors (cf. also \cite{Schoutens:2015uia}).} To do so, one has to study the representation theory of \cref{eq:noninv-KM-2d}. The representation theory of non-invertible symmetries was recently studied in \cite{Bartsch:2022mpm,Bartsch:2022ytj,Lin:2022xod,Bhardwaj:2022lsg,Cordova:2024vsq}. In order to study the representation theory of our non-invertible current algebra it is necessary to extend these results to continuous and centrally extended non-invertible symmetries. We will not attempt to do that in this paper, but we will return to it in future work.

\paragraph*{\(\boldsymbol{p=1.}\)} In this case we land on the orbifold branch of four-dimensional Maxwell theory, i.e. \(\O(2)\) gauge theory. The central extension appears when one fuses chiral with anti-chiral operators:
\begin{align}\label{eq:noninv-km-4d}
	D\qty(\Lambda_1^-,0)\otimes  D\qty(0,\Lambda_4^+) &= \exp(\frac{\k}{2} \int_{\Sigma_3} \Lambda_1^-\w\dd{\Lambda_4^+})D\qty(\Lambda_1^-,\Lambda_4^+) \nn
	&\oplus\exp(-\frac{\k}{2} \int_{\Sigma_3} \Lambda_1^-\w\dd{\Lambda_4^+}) D\qty(\Lambda_1^-,-\Lambda_4^+),
\end{align}
and similarly for \(\Lambda_2^+\) with \(\Lambda_3^-\). The chiral-chiral and anti-chiral-anti-chiral channels do not see the central extension:
\begin{equation}\label{eq:noninv-km-4d'}
	D\qty(\Lambda_1^-, 0)\otimes D\qty(\Lambda_3^-, 0) = D\qty(\Lambda_1^-+\Lambda_3^-,0)\oplus D\qty(\Lambda_1^--\Lambda_3^-,0),
\end{equation}
and similarly for \(\Lambda_2^+\) and \(\Lambda_4^+\). We will see later, in \cref{sec:nonlocal state-op}, that the current algebra \cref{eq:p-KM} is is spectrum-generating, i.e. we can solve the underlying theory by considering its representations (up to contributions from a decoupled neutral sector). In analogy with the comments on the \(p=0\) case, \cref{eq:noninv-km-4d,eq:noninv-km-4d'}, contains in principle all the information to solve \(\O(2)\) gauge theory. Again, we will come back to doing so in the future.

\section{Two-dimensional CFTs and the local state-operator correspondence}\label{sec:local state-op}

In this section we will review the usual state-operator correspondence for local operators. We will therefore consider a two-dimensional CFT with a zero-form \(\U(1)\) symmetry. This corresponds to the case \(p=0\) in the notation of \cref{sec:photonisation-higherKM} and the corresponding Kac--Moody algebra can be represented as a compact free scalar \cite{Dijkgraaf:1987vp,Ginsparg:1987eb,Bardakci:1987ee,Ginsparg:1988ui}. We will therefore be rederiving the standard state-operator correspondence for the compact free scalar, in order to build some muscle towards \cref{sec:nonlocal state-op}. Along the way we will remind the reader of some standard facts in two-dimensional conformal field theories, in order to simplify and compare with the discussion of the four-dimensional case, in \cref{sec:nonlocal state-op}.

\subsection{Partition function and the spectrum}\label{ssec:Z 2d}

\subsubsection*{The view from the path integral}
Consider a compact free boson, \(\phi\sim \phi+2\pi\), on a closed Riemann surface, \(X\). Let us first review the path integral of the compact scalar. Its action reads
\begin{equation}
	S[\phi]\coloneqq \frac{1}{2\g^2}\int_X f^\phi \w\star f^\phi,
\end{equation}
where \(f^\phi \coloneqq f^\phi_\t{harm}+\dd{\phi}\) is the curvature of the free scalar, with \(f^\phi_\t{harm}\in \harm^1(X)\). In this notation, \(\phi\) is a well-defined, single-valued function (a zero-form), subject to the compactness condition, and all the winding has been passed on to the harmonic piece of its curvature, \(f^\phi_\t{harm}\). The coupling constant, \(\g\), is related to the radius of the compact scalar as \(R^2 = \frac{4\pi}{\g^2}\), as can be easily seen by rescaling \(\phi\) to \(\Phi = R\, \phi\), so that \(\Phi\sim \Phi + 2\pi R\).\footnote{We use conventions in which the action for \(\Phi\) is normalised as \(S[\Phi]=\frac{1}{8\pi}\int\dd[2]{z} \pd \Phi \conj{\pd} \Phi\).} The harmonic form, \(f^\phi_\t{harm}\) can always be chosen uniquely to be orthogonal to \(\dd{\phi}\), so the action splits into a harmonic piece and an oscillator piece:
\begin{equation}\label{eq:scalar-action-split}
	S[\phi] = \frac{1}{2\g^2}\int_X f^\phi_\t{harm} \w\star f^\phi_\t{harm} + \frac{1}{2\g^2}\int_X \dd{\phi}\w\star\dd{\phi}.
\end{equation}
Let us evaluate the partition function of the compact scalar, i.e. the path integral
\begin{equation}
	\parti[X] = \int\DD{f^\phi_\t{harm}}\DD{\phi} \ex{-S[\phi]} = \parti_\t{harm}[X]\parti_\t{osc}[X].
\end{equation}

We begin by analysing the harmonic piece. In our notation, integrality of the winding of the compact scalar is expressed as
\begin{equation}
	\int_{\Sigma_1} f^\phi_\t{harm}\in 2\pi\Z,
\end{equation}
on any one-cycle \(\Sigma_1\) of \(X\). We can choose a convenient basis, \(\set{\tau_\sfi^{(1)}}_{\sfi=1}^{\b_1(X)}\) of harmonic one-forms of \(X\) such that, given a basis \(\set{C_{\sfi(1)}}_{\sfi=1}^{\b_1(X)}\) of one-cycles, it obeys
\begin{equation}\label{eq:topo basis 2d}
	\int_{C_{\sfi(1)}}\tau_\sfj^{(1)} = \delta_{\sfi\sfj}.
\end{equation}
Such a basis will be called hereafter the \emph{topological basis}. In the above, \(\b_1(X)\coloneqq \dim\H^1(X)\) is the first Betti number of \(X\). In this basis we can expand
\begin{equation}
	f^\phi_\t{harm} = 2\pi n^\sfi \tau_\sfi^{(1)}, \qquad n^\sfi\in\Z,
\end{equation}
where a sum over the repeated index, \(\sfi\), from \(1\) to \(\b_1(X)\) is implied. The harmonic piece of the action \cref{eq:scalar-action-split} can therefore be written
\begin{equation}
	S_\t{harm}[\phi] =\Half\qty(\frac{2\pi}{\g})^2 n^\sfi \qty[\bbG^{(1)}]_{\sfi\sfj} n^\sfj,
\end{equation}
where
\begin{equation}\label{eq:gram1 2d}
	\qty[\bbG^{(1)}]_{\sfi\sfj} \coloneqq \int_{X} \tau_\sfi^{(1)}\w\star \tau_\sfj^{(1)}
\end{equation}
is the Gram matrix of the above topological basis. The harmonic piece of the path integral then reads
\begin{equation}
	\parti_\t{harm}[X] = \sum_{\vec{n}\in\Z^{\b_1(X)}}\exp(-\Half\qty(\frac{2\pi}{\g})^2 \vec{n}\cdot\bbG^{(1)}\cdot\vec{n}),
\end{equation}
where we combined \(n^\sfi\) into a \(\b_1(X)\)-dimensional vector, \(\vec{n}\). The oscillator contribution is a straightforward Gaussian integral, yielding
\begin{equation}
	\parti_\t{osc}[X] = \frac{\vol_0}{\sqrt{\detp\lapl_0}} = \qty(\frac{\det\qty(\frac{2\pi}{\g^2} \bbG^{(0)})}{\detp\lapl_0})^{\frac{1}{2}},
\end{equation}
where \(\bbG^{(0)}\) is defined similarly as \(\bbG^{(1)}\), but with respect to the zero-form topological basis and \(\lapl_0 = \cdd\dd\) is the Laplacian acting on zero-forms. In total, the full partition function reads
\begin{equation}
	\parti[X] =  \qty(\frac{\det\qty(\frac{2\pi}{\g^2} \bbG^{(0)})}{\detp\lapl_0})^{\frac{1}{2}} \sum_{\vec{n}\in\Z^{\b_1(X)}}\exp(-\Half\qty(\frac{2\pi}{\g})^2 \vec{n}\cdot\bbG^{(1)}\cdot\vec{n}).
\end{equation}

In the next paragraph we will take a canonical approach and we will view the partition function as a thermal trace. In order to compare, let us write the answer for the torus partition function from the path integral. To that end, we now take \(X\) to be a torus \(X=\T^2_\uptau \coloneqq \C\big/\qty(\Z\oplus\uptau\, \Z)\), with \(\uptau\) in the upper-half plane, \(\uptau\in\bbH\). Using the standard homology basis of one-cycles of the torus in terms of the \(A\)- and \(B\)-cycles, we have that
\begin{equation}
	\bbG^{(1)} = \frac{1}{\Im\uptau}\mqty(1 & \Re\uptau \\ \Re\uptau & \abs{\uptau}^2).
\end{equation}
The harmonic zero-forms are simply the constant functions, so \(\bbG^{(0)} = \Im\uptau\). Finally, on the torus it is straightforward to calculate, using zeta-function regularisation that
\begin{equation}
	\detp\lapl_0 = (\Im\uptau)^2\abs{\upeta(\uptau)}^2,
\end{equation}
where \(\upeta(\uptau)\) is the Dedekind eta function. Altogether the torus partition function reads
\begin{equation}
	\parti\qty[\T^2_\uptau] = \sqrt{\frac{2\pi}{\g^2\Im\uptau}}\frac{1}{\abs{\upeta(\uptau)}^2}\sum_{n,m\in\Z}\exp(-\Half \qty(\frac{2\pi}{\g})^2 \frac{\abs{n + \uptau\, m}^2}{\Im\uptau}).
\end{equation}
Finally, we can use the Poisson summation formula on the \(n\) sum to obtain
\begin{equation}\label{eq:T2-parti}
	\parti\qty[\T^2_\uptau] = \frac{\Theta\qty(q,\g)}{\abs{\upeta(q)}^2}
\end{equation}
where
\begin{equation}
	\Theta\qty(q,\g) \coloneqq \sum_{n,m\in\Z} q^{\frac{\pi}{2}\qty(\frac{m}{\g}+\frac{n \g}{2\pi})^2}\; \overline{q}^{\frac{\pi}{2}\qty(\frac{m}{\g}-\frac{n \g}{2\pi})^2},
\end{equation}
is the Siegel--Narain theta function, with \(q\coloneqq\ex{2\pi\ii \uptau}\) being the nome, \(\overline{q}\) its complex conjugate, and we wrote the eta function as a function of the nome, instead of the complex structure.

\subsubsection*{The view from the algebra}

Let us now match \cref{eq:T2-parti}, starting from the Kac--Moody algebra, \cref{eq:2d-KM} and considering its representations. For simplicity we will discuss the quantisation on a rectangular torus, with complex structure, \(\uptau=\ii\beta\). We will reinstate the generic complex structure at the end of the section. The Hamiltonian of the free scalar, on the spatial \(\S^1\), takes the Sugawara form
\begin{equation}
	H = L_0^+ + E_0^+ + L_0^- + E_0^-,
\end{equation}
with
\begin{equation}
	L_0^\pm = \frac{1}{\k} \sum_{\sfn\geq 0} J_{-\sfn}^\pm\, J_\sfn^\pm \qq{and}
	E_0^\pm = \frac{1}{2}\upzeta(-1) = -\frac{1}{24},
\end{equation}
where \(\upzeta(s)\) is the Riemann zeta function, regularising the zero-point energy, and \(\k\), the Kac--Moody level, is \(\frac{\g^2}{2}\). With this Hamiltonian, the non-zero-modes of the Kac--Moody algebra, \(J^\pm_{\sfn\neq 0}\) are just ladder operators, as can be seen by computing their commutation relations with the Hamiltonian. The negative modes are creation operators, while the positive modes annihilation operators:
\begin{align}
	\comm{H}{J^\pm_{-\sfn}} & = \phantom{+}\sfn\, J^\pm_{-\sfn} \qquad \sfn>0 \\
	\comm{H}{J^\pm_{\sfn}}  & = -\sfn\, J^\pm_{\sfn} \qquad \sfn>0.
\end{align}

The zero-modes, \(J_0^\pm\),  commute among themselves and with the Hamiltonian, and they label the various fixed momenutm and winding sectors. Their eigenstates,
\begin{equation}
	J_0^\pm \ket{j^+,j^-} = j^\pm \ket{j^+,j^-},
\end{equation}
are primary states, or in other words, the ground state on each sector:
\begin{equation}
	J^\pm_{\sfn}\ket{j^+,j^-} = 0,\qquad \sfn>0.
\end{equation}
To obtain the charges, \(j^\pm\), we simply have to invoke momentum and winding flux quantisation on the spatial \(\S^1\), i.e.
\begin{equation}\label{eq:flux quant 2d}
	\int_{\S^1} f^\phi \in 2\pi\Z \qq{and} \int_{\S^1}  \dual{f}^\phi \in 2\pi\Z,
\end{equation}
where \(\dual{f}^\phi\) is the widning, or magnetic dual of \(f^\phi\), defined as \(\dual{f}^\phi = \frac{\pi\ii}{\k}\star f^\phi\).\footnote{The factor of \(\ii\) is so that the action, written in the dual frame is positive semidefinite in Euclidean signature.} With this, we immediately get, with a convenient overall normalisation of the current:
\begin{equation}
	j^\pm = \frac{1}{\sqrt{8\pi}}\qty(2\pi m \pm \g^2 n), \qquad n,m\in\Z.
\end{equation}
The ground states are, then, labelled by two integers, \(n\) and \(m\), and we will denote them as 
\begin{equation}
	\ket{j^+,j^-}\eqqcolon \ket{n,m}.
\end{equation}

Over each of the two families of ground states sits a Verma module, \(\cV_{n,m}^\pm\), generated by acting with the creation operators \(J^\pm_{-\sfn}\). To each Verma module corresponds a character,
\begin{equation}
	\ch_{\cV_{n,m}^\pm}[q] \coloneqq \tr_{\cV_{n,m}^\pm} q^H.
\end{equation}
Having the explicit form of the Hamiltonian, and the algebra in our disposal, it is trivial task to compute the characters. They read:
\begin{equation}
	\ch_{\cV_{n,m}^\pm}[q] = q^{h^\pm-\frac{1}{24}} \qty(\prod_{\sfn = 1}^\infty\sum_{N_\sfn = 0}^\infty q^{\sfn\, N_\sfn }) = \frac{q^{h^\pm}}{\upeta(q)},
\end{equation}
where we recognise the form of the Dedekind eta function, upon performing the geometric sum. In the above, \(h^\pm = \qty(j^\pm)^2 \), is the eigenvalue of \(L_0^\pm\) acting on the ground states.

On each sector we have left- and right-moving states, i.e. a generic state on top of \(\ket{n,m}\) lives in \(\cV_{n,m}^+\otimes \cV_{n,m}^-\). The full Hilbert space of the theory is, then, a direct sum over all possible sectors:
\begin{equation}\label{eq:H circle}
	\cH_{\S^1} = \bigoplus_{n,m\in\Z} \cV_{n,m}^+\otimes\cV_{n,m}^-.
\end{equation}
A generic state of this Hilbert space is
\begin{equation}
	\hspace{-2em}\ket{n,m;\set{N_{\sfn}^+,N_{\sfm}^-, \cdots}} \coloneqq \qty(J_{-\sfn}^+)^{N_{\sfn}^+}\qty(J_{-\sfn}^-)^{N_{\sfm}^-}\cdots  \ket{n,m}.
\end{equation}
As such, we see that taking a trace, i.e. computing an extended character of all the Verma modules, lands us on the torus partition function, \cref{eq:T2-parti} (with \(\uptau=\ii\beta\)):
\begin{equation}\label{eq:xch 2d}
	\xch[q] \coloneqq \sum_{n,m\in\Z} \ch_{\cV_{n,m}^+\otimes \cV_{n,m}^-}[q] = \frac{1}{\abs{\upeta(q)}^2}\sum_{n,m\in\Z} q^{\Delta_{n,m}},
\end{equation}
with
\begin{equation}
	\Delta_{n,m} = h^+ + h^- = \frac{\g^2 n^2}{4\pi}+\frac{m^2 \pi}{\g^2}.
\end{equation}
Reinstating the generic complex structure, \(\uptau\in\bbH\), we obtain the well-appreciated, yet remarkable result:
\begin{equation}\label{eq:2d Z=ch}
	\parti\qty[\T^2_\uptau] = \xch[q] \qq{with} q = \ex{2\pi\ii \uptau}.
\end{equation}

\subsection{The state-operator correspondence}\label{sec:local-state-op}

As a preparatory exercise for the four-dimensional story of the later sections, we will review here the well-established state-operator correspondence for local operators. In particular we will show that the states in \(\cH_{\S^1}\), as in \cref{eq:H circle}, can be prepared by a path integral, with insertions of local operators, on the disc, \(\D^2\), whose boundary is the spatial \(\S^1\). We will do so for the compact scalar, where both sides of the state operator correspondence can be explicitly identified and checked. Moreover, a crucial role is played by the Kac--Moody algebra of the compact scalar, organising the local operators into primaries and descendants. This will serve as an analogy for the later sections, where, as we will see, the ideas will be similar.

\begin{figure}[!htb]
	\centering
	\def\svgwidth{0.8\textwidth}
	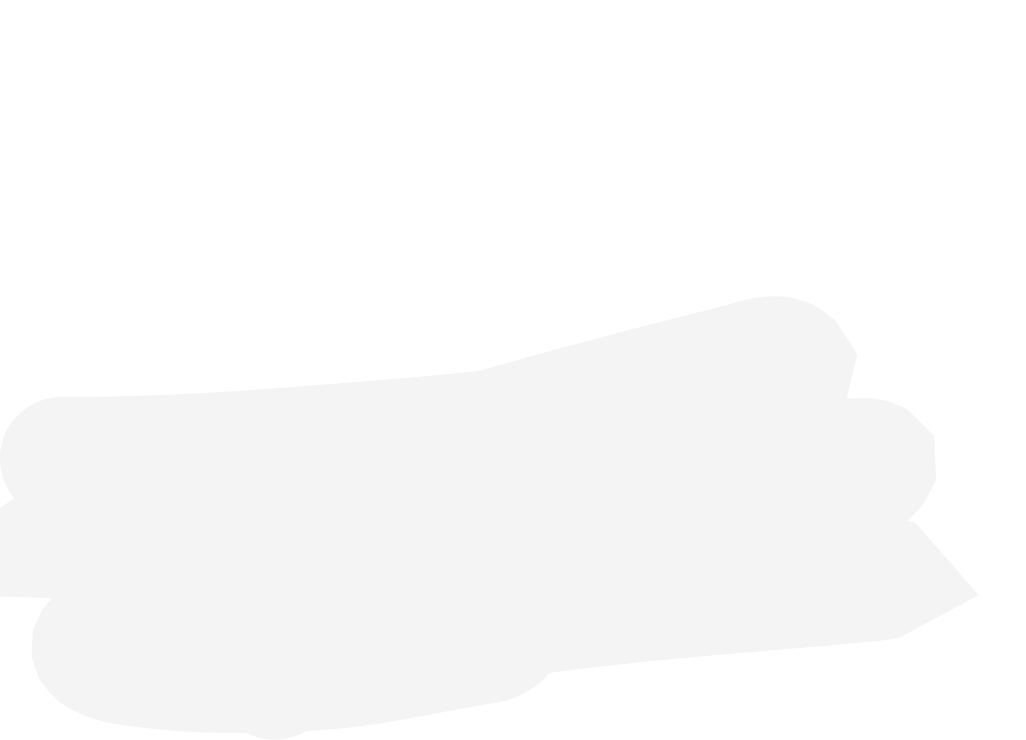
	\caption{The state-operator correspondence. Any state on \(\S^{d-1}\) can be prepared by a path integral on \(\B^{d}\) with a local operator inserted in the centre. The state then evolves in time on the Lorentzian cylinder \(\R\times\S^{d-1}\).}
	\label{fig:state-op fig}
\end{figure}

The general strategy towards a state-operator map for local operators (illustrated in \cref{fig:state-op fig}), common to all CFTs is the following. We take a spherical slice (here \(\S^1\)), of the Euclidean cylinder (here \(\R\times\S^1\)) at \(t=0\). Each state on this slice is prepared by a choice of boundary conditions at \(t=-\infty\). Mapping the Euclidean cylinder to the disc (rather, \(d\)-ball in higher dimensions) with a Weyl transformation, this choice gets mapped to a boundary condition at the centre of the disc.\footnote{We will parametrise the disc by a radial coordinate, \(r\in\closed{0}{1}\), and an angle \(\theta\in\ropen{0}{2\pi}\), i.e. \(\dd{s^2_{\D^2}} = \dd{r}^2 + r^2\dd{\theta^2}\).} This boundary condition can be satisfied by the insertion of a local operator. Then the path integral on the disc, with this operator insertion prepares the state that we are after. More precisely, the path integral on the disc, with boundary conditions
\(\Phi\qty(\pd\D^2) = \Phi_\pd\) (where \(\Phi\) denotes all dynamical fields) and an insertion \(\cO\qty[\Phi(0)]\) at the centre of the disc produces a wavefunctional:
\begin{equation}
	\Psi_\cO[\Phi_\pd] = \int_{\cC\qty[\Phi_\pd]} \DD{\Phi} \ex{-S[\Phi]}\ \cO[\Phi(0)],
\end{equation}
where \(\cC\qty[\Phi_\pd]\) denotes an appropriate functional space over the disc, with boundary conditions \(\Phi_\pd\). We will write the state produced by \(\cO\) as
\begin{equation}
	\ket{\cO} = \int_{\cC\qty[\,\cdot\,]} \DD{\Phi} \ex{-S[\Phi]}\ \cO[\Phi(0)],
\end{equation}
to indicate that we need to provide a boundary condition to get out a number.

Here we note that it would be much easier to work in complex coordinates, and exploit the power of Cauchy's theorem, as is commonly done in two-dimensional conformal field theories. We will not do so, however, as a preparatory exercise for the four-dimensional case where we cannot afford that luxury. For a discussion in a very similar vein, written in complex variables see e.g. \cite{Tong:2009np}.

Before discussing the states and the operators, it will be useful to obtain an integral expression for the ladder operators, in terms of the currents, that we can insert at an intermediate point in the path integral. To that end, let \(i_{\S^1_r}:\S^1_r\hookrightarrow \D^2\) be the map embedding a circle of radius \(r\) into the disc. The ladder operators can be expressed as:
\begin{equation}\label{eq:ladder-r}
	J^\pm_\sfm = \frac{\mp\ii r^{\sfm+1}}{\sqrt{2\pi}} \int_{\S^1_r} \ex{\pm\ii \sfm \theta} \ i_{\S^1_r}^* J^\pm,
\end{equation}
where \(i_{\S^1_r}^* J^\pm\) is the pullback of \(J^\pm\in \Omega^2\qty(\D^2)\) along the map \(i_{\S^1_r}\). This is arrived at by expanding \(J^\pm\) in Fourier modes, solving the closedness condition, \(\dd J^\pm = 0\) and then inverting the Fourier transform. As a sanity check, we can verify that \(J^\pm_\sfm\), given by \cref{eq:ladder-r}, obey the Kac--Moody algebra, \cref{eq:2d-KM}:
\begin{equation}
	\comm{J^\pm_\sfn}{J^\pm_\sfm} = \k\,\sfn\, \delta_{\sfn+\sfm,0},
\end{equation}
while the mixed commutators are zero.

Let us start by discussing the ground states, \(\ket{n,m}\). They are prepared by inserting a vertex operator, \(\t{V}_{n,m}\), of momentum \(n\) and winding \(m\):
\begin{equation}
	\t{V}_{n,m}(x) \coloneqq \exp(\ii\, n\,\phi(x) + \ii\, m\, \dual{\phi}(x)),
\end{equation}
where \(x\) is a point on the disc and \(\dual{\phi}\) is the winding dual of \(\phi\), i.e. the field whose curvature is \(\dual{f}^\phi\), as defined around \cref{eq:flux quant 2d}. The states are then obtained as:
\begin{equation}\label{eq:ground-states-PI 2d}
	\ket{n,m} = \int_{\cC[\,\cdot\,]} \DD{\phi} \ex{-S[\phi]}\ \t{V}_{n,m}(0).
\end{equation}
In this case, \(\cC[\,\cdot\,]\) is the space of smooth functions on the disc. To verify that these are indeed the ground states, it suffices to show that they are indeed annihilated by the lowering operators, \(J^\pm_{\sfn>0}\). To illustrate the argument, let us first focus on the state \(\ket{0,0}\), which is the zero-momentum, zero-winding state, corresponding to the identity operator. There we have
\begin{equation}\label{eq:2d J|0>}
	J^\pm_{\sfn}\ket{0,0} \sim \lim_{r\to 0}\int_{\cC[\,\cdot\,]} \DD{\phi}\ex{-S[\phi]}\qty(\int_{\S^1_r} r^{\sfn+1} \ex{\pm\ii\sfn \theta} \ i_{\S^1_r}^* J^\pm),
\end{equation}
where the twiddle indicates that we are ignoring purely numerical factors. Since we are integrating over smooth functions, and \(J^\pm=\bbP_\pm f^\phi\), the integrand is smooth inside the disc, for \(\sfn\geq 0\). Therefore, we have that
\begin{equation}
	J_\sfn^\pm\ket{0,0} = 0, \qquad \fall\, \sfn>0.
\end{equation}
In this case, we also have that \(J_0^\pm\ket{0,0} = 0\), as \(\ket{0,0}\) is a state of zero momentum and winding. To show that the states \(\ket{n,m}\) are annihilated by all positive modes the argument is essentially the same, save for the fact that before assessing the integrand one needs to look at the operator product between \(i_{\S^1_r}^* J^\pm\) and \(\t{V}_{n,m}\). Since \(J^\pm\) is a current and \(\t{V}_{n,m}\) is a primary operator, their OPE goes like \(\frac{1}{r}\). Therefore, \(\ket{n,m}\) are annihilated by all positive modes, but not by the zero-modes, as expected.

We have, so far, argued, that \cref{eq:ground-states-PI 2d} produces all the highest-weight states, of our Verma module discussion in \cref{ssec:Z 2d}. To complete the picture, what remains is a construction of the descendants. It suffices to consider the states \(\ket{n,m;\set{1_\sfn^\pm}}\); a generic state is straightforward to arrive at, afterwards. The claim is, then, that these states are prepared by:
\begin{equation}\label{eq:descendants PI 2d}
	\ket{n,m; \set{1_\sfn^\pm}} = \int_{\cC[\,\cdot\,]} \DD{\phi}\ex{-S[\phi]}\ \ \pd_\pm^\sfn \phi\,\t{V}_{n,m}(0), \qquad \sfn>0,
\end{equation}
where \(\pd_\pm\) is the \textquote{(anti-)self-dual derivative}\footnote{Intuitively \(\pd_+\) is just the holomorphic derivative and \(\pd_-\) the antiholomorphic derivative. If we insist on avoiding complex analysis language we can also define them as follows. First define bases of the (anti-)self-dual one-forms, \(\dd{x}^\pm \coloneqq \ex{\ii \theta}\qty(\dd{r}\pm\ii r\dd{\theta})\). Then dualise these to vectors fields, as \((\dd{x^\pm})^\sharp\) and take an interior product with the exterior derivative, so \(\pd_\pm \coloneqq \inp{\qty(\dd{x^\pm})^{\sharp}}{\dd}\).} and \(\pd_\pm^\sfn \phi \t{V}_{n,m}(0)\) denotes the operator obtained upon performing the OPE of \(\pd_\pm^\sfn \phi\) with \(\t{V}_{n,m}\) and placing it at \(0\). To verify the claim \cref{eq:descendants PI 2d}, observe what happens if we act with a lowering operator. As before, we treat the descendants of the state dual to the identity operator, \(\ket{0,0}\), first.
\begin{equation}
	J^\pm_{\sfm}\ket{0,0; \set{1_\sfn^\pm}} \sim \lim_{r\to 0}\int_{\cC[\,\cdot\,]} \DD{\phi}\ex{-S[\phi]}\ \qty(\int_{\S^1_r} \dd{\theta} r^{\sfm+1} \ex{\pm\ii(\sfm+1) \theta} \  \pd_\pm \phi(r,\theta)\, \pd_\pm^\sfn \phi(0)),
\end{equation}
where we have also used the fact that, for all that matters here, \(i_{\S^1_r}^* J^\pm\) is proportional to \(\ex{\pm\ii \theta}\pd_{\pm} \phi \dd{\theta}\). We recognise the insertion as the \((\sfn-1)\)-th (anti-)self-dual derivative of a current. So we can invoke the current-current OPE \cite{DiFrancesco:1997nk}, to get
\begin{equation}
	\pd_\pm \phi(r,\theta)\; \pd_\pm^\sfn \phi(0) \sim \frac{\k\,\sfn!}{r^{\sfn+1}}\ \ex{\mp\ii(\sfn+1)\theta}.
\end{equation}
This, in turn, implies:
\begin{equation}
	J^\pm_{\sfm}\ket{0,0; \set{1_\sfn^\pm}} \sim \lim_{r\to 0}\int_{\cC[\,\cdot\,]} \DD{\phi}\ex{-S[\phi]}\ \qty(\int_{0}^{2\pi} \dd{\theta} r^{\sfm-\sfn} \ex{\pm\ii (\sfm-\sfn)\theta}) \sim \k\; \delta_{\sfm,\sfn}\ket{0,0}.
\end{equation}
The factor of \(\sfn\) in the Kac--Moody algebra \cref{eq:2d-KM} is absorbed into the proportionality symbol, and is related to a rescaling of the operator insertion by \(\frac{1}{(\sfn-1)!}\). However, we kept track of the Kac--Moody level, since it stems from the OPE of the currents.

Inserting back \(\t{V}_{n,m}\), it is now a matter of Wick contractions and restating the above arguments to show that
\begin{equation}
	J^\pm_\sfm\ket{n,m;\set{1_\sfn^\pm}} = \k\, \sfm\, \delta_{\sfm,\sfn}\ket{n,m}.
\end{equation}
This concludes the proof that the operator \(\t{V}_{n,m}[\phi]\ \pd_\pm^\sfn \phi\) prepares the state \(\ket{n,m;\set{1_\sfn^\pm}}\). We therefore have all the ingredients to write down the generic state. It is given by:
\begin{empheq}[box=\obox]{equation}
	\ket{n,m;\set{N_{\sfn}^+,N_{\sfm}^-,\cdots}} = \int_{\cC[\,\cdot\,]} \DD{\phi}\ex{-S[\phi]}\ \qty{\qty(\pd_+^\sfn \phi)^{N_\sfn^+} \qty(\pd_-^\sfm \phi)^{N_\sfm^-}\cdots\t{V}_{n,m}(0)}.
\end{empheq}

\section{Four-dimensional Maxwell theory}\label{sec:4d}

Having dealt with the two-dimensional case in sufficient detail we now turn to the main focus of this paper: 4d and the nonlocal state operator correspondence. We will illustrate this in four dimensions, although the results extend to generic even dimensions easily. We therefore, consider a unitary four-dimensional CFT, with a \(\gf{\U(1)}{1}\) symmetry. As we have explained in \cref{sec:photonisation-higherKM}, this theory has necessarily a free photon description. We will, therefore focus on free Maxwell theory. We will mimick the structure of \cref{sec:local state-op}: we will first discuss the path integral on arbitrary compact manifolds; then we will use the \(p=1\) form of the current algebra \cref{eq:p-KM}, study its representations and show that they explicitly reproduce the path integral expressions. This will then lead us to the nonlocal state-operator correspondence.

\subsection{The Maxwell path integral}\label{ssec:Z 4d}

We consider Maxwell theory on a closed, compact four-dimensional manifold \(X\).\footnote{We will take \(X\) to be torsion-free, throughout. All results can be generalised to manifolds with torsion, following, for example, the arguments in \cite{Donnelly:2016mlc}.} Since we are interested in topologically non-trivial manifolds we must also allow for a theta angle. The action is:
\begin{equation}
	S[a]\coloneqq \frac{1}{2 \g^2}\int_X f \w\star f - \frac{\ii \theta}{16\pi\,^2}\int_X f\w f, \label{eq:act-maxwell}
\end{equation}
where the closed two-form \(f\) is the curvature of a \(\U(1)\) gauge field, \(a\in \Omega^1(X)\). We will often abbreviate the first term as \(\norm{f}^2\), where the norm is with respect to the usual Hodge inner-product. Hodge decomposition instructs us to write \(f=f_\t{harm}+\dd{a}\), where \(f_\t{harm}\) is a harmonic representative of the second cohomology class of \(X\), chosen uniquely to be orthogonal to \(\dd{a}\) in the Hodge inner product. Moreover, the theta term does not see the topologically trivial piece, so the action reads, with this decomposition
\begin{equation}
	S[a] = \underset{\blue{\smqty{\rotcoloneqq \\ S^{\inst}[f_\t{harm}]}}}{\blue{\underbrace{\textcolor{black}{\frac{1}{2 \g^2}\norm{f_\t{harm}}^2 - \frac{\ii \theta}{16\pi\,^2}\int_X f_\t{harm}\w f_\t{harm}}}}} + \underset{\green{\smqty{\rotcoloneqq \\ S^{\osc}[a]}}}{\green{\underbrace{\vphantom{\int_X}\textcolor{black}{\frac{1}{2 \g^2}\norm{\dd{a}}^2}}}}.
\end{equation}

With a theta angle turned on, it is useful to introduce the complexified coupling constant:
\begin{equation}
	\tt \coloneqq \frac{\theta}{4\pi}+\frac{2\pi\ii}{\g^2}.
\end{equation}
We refrain from using the more common \(\uptau\) for the complexified coupling constant to avoid confusion with the two dimensional case, where \(\uptau\) represents the complex structure of the base space. Sometimes we will use
\begin{equation}
	\tt^\pm\coloneqq\Re\tt\pm\ii\Im\tt,
\end{equation}
to write some of our formulas more compactly.

Passing to the path integral, the partition function of Maxwell theory splits into an instanton contribution and an oscillator contribution
\begin{equation}
	\parti[X;\tt] = \int\frac{\DD{f_\t{harm}}\DD{a}}{\vol\cG} \exp(-S^\inst[f_\t{harm}]-S^\osc[a]) = \parti^\inst[X;\tt]\parti^\osc[X;\tt],
\end{equation}
where \(\cG\) is the group of gauge transformations; shifts of \(a\) by a flat connection. We will compute each piece separately and discuss the respective integration measures in detail.

Let us first focus on the oscillator part. By Hodge decomposition, and upon employing the Faddeev--Popov procedure, we can gauge-fix \(a\) to be coclosed. The Faddeev-Popov ghosts are, in this case, zero-forms with fermionic coefficients. The oscillator part of the action becomes then
\begin{equation}
	S^\osc[a] = \frac{1}{2\g^2} \ip{a}{\tlapl_1\; a},
\end{equation}
where \(\tlapl_p\!\coloneqq\!\eval{\lapl_p}_{\ker\cdd}\), is the transversal Laplacian, i.e. the Hodge Laplacian, \(\lapl_p=\cdd\dd+\dd\cdd\), restricted to coclosed forms. Here we can expand the action in terms of eigen-one-forms of the transversal Laplacian,
\begin{equation}
	a = \sum_{\lambda\in\spec_X(\tlapl_1)} c_\lambda a_\lambda.
\end{equation}
In the above, \(\lambda=0\), and \(a_0\) collects all the zero-mode spectrum of the Laplacian, \(\tlapl_1 a_0 = 0\). We can shift all the \(\g\)-dependence on the zero-mode path integral by normalising the measure as
\begin{equation}
	\DD{a}\coloneqq \prod_{\lambda\in\spec_X(\tlapl_1)} \frac{\dd{c_\lambda}}{\sqrt{2\pi\,}\g}.\label{eq:a-measure}
\end{equation}
Both the gauge field and the ghost path integrals are Gaussian and can be evaluated directly to give
\begin{equation}
	\parti^{{\osc}}[X;\tt] = \frac{\vol(\H^1(X;\Z))}{\vol(\H^0(X;\Z))}\; \qty(\frac{\detp\tlapl_0}{\detp\tlapl_1})^{1/2}.
\end{equation}
Using the specrtral properties of the Hodge Laplacian, we can rewrite this in terms of the full Laplacian as
\begin{equation}
	\parti^{{\osc}}[X;\tt] = \frac{\vol(\H^1(X;\Z))}{\vol(\H^0(X;\Z))}\; \frac{\detp\lapl_0}{\sqrt{\detp\lapl_1}}. \label{eq:parti-osc}
\end{equation}
In the above, \(\vol(\H^1(X;\Z))\), and \(\vol(\H^0(X;\Z))\) is the zero-mode volume of the gauge fields and the ghosts, respectively, computed with the measure \cref{eq:a-measure}. To see that, note that the zero-modes of \(\lapl_1\) are precisely flat connections, or in other words elements of \(\H^1(X)\). Since we identify two connections under large gauge transformations (these are packaged in the instanton sum), we must look at \(\H^1(X;\Z)\). We have assumed that the manifold, \(X\), is torsion-free and thus \(\H^1(X;\Z)\cong\Z^{\b_1(X)}\). We can therefore use the topological basis of the harmonic one-forms, \(\set{\tau_\sfi^{(1)}}_{\sfi=1}^{\b_1(X)}\), defined similarly as in \cref{eq:topo basis 2d} to write
\begin{equation}\label{eq:gram1 4d}
	\qty[\bbG^{(1)}]_{\sfi\sfj} \coloneqq \int_{X} \tau_\sfi^{(1)}\w\star \tau_\sfj^{(1)}.
\end{equation}
With that we can integrate over \(\H^1(X;\Z)\) with the measure \cref{eq:a-measure}, to get
\begin{equation}
	\vol(\H^1(X;\Z)) = \det(\frac{2\pi\,}{\g^2}\bbG^{(1)})^{1/2}.
\end{equation}
For the ghosts we have similarly
\begin{equation}
	\vol(\H^0(X;\Z)) = \det(\frac{2\pi\,}{\g^2}\bbG^{(0)})^{1/2},
\end{equation}
where \(\bbG^{(0)}\) is defined analogously by the topological basis of \(\H^0(X;\Z)\). If \(\b_1(X)=0\), the Laplacian \(\tlapl_1\)  has no zero-modes and hence we compute the full determinant \(\det\tlapl_1\).

It is convenient to disentangle base-space and target-space quantities to keep track of the modular properties and of electromagnetic duality separately. The only place where \(\g\) appears is in the zero-mode volumes and therefore we have that
\begin{equation}
	\parti^{\osc}[X;\tt] = \qty(\frac{2\pi\,}{\g^2})^{\half\qty(\b_1(X)-\b_0(X))}\ \qty(\frac{\det\bbG^{(1)}}{\det\bbG^{(0)}})^{1/2} \frac{\detp\lapl_0}{\sqrt{\detp\lapl_1}}.
\end{equation}
For the rest, we are only interested connected manifolds, therefore \(\b_0(X)=1\) and \(\det\bbG^{(0)}=\vol(X)\). Furthermore, note that the prefactor is just \(\Im\tt\). So, all in all, the oscillator piece reads, finally
\begin{equation}
	\parti^{\osc}[X;\tt] = \qty(\Im \tt)^{\half\qty(\b_1(X)-1)}\ \qty(\frac{\det\bbG^{(1)}}{\vol(X)})^{1/2}\frac{\detp\lapl_0}{\sqrt{\detp\lapl_1}}. \label{eq:partiosc}
\end{equation}

We now turn to the instanton contribution. Similarly as in \cref{sec:local state-op}, \(f_\t{harm}\) is integrally quantised on two-cycles,
\begin{equation}
	\int_{\Sigma_2} f_\t{harm} \in 2\pi\Z,
\end{equation}
so we can expand $f_\t{harm} = 2\pi n^\sfi \tau_\sfi^{(2)}$ with $n^\sfi\in\Z$, and\(\set{\tau_\sfi^{(2)}}_{\sfi=1}^{\b_2(X)}\) is the topological basis of hamronic two-forms. The first term of the instanton piece of the action becomes, then:
\begin{equation}
	\frac{1}{2\g^2}\norm{f_\t{harm}}^2=\frac{2\pi^2}{\g^2} n^\sfi\; \qty[\bbG^{(2)}]_{\sfi\sfj}\; n^\sfj, \qquad n^\sfi\in\Z,
\end{equation}
where \(\bbG^{(2)}\), is defined similarly as in \cref{eq:gram1 4d}. For the theta term we make use of the intersection bilinear on \(X\):
\begin{equation}
	\bbQ_{\sfi\sfj} \coloneqq \int_X \tau_\sfi^{(2)}\w \tau_\sfj^{(2)},
\end{equation}
to write
\begin{equation}
	\frac{\ii \theta}{16\pi\,^2} \int_X f_\t{harm}\w f_\t{harm} = \frac{\ii \theta}{4} n^\sfi\; \bbQ_{\sfi\sfj}\; n^\sfj.
\end{equation}
Finally, the measure, \(\DD{f_\t{harm}}\), becomes a sum over (vectors of) integers, \(\vec{n}\in\Z^{\b_2(X)}\). Putting it all together, the instanton contribution reads
\begin{equation}
	\parti^{{\inst}}[X;\tt] = \sum_{\vec{n}\in\Z^{\b_2(X)}} \exp(\pi\,\ii\ \vec{n}\cdot \bbB(t) \cdot \vec{n}), \label{eq:partiinst}
\end{equation}
where we have defined, for conciseness, the matrix
\begin{equation}
	\bbB(\tt) \coloneqq \Re \tt\; \bbQ + \ii\Im \tt\; \bbG^{(2)} = \frac{2\pi\,\ii}{\g^2} \bbG^{(2)} + \frac{\theta}{4\pi\,} \bbQ.
\end{equation}

Combining the oscillator and the instanton contributions we have in total the partition function of free Maxwell theory on a general, torsion-free, connected, closed four-manifold \(X\):
\begin{equation}
	\parti[X;\tt] = \qty(\Im\tt)^{\half\qty(\b_1(X)-1)}\ \qty(\frac{\det\bbG^{(1)}}{\vol(X)})^{1/2}\frac{\detp\lapl_0}{\sqrt{\detp\lapl_1}} \sum_{\vec{n}\in\Z^{\b_2(X)}} \exp(\pi\,\ii\ \vec{n}\cdot \bbB(\tt)\cdot \vec{n}). \label{eq:partimaxwellfull}
\end{equation}

\subsubsection*{\(\SL(2,\Z)\) duality}

Before moving on, we briefly comment on some of the duality properties of this theory. It is well-known, that Maxwell theory enjoys an \(\SL(2,\Z)\) duality group, generated by S-duality, \(\tt\mapsto-1/\tt\) and T-duality, \(\tt\mapsto \tt+1\). It is, however, also known that the \(\SL(2,\Z)\) duality group is afflicted with an anomaly \cite{Witten:1995gf,Seiberg:2018ntt,Hsieh:2019iba}. On a generic manifold, one then either needs to sacrifice one of the two generators, or couple the system to a five-dimensional SPT phase, carrying the duality anomaly (as is done, e.g. in \cite{Hsieh:2019iba}). In this section we calculate an instantiation of the duality anomaly using the form \cref{eq:partimaxwellfull}, and discuss ways to guarantee that either S- or T-duality holds on the nose. We also show that, in fact, the manifolds required to obtain our state-operator map, are free from the duality anomaly.

Key to the duality properties, will be a version of the Poisson summation formula:
\begin{equation}
	\sum_{\vec{n}\in\Z^N} \ex{-\vec{n}\cdot\bbA\cdot\vec{n}} = \frac{\pi^{N/2}}{\sqrt{\det\bbA}}\sum_{\vec{n}\in\Z^N}\ex{-\pi^2 \vec{n}\cdot\inv{\bbA}\cdot\vec{n}}.
\end{equation}
The form \cref{eq:partimaxwellfull} is particularly handy to perform Poisson resummation. In particular, we can check that \(-\trans{\bbQ}\bbB(-1/\tt)\trans{\bbQ}\) is the inverse of \(\bbB(\tt)\), and the partition function can be expressed as:
\begin{equation}
	\parti[X;\tt] = \qty(\frac{\qty(\Im\tt)^{\qty(\b_1(X)-1)}}{\det(-\ii \bbB(\tt))})^{1/2} \qty(\frac{\det\bbG^{(1)}}{\vol(X)})^{1/2} \frac{\detp\lapl_0}{\sqrt{\detp\lapl_1}} \sum_{\vec{n}\in\Z^{\b_2(X)}}\exp(\pi\,\ii\; \vec{n}\cdot \bbB(-1/\tt)\cdot \vec{n}).
\end{equation}
Let us now make some simplifying assumptions for the rest of the calculation. Since \(\H^2(X)\) is a middle cohomology group for a four-dimensional manifold its elements can be decomposed into self-dual and anti-self-dual pieces. The dimensions of the (anti)-self-dual parts of the cohomology group are \(\b_2^\pm(X)\), such that \(\b_2^+(X)+\b_2^-(X) = \b_2(X)\) and \(\b_2^+(X)-\b_2^-(X)= \sigma\), with \(\sigma\) being the Hirzebruch signature of \(X\). For all the manifolds that we are interested in (namely products of spheres of different dimensions), \(\sigma=0\), therefore \(\b_2^+(X) = \b_2^-(X) = \half \b_2(X)\). We will assume this for the following, although it is straightforward to generalise the discussion to manifolds with \(\sigma\neq 0\).

The determinant of \(\bbB(\tt)\) can be calculated to be \(\det(-\ii \bbB(\tt)) = \abs{\tt}^{\b_2(X)}\). This follows immediately from the fact that the eigenvalues of \(\bbG^{(1)}\inv{\bbQ}\) are \(\pm 1\), with multiplicity \(\half \b_2(X)\). Furthermore using \(\Im(-1/\tt) = \abs{\tt}^{-2}\Im(\tt)\), we get the S-dual form of the partition function:
\begin{equation}
	\parti\qty[X;-\frac{1}{\tt}]= \abs{\tt}^{\frac{1}{2}\upchi(X)}\parti[X;\tt], \label{eq:S-duality}
\end{equation}
where \(\upchi(X)\) is the Euler characteristic of \(X\). Moreover, we can immediately see that if \(\bbQ\) is even
\begin{equation}\label{eq:T-duality4D}
	\parti[X;\tt+1] = \parti[X;\tt],
\end{equation}
since its effect on \cref{eq:partimaxwellfull} is to shift the exponent by \(2\pi\times\t{integer}\), whereas if \(\bbQ\) is odd, \(\parti[X;\tt+2] = \parti[X;\tt]\).\footnote{This can be corrected for, by adding an extra, topological term, to the action, as in \cite{Seiberg:2018ntt}} \cref{eq:S-duality} and \cref{eq:T-duality4D}, correspond to the action of the S and T generators of the \(\SL(2,\Z)\) duality group. In its current form, the partition function of Maxwell theory favours T-duality, while the S-duality transformation, seemingly suffers from the aforementioned anomaly.

However, this is deceiving. Namely, we can add to the action a counterterm of the form
\begin{equation}
	S_\t{ct} = \frac{1}{32\pi^2} \int_X f(\tt)\, \epsilon^{abcd} R_{ab}\w R_{cd} = f(\tt)\, \upchi(X),
\end{equation}
where the last equation is only true if \(f(\tt)\) does not depend on \(X\). Choosing  \scalebox{0.97}{\(f(\tt)\!=\!-\frac{1}{4}\!\log\Im\tt\)}, the partition function gets modified to
\begin{equation}
	\widetilde{\parti}[X;\tt] \coloneqq \qty(\Im\tt)^{\frac{1}{4}\upchi(X)} \parti[X;\tt]. \label{eq:parti-anomaly-free}
\end{equation}
It is straightforward to check that indeed, both transformations are healthy, and
\begin{equation}
	\widetilde{\parti}\qty[X;\gamma\cdot\tt] = \widetilde{\parti}\qty[X;\tt], \qquad \fall\gamma\in\SL(2,\Z).
\end{equation}
Written in terms of \cref{eq:partimaxwellfull}, the duality-corrected version of the partition function reads
\begin{empheq}[box=\obox]{equation}
	\widetilde{\parti}\qty[X;\tt] = \qty(\Im\tt)^{\frac{\b_2(X)}{4}}\ \qty(\frac{\det\bbG_1}{\vol(X)})^{1/2}\frac{\detp\lapl_0}{\sqrt{\detp\lapl_1}} \sum_{\vec{n}\in\Z^{\b_2(X)}} \exp(\frac{\pi\ii}{2}\ \vec{n} \cdot \bbB(\tt)\cdot\vec{n}).
\end{empheq}

Finally, we mention for completeness, that for manifolds of non-vanishing Hirzebruch signature, absorbing the S-duality non-invariance into a counterterm is not for free. Instead it is now the T-transformation that acts non-trivially \cite{Seiberg:2018ntt}:
\begin{equation}\label{eq:partimaxwellfull1}
	\widetilde{\parti}\qty[X;\tt+1] = \ex{-\frac{1}{3}\pi\ii \sigma} \widetilde{\parti}\qty[X;\tt].
\end{equation}
Thus in those cases the \(\SL(2,\Z)\) duality group indeed suffers from an anomaly. In what follows we will drop the tilde from the partition function, to simplify the notation. In most cases it will not even make a difference, since we will be mainly interested in manifolds with a circle factor where the tilded and the untilded partition functions coincide.

\subsection{\texorpdfstring{Partition functions on \(\S^1\times \Sigma\)}{Partition functions on S¹×Σ}}

Of special importance to our discussion are manifolds of the form \(X=\S^1\times \Sigma\), where \(\Sigma\) is some connected, closed, torsion-free, orientable three-manifold. These are the types of manifolds on which the partition function admits a trace interpretation and the radius \(\beta\) of the circle is the inverse temperature. We will therefore specialise the above discussion to those manifolds and obtain more explicit formulas for the partition functions.

We will make frequent use of various topological characteristics of such manifolds, so we outline those here. First of all, note that for these manifolds, the Künneth formula and Poincaré duality, imply that that all their Betti numbers are determined by just one of them. Namely:
\begin{equation}
	\begin{aligned}
		 & \b_0\qty(\S^1\times \Sigma)  = \b_4\qty(\S^1\times \Sigma)  = 1                                           \\
		 & \b_1\qty(\S^1\times \Sigma) = \b_3\qty(\S^1\times \Sigma)   = 1 + \b_1\qty(\Sigma) = 1 + \b_2\qty(\Sigma) \\
		 & \b_2\qty(\S^1\times \Sigma)  = 2 \b_1\qty(\Sigma) = 2 \b_2(\Sigma).
	\end{aligned}
\end{equation}
The topological basis of harmonic one-forms and two-forms on \(\S^1\times \Sigma\) will then be induced by the respective basis on \(\Sigma\). For the one-forms, a basis is given by the unique normalised harmonic one-form on the circle, \(\frac{\dd{\tau}}{2\pi \beta}\) (in local coordinates),  and the topological basis of harmonic one-forms on \(\Sigma\). As such, the Gram matrix of this basis becomes:
\begin{equation}
	\bbG^{(1)} = \operatorname{diag}\!\qty(\frac{\vol(\Sigma)}{2\pi \beta}, 2\pi \beta\, \bbG^{(1)}_\Sigma),
\end{equation}
where \(\bbG^{(1)}_\Sigma\) is the Gram matrix associated with the topological basis on \(\Sigma\). For the two-forms, we have similarly
\begin{equation}\label{eq:G2}
	\bbG^{(2)} = \operatorname{diag}\!\qty(\beta\,\bbE, \inv{\qty(\beta\,\bbE)}),
\end{equation}
where \(\bbE = 2\pi\bbG^{(2)}_\Sigma\) and we have also used Poincaré duality to bring it in this, symmetric, form. As indicative from the choice of notation, the matrix \(\bbE\) will bear the interpretation of an energy, namely that carried by Wilson and 't Hooft loops placed on various spatial one-cycles. Finally, the intersection matrix is given by
\begin{equation}
	\bbQ = \mqty(\admat{\id_{\b_1(\Sigma)},\id_{\b_1(\Sigma)}}).
\end{equation}

With this in mind, we now go to compute the partition function on \(X=\S^1_\beta\times \Sigma\), starting from \cref{eq:partimaxwellfull1}. First, the instanton contribution, \cref{eq:partiinst}, is
\begin{equation}
	\parti^\inst\qty[\S^1\times \Sigma;\tt] = \sum_{\vec{n},\vec{m}\in\Z^{b^1(\Sigma)}}  \exp(\pi \ii\;
	{\mqty(\trans{\vec{n}}       & \trans{\vec{m}})
	\mqty(\ii \Im \tt\ \beta\,\bbE & \Re \tt\ \id       \\ \Re \tt\ \id & \ii \Im \tt\ \inv{\qty(\beta\,\bbE)})
	\mqty(\vec{n}                                   \\ \vec{k})}
	),
\end{equation}
which becomes
\begin{equation}
	\parti^\inst\qty[\S^1\times \Sigma;\tt] = \qty(\Im\tt)^{-\frac{\b_1(\Sigma)}{2}} \beta^{\frac{\b_1(\Sigma)}{2}} \qty(\det\bbE)^{\frac{1}{2}} \sum_{\vec{n},\vec{m}\in\Z^{\b_1(\Sigma)}} q^{\Delta_{\vec{n},\vec{m}}},
\end{equation}
after a Poisson resummation. In the above we have defined, \(q \coloneqq \ex{-\beta}\), and
\begin{equation}\label{eq:Dnm PI}
	\Delta_{\vec{n},\vec{m}} \coloneqq \frac{1}{2}\frac{\adj{\qty(\vec{n} + \tt\, \vec{m})} \bbE \qty(\vec{n} + \tt\, \vec{m})}{\Im\tt}.
\end{equation}
Adding to that the oscillator contribution, \cref{eq:parti-osc}, and using that \(\bbG^{(1)}_\Sigma = \inv{\qty[\bbG^{(2)}_\Sigma]}\), by Poincaré duality, we get in total
\begin{equation}
	\parti\qty[\S^1\times \Sigma;\tt] = \beta^{\b_1(\Sigma)-1}\frac{\detp\lapl_0}{\sqrt{\detp\lapl_1}}\ \sum_{\vec{n},\vec{m}\in\Z^{b^1(\Sigma)}} q^{\Delta_{\vec{n},\vec{m}}}.
\end{equation}

As far as the determinants are of concern, we have the following:
\begin{equation}
	\spec(\lapl_0,\S^1\times \Sigma) = \set{\frac{(2\pi )^2 k^2}{\beta^2}+\lambda_{\sfn_0},\ k\in\Z, \sfn_0\in\sN_0},
\end{equation}
where \(\lambda_{\sfn_0}\) are the eigenvalues of the scalar laplacian on \(\Sigma\) and \(\sN_0\) is a countable index set. It is then straightforward to calculate, upon zeta-function regularising the determinants and employing Euler's product formula for the hyperbolic sine:
\begin{equation}
	\detp_{\S^1\times \Sigma}\lapl_0 = \beta^{2\b_0(\Sigma)} \prod_{\sfn_0 \in\sN_0^*}\sinh[2](\frac{1}{2}\beta\sqrt{\lambda_{\sfn_0}}),
\end{equation}
where \(\sN_0^*\coloneqq\sN_0\setminus\set{\sfn_0 \suchthat \lambda_{\sfn_0}=0}\). Similarly, the spectrum of the one-form Laplacian is
\begin{align}
	\spec(\lapl_1,\S^1\times \Sigma) = \set{\smash[b]{\underbrace{\frac{(2\pi )^2 n^2}{\beta^2}+\lambda_{\sfn_1}}_{\mathclap{\smqty{\rotcoloneqq \\ \lambda_{n,\sfn_1}}}}},\ n\in\Z, \sfn_1\in\sN_1}&\cup\nn
	 & \hspace*{-7em}\cup\set{\smash[b]{\underbrace{\frac{(2\pi )^2 k^2}{\beta^2}+\lambda_{\sfn_0}}_{\mathclap{\smqty{\rotcoloneqq               \\ \lambda_{k,\sfn_0}}}}},\ k\in\Z, \sfn_0\in\sN_0},
	\vphantom{\underbrace{\frac{(2\pi )^2 n^2}{\beta^2}+\lambda_{\sfn_1}}_{{\smqty{\rotcoloneqq                                                  \\ \lambda_{n,\sfn_1}}}}}
\end{align}
with \(\lambda_{\sfn_1}\) being the eigenvalues of \(\lapl_1\) on \(\Sigma\) and \(\sN_1\) a countable index set, counting the eignevalues of the one-form Hodge Laplacian on \(\Sigma\). The determinant of \(\lapl_1\) is then:
\begin{equation}
	\detp_{\S^1\times \Sigma}\lapl_1 = \beta^{2\qty(\b_1(\Sigma) + b_0(\Sigma))}\qty[\prod_{\sfn_1 \in\sN_1^*}\sinh[2](\frac{1}{2}\beta\sqrt{\lambda_{\sfn_1}})]\ \qty[\prod_{\sfn_0 \in\sN_0^*}\sinh[2](\frac{1}{2}\beta\sqrt{\lambda_{\sfn_0}})],
\end{equation}
with \(\sN_1^*\) defined analogously as \(\sN_0^*\). Therefore (recall \(\b_0(\Sigma)=1\))
\begin{align}
	\frac{\detp\lapl_0}{\sqrt{\detp\lapl_1}} & = \beta^{1-\b_1(\Sigma)} \prod_{\smqty{\sfn_0\in\sN_0^*                                                                               \\ \sfn_1\in\sN_1^*}} \frac{\sinh(\frac{1}{2}\beta\sqrt{\lambda_{\sfn_0}})}{\sinh(\frac{1}{2}\beta\sqrt{\lambda_{\sfn_1}})} \nonumber \\
	                                         & = \beta^{1-\b_1(\Sigma)}\qty[\prod_{\sfn\in\sN_\perp^*} q^{-\frac{1}{2}\sqrt{\lambda_{\sfn}}}\qty(1-q^{\sqrt{\lambda_{\sfn}}})]^{-1},
\end{align}
where \(\lambda_\sfn\) are the eigenvalues of the transversal Laplacian on one-forms on \(\Sigma\), indexed by \(\sfn\in\sN_\perp\), and \(\sN_\perp^*\) collects the non-zero modes. To reach the last equality we have used the spectral properties of the Hodge Laplacian. We will give this quantity a special name:
\begin{equation}
	\upeta_\Sigma(q)\coloneqq \qty[\prod_{\sfn\in\sN_\perp^*} q^{-\frac{1}{2}\sqrt{\lambda_{\sfn}}}\qty(1-q^{\sqrt{\lambda_{\sfn}}})]^{1/2}, \label{eq:eta-def}
\end{equation}
as it serves as a four-dimensional generalisation of the Dedekind eta function. Putting everything together, the partition function takes the very simple form:
\begin{empheq}[box=\obox]{equation}
	\parti\qty[\S^1\times \Sigma;\tt] = \frac{\Theta_\Sigma\qty(q,\tt)}{\upeta_\Sigma(q)^2}, \label{eq:parti-S1-Sigma-pathint}
\end{empheq}
where
\begin{equation}\label{eq:theta-def}
	\Theta_\Sigma\qty(q,\tt) \coloneqq \sum_{\vec{n},\vec{m}\in\Z^{\b_1(\Sigma)}} q^{\Delta_{\vec{n},\vec{m}}},
\end{equation}
is a generalisation of the Siegel--Narain theta function (that depends on the topology of \(\Sigma\)). Observe that this form of the partition function exhibits a remarkable similarity with the two-dimensional formulas of \cref{sec:local state-op} (in particular with \cref{eq:xch 2d})! We will see, in the following section, that this partition function can, too, be interpreted as an extended character; an extended character of the higher-dimensional current algebras of \cref{sec:photonisation-higherKM}.

\subsection{Current algebra and the spectrum}\label{ssec:current-Maxwell}

We now shift gears from the path integral and turn to the current algebras of \cref{sec:photonisation-higherKM}. In keeping with the previous section, we will focus on the \(p=1\) incarnation of the current algebra \cref{eq:p-KM} and expand it in modes, to obtain the algebra of the individual modes. This will allow us to construct representations of the current algebra, and will lead, eventually, to identifying the exact Hilbert space of Maxwell theory on any compact spatial slice in terms of the representations of \cref{eq:p-KM}. Finally, we will compute the characters of the aforementioned representations and show that they reproduce exactly the path integral expressions for the partition functions of Maxwell theory.

The starting point is the current algebra \cref{eq:1-KM}:\footnote{Having allowed for a topological theta term, the phase space variables that led to this algebra are slightly different. It is straightforward, however, to repeat the calculation with a theta term to show that the algebra is unchanged.}
\begin{equation}\label{eq:comm-QL}
	\comm{Q_{\Lambda_1}^\pm}{Q_{\Lambda_2}^\mp}_{\Sigma} = \pm \k \int_{\Sigma} \Lambda_1^{\mp}\w \dd{\Lambda_2^{\pm}},
\end{equation}
where we remind the reader the definition of the codimension-one charges:
\begin{equation}
	Q_\Lambda^\pm[\Sigma] \coloneqq \int_{\Sigma} J^\pm \w \Lambda^\mp,
\end{equation}
with \(J^\pm\) being the self- and anti-self-dual conserved two-form currents, \cref{eq:self-dual-current}, and \(\Lambda^\mp\) the chiral and anti-chiral one-forms, \cref{eq:chiral-forms}, while the (not necessarily integer) level \(\k\) is related to the coupling constant of Maxwell theory as
\begin{equation}
	\k = \frac{\g^2}{2} = \frac{\pi}{\Im\tt}.
\end{equation}
The strategy to obtain a mode algebra is to expand the (pullbacks on \(\Sigma\) of) the forms \(J^\pm\) and \(\Lambda^\mp\) in terms of a suitable basis. Such a basis is given by the eigenforms of the Hodge Laplacian, \(\lapl_p\). However, since \(J^\pm\) is closed, it will only have overlap with the longitudinal eigenform components of the two-form Hodge Laplacian. Relatedly, due to the gauge invariance in defining the charges, up to a shift of \(\Lambda^\mp\) by a closed form, \(\Lambda^\mp\) has overlap only with the transversal eigenforms of the one-form Hodge Laplacian. As such, we consider the (orthonormalised) basis of transversal one-forms on \(\Sigma\), given as
\begin{equation}
	\sB^1_\perp(\Sigma) \coloneqq \set{\phi_\sfn\in \Omega^1(\Sigma) \suchthat \tlapl_1 \phi_\sfn = \lambda_\sfn \phi_\sfn}_{\sfn\in\sN_\perp},
\end{equation}
with \(\tlapl_1\) and \(\sN_\perp\) defined like in \cref{ssec:Z 4d}. It is easy to see, that a basis for the longitudinal two-forms is given in terms of \(\star_\Sigma \phi_\sfn\):
\begin{equation}
	\sB^2_\parallel(\Sigma) \coloneqq \set{\star_\Sigma\phi_\sfn\in \Omega^2(\Sigma) \suchthat \tlapl_1 \phi_\sfn = \lambda_\sfn \phi_\sfn}_{\sfn\in\sN_\perp}.
\end{equation}
Therefore we expand:
\begin{align}
	i^*_\Sigma\Lambda^\mp & = \sum_{\sfn\in\sN_\perp} \Lambda^\mp_\sfn \phi_\sfn         \\
	i^*_\Sigma J^\pm      & = \sum_{\sfn\in\sN_\perp} J^\pm_\sfn \star_\Sigma \phi_\sfn.
\end{align}
In this basis, the current modes define individually conserved currents. In other words:
\begin{equation}
	Q_\sfn^\pm \coloneqq Q_{\Lambda_\sfn^\mp \phi_\sfn}^\pm = J_\sfn^\pm.
\end{equation}
These current, or charge modes obey the following algebra, folowing from \cref{eq:comm-QL}:
\begin{equation}\label{eq:mode-alg Q}
	\comm{Q^\pm_{\sfn}}{Q^\mp_{\sfm}} = \pm\k \ip{\phi_{\sfn}}{\star\dd{\phi_{\sfm}}}_\Sigma.
\end{equation}

There exists a basis which diagonalises the above mode algebra. Let us focus, momentarily, on the non-zero part of the spectrum. On a three-dimensional closed manifold, \(\tlapl_1\) is the square of another self-adjoint operator, the Beltrami operator: \(\star\dd\). The spectrum of \(\star\dd\) consists of \(\pm\sqrt{\lambda_\sfn}\). The eigen-one-forms of \(\star\dd\), and their Hodge stars provide bases for coclosed one-, and closed two-forms respectively, which simultanesously diagonalise \(\tlapl_1\) and \(\conj{\tlapl}_2 \coloneqq\eval{\lapl_2}_{\ker\dd}\). These two bases and can be used to diagonalise \cref{eq:mode-alg Q}.   While, generically, the spectrum of \(\tlapl_1\) is simple, meaning that there are no degeneracies of eigenvalues \cite{encisoNondegeneracyEigenvaluesHodge2010}, for many of the physical applications we have in mind, and more importantly, in the case of this paper, for the state-operator correspondence of \cref{sec:nonlocal state-op}, we are interested in products of spheres. In such cases the spectrum of \(\tlapl_1\) is actually twofold degenerate (cf. \cref{app:Laplacian}). Equivalently, both signs of \(\pm\sqrt{\lambda_\sfn}\) appear in the spectrum of \(\star\dd\). We proceed in the main text with this assumption, in order to connect smoothly with \cref{sec:nonlocal state-op}, and we treat the generic case in \cref{app:KM-simple-spectrum}, where we show that the results of this section remain unchanged. With this disclaimer, we proceed with the basis:
\begin{equation}
	\sV^1_\perp(\Sigma) = \set{\phi_{0\sfi},\phi_{\sfn \sigma}\in \Omega^1(\Sigma) \suchthat \star\dd \phi_{\sfn \sigma} = \sigma\sqrt{\lambda_\sfn} \phi_{\sfn \sigma}}_{\sfn\in\sN_\perp^*}^{\sigma=\pm},
\end{equation}
and its Hodge dual. In the above we have also included all the zero-modes of \(\tlapl_1\), \(\phi_{0\sfi}\), induced by those of \(\lapl_1\). The number of zero-modes is \(\dim\ker\lapl_1 = \b_1(\Sigma) = \b_2(\Sigma)\). Moreover, we have taken this basis to be orthonormalised. In this basis, \(i^*_\Sigma J^\pm\) and \(i^*_\Sigma \Lambda^\mp\) read
\begin{align}
	i^*_\Sigma J^\pm       & = \sum_{\sfi=1}^{\b_2(\Sigma)} J^\pm_{0\sfi} \star_{\Sigma}\phi_{0\sfi} + \sum_{\substack{\sfn\in\sN_\perp^* \\ \sigma=\pm}} J^\pm_{\sfn \sigma} \star_\Sigma \phi_{\sfn \sigma}, \label{eq:J-modeexp} \\
	i^*_\Sigma \Lambda^\mp & = \sum_{\sfi=1}^{\b_1(\Sigma)} \Lambda^\mp_{0\sfi} \phi_{0\sfi} + \sum_{\substack{\sfn\in\sN_\perp^*         \\ \sigma=\pm}} \Lambda^\mp_{\sfn \sigma} \star_\Sigma \phi_{\sfn \sigma},
\end{align}
where \(\sN_\perp^*\) denotes the non-zero part of the spectrum. As before:
\begin{align}
	Q_{0\sfi}^\pm   & \coloneqq Q^\pm_{\Lambda_{0\sfi}^\mp \phi_{0\sfi}} = J^\pm_{0\sfi}, \qq{and}          \\
	Q_{\sfn \sigma} & \coloneqq Q^\pm_{\Lambda^\mp_{\sfn \sigma} \phi_{\sfn \sigma}} = J^\pm_{\sfn \sigma}.
\end{align}
The algebra of the modes \(Q^\pm_{\sfn \sigma}\) is
\begin{empheq}[box=\obox]{equation}
	\comm{Q^\pm_{\sfn \sigma}}{Q^\mp_{\sfm \tau}} = \pm\k\,\sigma\,\sqrt{\lambda_\sfn} \delta_{\sfn\sfm} \delta_{\sigma \tau}. \label{eq:km4-modes}
\end{empheq}
This is a direct four-dimensional analogue of the two-dimensional \(\widehat{\fu}(1)\times\widehat{\fu}(1)\) Kac--Moody algebra \cref{eq:2d-KM}. Before moving on, let us note that we can redefine the modes as\footnote{Note that in Lorentzian signature, the projectors take the form \(\bbP_\pm^\t{L}=\half\qty(\id\pm\ii\star)\). Demanding that the field strength, \(f\), be real implies that \(\conj{J^\pm} = J^\mp\), where the overline denotes complex conjugation. Therefore, \(Q^\mp_{\sfn \sigma}\) is indeed the Hermitian conjugate \(Q^\pm_{\sfn \sigma}\).}
\begin{equation}
	\cA_{\sfn\pm} \coloneqq Q^\pm_{\sfn \pm}, \qq{and} \cA^\dagger_{\sfn\pm} \coloneqq Q^\mp_{\sfn \pm}.
\end{equation}
This redefinition reduces the current algebra to a collection of harmonic oscillators:
\begin{equation}\label{eq:KM-A}
	\comm{\cA_{\sfn \sigma}}{\cA^\dagger_{\sfm \tau}} = \k\,\sqrt{\lambda_\sfn} \delta_{\sfn\sfm}\delta_{\sigma\tau},
\end{equation}
and makes the quantisation of Maxwell theory on arbitrary spatial topology an exercise in quantum mechanics.

We now proceed with quantisation. The Hamiltonian on \(\Sigma\) is simply
\begin{equation}\label{eq:Hamiltonian1}
	H_\Sigma = \frac{1}{2\g^2}\qty(\norm{\vec{E}}_\Sigma^2 + \norm{\vec{B}}_\Sigma^2),
\end{equation}
where the electric and magnetic fields, \(\vec{E}\) and \(\vec{B}\), are defined in terms of pullbacks on \(\Sigma\), of \(\star f\) and \(f\), respectively, in the usual fashion. The underlying current algebra of Maxwell theory allows us to write the Hamiltonian in a Sugawara form. In particular, expressing \(\vec{E}\) and \(\vec{B}\) in terms of the modes \(\cA_{\sfn \sigma}^{(\dagger)}\), we have
\begin{equation}
	H_\Sigma = \frac{1}{\k}\sum_{\sfi=1}^{\b_2(\Sigma)} Q_{0\sfi}^+ Q_{0\sfi}^- + \frac{1}{\k}\sum_{\substack{\sfn\in\sN_\perp^* \\ \sigma=\pm}} \cA^\dagger_{\sfn \sigma}\cA_{\sfn \sigma} + E_0,
\end{equation}
where \(E_0 = \frac{1}{2}\sum_{\sfn,\sigma} \sqrt{\lambda_\sfn}\), is the (potentially infinite) vacuum energy. The Hamiltonian is, then, that of a countably infinite collection of harmonic oscillators, with creation (resp. annihilation) operators \(\cA_{\sfn\pm}^{\dagger}\) (\(\cA_{\sfn\pm}\)), raising (lowering) the energy by \(\sqrt{\lambda_\sfn}\). From here on, the story follows closely its two dimensional analogue of \cref{sec:local state-op}, with slight differences that can all be traced back to the different self-duality properties of middle forms on \(4n\)- versus \((4n+2)\)-dimensional manifolds.

The zero-modes, \(Q^\pm_{0\sfi}\), commute with the Hamiltonian and their eigenstates have a distinguished role. Let \(\ket{j}\), be such a state, with
\begin{equation}\label{eq:primary1}
	Q^\pm_{0\sfi} \ket{j} = j^\pm_\sfi \ket{j}, \qquad \sfi\in\set{1,\cdots, \b_2(\Sigma)},
\end{equation}
that is of highest-weight. Namely, all lowering operators should annihilate the state:
\begin{equation}\label{eq:primary2}
	\cA_{\sfn\pm}\ket{j} \demeqq 0, \qquad \fall\sfn\in\sN^*.
\end{equation}
These states are, then, the \emph{primary states} of the current algebra \cref{eq:km4-modes}. Similarly to the two-dimensional case, flux quantisation determines the eigenvalues \(j^\pm_\sfi\). For any 2-cycle \(C_{(2)\sfi}\in\H_2(\Sigma)\), the magnetic and electric fluxes are quantised:
\begin{align}
	\Phi_\t{mag}  & \coloneqq \int_{C_{(2)\sfi}} f \in 2\pi\,\Z,         \\
	\Phi_\t{elec} & \coloneqq \int_{C_{(2)\sfi}} \check{f} \in 2\pi\,\Z.
\end{align}
In the above, \(\check{f}\) is the magnetic dual field strength, which, in the presence of a theta term, is given by \cite{Verlinde:1988sn}:
\begin{equation}
	\check{f} \coloneqq \Re\tt\ f + \ii \Im\tt\ \star f.
\end{equation}
At the quantum level, we demand that the flux quantisation holds inside expectation values, i.e. \(\ev{\Phi_\t{mag}}{j}\in 2\pi\Z\) and \(\ev{\Phi_\t{elec}}{j} \in 2\pi\Z\). It is clear that only the zero-modes contribute to that expectation value. This fixes exactly the charges, \(j^\pm_\sfi\). A few lines of linear algebra, to switch from the orthonormal basis with which \(J^\pm_{0\sfi}\) were obtained to the topological basis, which is the natural basis for harmonic forms, show that
\begin{equation}\label{eq:jpm-eigenvalue}
	\vec{j}^\pm = \pm \pi\ii\ \frac{\bbJ\qty(\vec{n}+\tt^\mp \vec{m})}{\Im\tt}.
\end{equation}
In the above we combined \(j^\pm_\sfi\) into a \(\b_2(\Sigma)\)-dimensional vector \(\vec{j}^\pm\). Moreover, \(\vec{n}\) and \(\vec{m}\) are \(\b_2(\Sigma)\)-dimensional vectors of integers, \(\tt^\pm \coloneqq \Re\tt \pm \ii \Im\tt\), and the matrix \(\bbJ\) satisfies \(\bbJ^2 = \bbG_\Sigma^{(2)}=\frac{1}{2\pi}\bbE\), in terms of the ``energy'' matrix, defined below \cref{eq:G2}.\footnote{Since \(\bbJ\) is a square root of the energy matrix, the reader might worry about the sign of the square root of each diagonal entry. This is immaterial, however, since the sign can be absorbed in a sign-flip of the components of \(\vec{n}\) and \(\vec{m}\), which simply results in a reshuffling/relabelling of the different sectors of the Hilbert space.} Since the eignevalues of the zero-modes are determined in terms of the vectors of integers, \(\vec{n}\) and \(\vec{m}\), we will henceforth denote the primary states, \(\ket{j}\),  as \(\ket{\vec{n},\vec{m}}\). These states have an energy, above the vacuum energy, that can simply be determined by acting with the Hamiltonian to be
\begin{equation}
	H_\Sigma\ket{\vec{n},\vec{m}} = \Delta_{\vec{n},\vec{m}} \ket{\vec{n},\vec{m}},
\end{equation}
with
\begin{equation}\label{eq:Dnm KM}
	\Delta_{\vec{n},\vec{m}} = \frac{1}{2}\frac{\adj{\qty(\vec{n}+\tt\,\vec{m})}\;\bbE\;\qty(\vec{n}+\tt\,\vec{m})}{\Im\tt}.
\end{equation}
Note that this is the exact same formula as \cref{eq:Dnm PI}. This is not a coincidence. We will clarify the relation between the instanton sectors and the current algebra primaries, in the next section. But first we should finish constructing the Hilbert space.

The remaining states are obtained by acting on \(\ket{\vec{n},\vec{m}}\) with the raising operators, \(\cA^\dagger_{\sfn\pm}\). This results in a Verma module, \(\cV_{\vec{n},\vec{m}}\), spanned by \(\ket{\vec{n},\vec{m}}\) and its descendants:
\begin{equation}\label{eq:generic state}
	\ket{\vec{n},\vec{m};\set{N_{\sfn \sigma}}_{\sfn\in\sN_\perp^*}^{\sigma=\pm}} \coloneqq \prod_{\substack{\sfn\in\sN_\perp^* \\ \sigma=\pm}}\qty(\cA^\dagger_{\sfn \sigma})^{N_{\sfn \sigma}} \ket{\vec{n},\vec{m}},
\end{equation}
where \(N_{\sfn \sigma}\) are positive integers, indicating how many times each creation operator acts, to obtain the desired state. The energy gap between these states and the ground states, at the bottom of the Verma module, is
\begin{equation}
	E_{\set{N_{\sfn \sigma}}} = \sum_{\substack{\sfn\in\sN_\perp^* \\ \sigma=\pm}}N_{\sfn \sigma}\sqrt{\lambda_{\sfn}}.
\end{equation}
Finally, as a matter of convention, we choose the states such that states at different level are orthogonal, and each state has unit norm, \(\braket{\vec{n},\vec{m};\set{N_{\sfn \sigma}}}{\vec{n},\vec{m};\set{M_{\sfn \sigma}}} = \delta_{\set{N_{\sfn \sigma}},\set{M_{\sfn \sigma}}}\). Moreover, states at different topological sectors are orthogonal by construction. Finally, the full Hilbert space of the theory, is simply a direct sum of the Verma modules:
\begin{equation}
	\cH_\Sigma = \bigoplus_{\vec{n},\vec{m}\in\Z^{\b_2(\Sigma)}} \cV_{\vec{n},\vec{m}}.
\end{equation}

To complete the study of the representation theory of our current algebra, what remains is to compute the characters of each of the Verma modules: \(\ch_{\cV_{\vec{n},\vec{m}}}[q]\coloneqq \tr_{\cV_{\vec{n},\vec{m}}} q^H\). We have already performed the difficult task of calculating the energies, so all we have to do now is sum them. This yields, at first:
\begin{equation}
	\ch_{\cV_{\vec{n},\vec{m}}}[q] = q^{\Delta_{\vec{n},\vec{m}}}\ \qty(\prod_{\sfn\in\sN_\perp^*} q^{\half\sqrt{\lambda_{\sfn}}})\ \qty(\prod_{\sfn\in\sN_\perp^*} \sum_{N_\sfn = 0}^\infty q^{N_\sfn \sqrt{\lambda_\sfn}}),
\end{equation}
which, after resumming the geometric series, becomes
\begin{equation}
	\ch_{\cV_{\vec{n},\vec{m}}}[q] = \frac{q^{\Delta_{\vec{n},\vec{m}}}}{\upeta_\Sigma(q)^2}.
\end{equation}
In the last equation we used the definition \cref{eq:eta-def}, of the four-dimensional eta function. The \(\upeta_\Sigma(q)\) term is resumming the contribution of the descendants, while the \(q^{\Delta_{\vec{n},\vec{m}}}\) term is capturing the contribution of the primaries. Finally, summing over the Verma modules yields an extended character of our current algebra. All the descendants give the same contribution, while for the primaries we have to sum all their contributions. This gives:
\begin{equation}\label{eq:xch 4d}
	\xch[q] \coloneqq \sum_{\vec{n},\vec{m}\in\Z^{\b_2(\Sigma)}} \ch_{\cV_{\vec{n},\vec{m}}}[q] = \frac{\Theta_\Sigma(q;\tt)}{\upeta_\Sigma(q)^2}.
\end{equation}

Comparing with \cref{eq:parti-S1-Sigma-pathint}, we see a remarkable identification of the \(\S^1\times \Sigma\) partition function, as computed through the path integral, with an extended character of an infinite-dimensional algebra:
\begin{empheq}[box=\obox]{equation}\label{eq:Z=xch 4d}
	\parti\qty[\S^1_\beta\times \Sigma;\tt] = \xch[q], \qquad q=\ex{-\beta}.
\end{empheq}
Let us comment on this result. Identifying the nome with \(\ex{-\beta}\), gives a physical meaning to the character, \(\xch[q]\). It is simply a thermal trace over the Hilbert space over \(\Sigma\). From this point of view, it ought to reproduce the path integral expression. It is known, however, that there are subtleties in matching the two quantities. The compatibility of the manifestly covariant path integral formalism with the manifestly unitary canonical formalism has been a source of confusion \cite{Vassilevich:1991rt,Vassilevich:1992ad,Fukutaka:1992si,Vassilevich:1994cz}, ultimately resolved in \cite{Donnelly:2013tia} by arguing the form of the canonical partition function. Our formula \cref{eq:Z=xch 4d}, provides a more explicit manifestation of Donnelly and Wall's unitarity proof \cite{Donnelly:2013tia}. Recovering unitarity is an important point, as it was a crucial ingredient for the photonisation argument of \cref{sec:photonisation-higherKM}, which ultimately led to the higher-dimensional current algebras.

It is interesting to keep track of which quantity is mapped to what in the identification \(\cref{eq:parti-S1-Sigma-pathint}=\cref{eq:xch 4d}\). In \cref{eq:parti-S1-Sigma-pathint}, the Theta function stems from the sum over instantons. This counts, effectively, insertions of Wilson--'t Hooft operators in Maxwell theory \cite{Verlinde:1988sn}. On the other hand, on the canonical side, the Theta function comes from the primary states. Both of them are subsequently decorated with oscillators that give rise to the eta functions. This story is pointing at a special connection between the spectrum of operators and states in Maxwell theory, which we shall explicate in the following section.

\subsection{\texorpdfstring{The spectrum on \(\S^2\times\S^1\)}{The spectrum on S²×S¹}}\label{ssec:S2xS1}

Among all three-dimensional topologies, of particular interest to us is \(\Sigma\cong\S^2\times \S^1\), for reasons that will become evident in the next section. Here we briefly specialise the generic results of \cref{ssec:current-Maxwell} and collect the relevant formulas, for the case of \(\S^2\times\S^1\).

First, the geometry. Using a Weyl rescaling of the spacetime metric, we can set the radius of \(\S^1\) to be unity and the only parameter is the radius of the sphere, or, really, the dimensionless ratio of sphere over the circle radius, which we will denote as \(r_0\). We will refer to this spatial slice as \(\Sigma_{r_0}\). Its metric is:
\begin{equation}
	\dd{s^2_{\Sigma_{r_0}}} = r_0^2 \qty(\dd{\theta^2}+\sin^2\theta\dd{\varphi^2}) + \dd{\eta^2},
\end{equation}
where \(\theta\in\ropen{0}{\pi}\) and \(\varphi\in\ropen{0}{2\pi}\) are coordinates on the sphere and \(\eta\in\ropen{0}{2\pi}\) is the coordinate on the circle.

As we review in \cref{app:Laplacian}, on \(\S^2\times\S^1\) there is a single zero-mode of the one- or the two-form Hodge Laplacian. The rest of the modes are labelled by the angular and magnetic quantum numbers, \(\ell\) and \(m\),  on the sphere, and the momentum, \(k\)  on the circle, as well as the Beltrami label, \(\sigma=\pm\). The eigenvalue of the Laplacian of a mode labelled by \(\sfn=(\ell,m,k)\) is
\begin{equation}\label{eq:lambda(r0)}
	\lambda_\sfn(r_0) = \frac{\ell(\ell+1)}{r_0^2}+k^2.
\end{equation}
The Hamiltonian on this slice reads:
\begin{equation}
	H_{\Sigma_{r_0}} = \frac{1}{\k} Q_0^+\,Q_0^- + \frac{1}{\k}\sum_{\substack{\sfn=(\ell,m,k) \\ \sigma=\pm}} \cA^\dagger_{\sfn \sigma} \cA_{\sfn \sigma} + E_0.
\end{equation}

The primary states, are labelled by two integers: \(\ket{n,m}\), with \(n,m\in\Z\). The unique harmonic two-form on \(\Sigma_{r_0}\) (normalised as in \cref{eq:topo basis 2d}) is
\begin{equation}
	\tau^{(2)} = \frac{1}{4\pi}\sin^2 \theta \dd{\theta}\w\dd{\varphi},
\end{equation}
and yields \(\bbG_{\Sigma_{r_0}}^{(2)} = \inv{(2 r_0^2)}\). This gives, via \cref{eq:jpm-eigenvalue,eq:Dnm KM}, the charge and the energy of the of the primary states, as:
\begin{align}
	j^\pm        & = \pm \pi\ii \frac{1}{\sqrt{2} r_0}\frac{n+\tt^\mp m}{\Im\tt}, \qq{and} \label{eq:Jnm S2xS1} \\
	\Delta_{n,m} & = \frac{\pi}{2 r_0^2}\frac{\abs{n+\tt\,m}^2}{\Im\tt}, \label{eq:Dnm S2xS1}
\end{align}
respectively.\footnote{This formulas may look funny, on dimensional grounds, but remember that \(r_0\) is a dimensionless parameter. To reinstate dimensionful parameters, the energy is proportional to the Gram matrix, which, in this case, is \(\frac{\vol(\t{circle})}{\vol(\t{sphere})}\), which has the correct dimensions to be an energy.} The rest of the states are just obtained by acting with creation operators, like in \cref{eq:generic state}. For example, the first excited states are obtained by acting with \(\cA_{\sfn \sigma}^\dagger\), with \(\sfn=(1,m,0)\) or \(\sfn=(0,0,1)\) depending on the value of \(r_0\) and have energy
\begin{equation}
	E_{1,m,0}(r_0) = \frac{\sqrt{2}}{r_0} \qq{and} E_{0,0,1}(r_0) = 1,
\end{equation}
above the ground-state energy, respectively.

\section{The state-operator correspondence}\label{sec:nonlocal state-op}

We now have all the ingredients to arrive at the culmination of this paper; the nonlocal state-operator correspondence. While the matching of the partition function and the current algebra extended character is valid on any connected, closed, torsion-free, orientable three-manifold, for the state-operator correspondence we need to choose a specific manifold. In this paper, we will only treat the easiest case, deferring more complex cases for future investigation. We restrict ourselves with the topology \(\Sigma \cong \S^2\times\S^1\).

\begin{figure}[!htb]
	\centering
	\def\svgwidth{0.8\textwidth}
	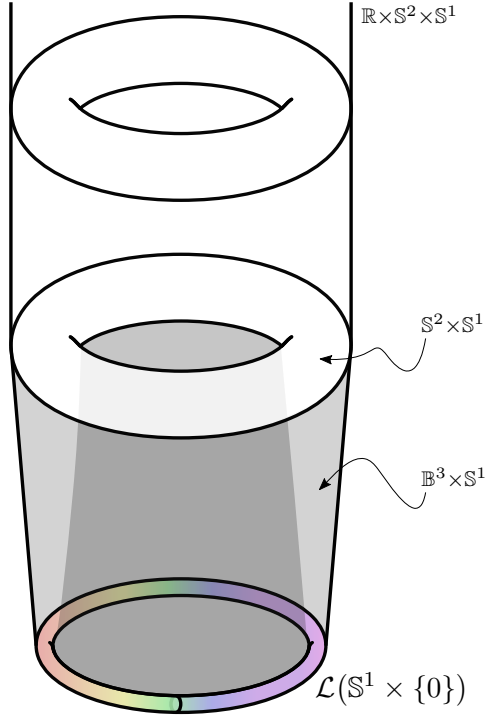
	\caption{The nonlocal state-operator correspondence. Any state on \(\S^{2}\times\S^1\) can be prepared by a path integral on \(\B^{3}\times\S^1\) with a line operator, \(\cL\qty(\S^1\times\set{0})\), at the centre of the ball, wrapping the \(\S^1\). The state then evolves in time on the Lorentzian ``cylinder'' \(\R\times\S^{2}\times\S^1\).}
	\label{fig:state-operator-nonlocal}
\end{figure}

The picture we have in mind here is the following. States on \(\cH_\Sigma\) are prepared by a path integral on a manifold whose boundary is \(\Sigma=\S^2\times\S^1\), with various operator insertions, which we will clarify momentarily. Besides local operators, four-dimensional Maxwell theory is host to a plethora of line operators (which we classify in \cref{ssec:line-ops}). We place these operators on the \(\S^1\). This leaves us with filling-in the \(\S^2\) to obtain a three-ball, \(\B^3\). As we will see, the path integral on \(Y\coloneqq \B^3\times\S^1\), with line operators inserted on \(\S^1\times\set{0}\), where by \(\set{0}\) we denote the origin on \(\B^3\), produces all the states, \cref{eq:generic state}, of \(\cH_\Sigma\) that we described above. See \cref{fig:state-operator-nonlocal}. In notation similar to \cref{sec:local state-op} we denote these states as
\begin{equation}\label{eq:state L}
	\ket{\cL} \coloneqq \int_{\cC[\,\cdot\,]} \DD{a} \ex{-S[a]}\ \cL\qty(\S^1\times\set{0}),
\end{equation}
where \(\cL\) is some line operator, and \(\cC[\,\cdot\,]\) is, again an appropriate functional space over \(Y\) (now encapsulating information about gauge equivalent configurations), pending additional input regarding boundary conditions on \(\pd Y\). Once boundary conditions, say \(a(\pd Y)=a_\pd\), are imposed, the path integral \cref{eq:state L} produces the wavefunctional
\begin{equation}
	\Psi_\cL\qty[a_\pd] = \braket{a_\pd}{\cL}.
\end{equation}
Here we will show that these states reproduce the Hilbert space of \cref{ssec:current-Maxwell} entirely.

\subsection{Line operators}\label{ssec:line-ops}

First we outline the possible line operators in our disposal. Since in pure Maxwell theory there are no local charged operators, the only gauge-invariant line operators must be supported either on closed loops or infinitely extending lines. In what follows we will only consider closed loops (which we will denote as \(\gamma\)), as we are dealing with compact spaces. Infinitely extending lines can be obtained as a limit. The basic kind of line operator there is, is a Wilson loop:
\begin{equation}\label{eq:wilson}
	\W_n(\gamma) \coloneqq \exp(\ii\, n\, \int_\gamma a), \qquad n\in\Z.
\end{equation}
These operators are electrically charged under the one-form symmetry of Maxell theory, which served as the starting point of photonisation (cf. \cref{sec:photonisation-higherKM}). In particular, if
\begin{equation}
	U^\t{elec}_C \coloneqq \exp(\ii \int_C \star f),
\end{equation}
is the generator of the electric one-form symmetry, where \(C\) is a closed two-dimensional surface, \(\W_n(\gamma)\) carries charge \(n\):
\begin{equation}
	U^\t{elec}_C\cdot\W_n(\gamma) = \exp(\frac{2\pi\,\ii\, n}{\Im\tt}\; \t{Link}\qty(C,\gamma))\W_n(\gamma),
\end{equation}
where \(\t{Link}\qty(C,\gamma)\) denotes the linking number between the closed surface \(C\) and the loop \(\gamma\).

Furthermore, there exist magnetically charged line operators. As we explained in \cref{sec:photonisation-higherKM}, and as is well-understood in Maxwell theory \cite{Gaiotto:2014kfa}, there is also a magnetic one-form symmetry, with generators (in the absense of a theta term)
\begin{equation}
	U^\t{mag}_C \coloneqq \exp(\ii \int_C f).
\end{equation}
The operators charged under this symmetry are 't Hooft loops. In the electric presentation of the theory these are described by excising a loop from spacetime, providing boundary conditions for the gauge-fields that fix the value of \(\int f\) on a thin tube surrounding the loop. They can be written more compactly, however, in the magnetic presentation, where we exchange the gauge-field for a magnetic gauge-field, \(\check{a}\), whose curvature is \(\check{f} = \Re\tt\, f + \ii \Im\tt \star f\). The 't Hooft loops become, in this presentation, Wilson loops for the magnetic photon:
\begin{equation}
	\t{'tH}_m(\gamma) \coloneqq \exp(\ii\, m\int_\gamma\check{a}), \qquad m\in\Z.
\end{equation}
and carry ``electric'' charge in the magnetic presentation:
\begin{equation}
	\exp(\ii \int_C \star \check{f})\cdot \t{'tH}_m(\gamma) = \exp(\frac{2\pi\,\ii\, m\, \abs{\tt}^2}{\Im\tt} \t{Link}\qty(C,\gamma))\ \t{'tH}_m(\gamma).
\end{equation}
In the original, electric, presentation, they carry magnetic charge:
\begin{equation}
	U^\t{mag}_C\cdot \t{'tH}_m(\gamma) = \exp(2\pi\,\ii\, m\;\t{Link}\qty(C,\gamma))\,\t{'tH}_m(\gamma).
\end{equation}
In the presence of a theta term, they also carry electric charge, due to the Witten effect. This is given by:
\begin{equation}
	U^\t{elec}_C\cdot \t{'tH}_m(\gamma) = \exp(2\pi\, \ii\, m\,\frac{\Re\tt}{\Im\tt}\;\t{Link}\qty(C,\gamma))\,\t{'tH}_m(\gamma).
\end{equation}

More generally, we can consider dyonic, or Wilson--'t Hooft, line operators labelled both by an electric and a magnetic charge:
\begin{equation}\label{eq:W-tH}
	\W_{n,m}(\gamma) \coloneqq \exp(\ii\, n \int_\gamma a + \ii\,m \int_\gamma \check{a}), \qquad n,m\in\Z.
\end{equation}
Simultaneously, we will combine the one-form symmetries into their irreducible, self- and anti-self-dual, incarnations, as in \cref{sec:photonisation-higherKM}:
\begin{equation}
	U^\pm_C \coloneqq \exp(\ii\int_C J^\pm).
\end{equation}
The Wilson--'t Hooft lines carry charge under \(U^\pm_C\) as:
\begin{equation}
	U^\pm_C\cdot \W_{n,m} = \exp(\pm 2\pi \frac{n+\tt^\pm m}{\Im\tt} \t{Link}(C,\gamma)) \W_{n,m}.
\end{equation}

\begin{figure}[!htbp]
	\centering
	\def\svgwidth{0.9\textwidth}
	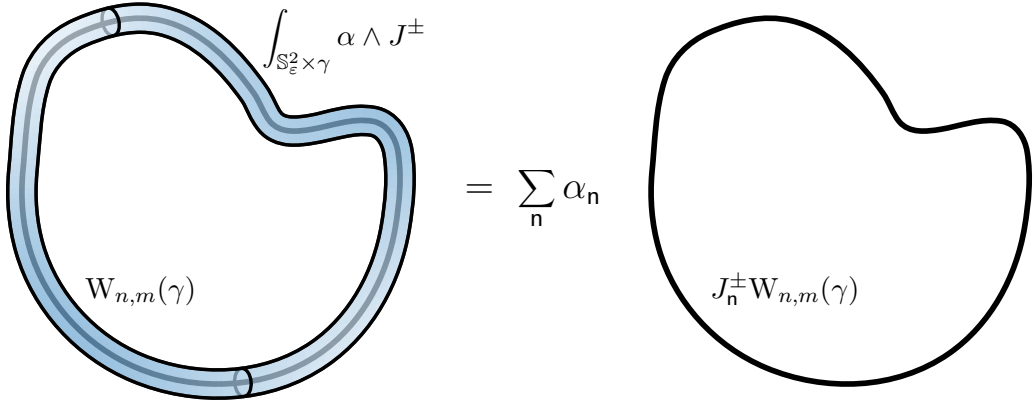
	\caption{An example of descendant line operators. On the left-hand-side of the equation, a small tube (depicted in \blue{blue}, modulated to indicate different ``density'' along the circle) of the operator \(J^\pm\) smeared over a function, \(\alpha\), surrounds the curve \(\gamma\) which supports a Wilson--'t Hooft line. On the right-hand-side is the result of shrinking the tube on the line, namely a series of new, descendant, line operators.}
	\label{fig:surrounding}
\end{figure}

These are the basic, or primary, in a sense to be made precise below, line operators. But this is not the complete story. We can obtain composite, or descendant (in the same to-be-made-precise sense) line operators by smearing functions of gauge-invariant local operators on \(\gamma\). The only gauge-invariant local operators in our disposal are polynomials of (derivatives of) \(J^\pm\). An unambiguous way to smear the local operators over the basic line operators, is to excise an infinitesimal tube around \(\gamma\), i.e. an \(\S^2_\varepsilon\times \gamma\), on which we spread the desired local operator, and shrink it onto \(\gamma\) by taking the limit \(\varepsilon\to 0\). This is completely analogous to how disorder operators, including the aforementioned basic 't Hooft loops (in the electric presentation), are defined in quantum field theory. Using the mode expansion of \(J^\pm\) on a three-dimensional surface \cref{eq:J-modeexp}, and the OPE of the currents we can convert the complicated local operator into sums of powers of the modes, \(J^\pm_\sfn\), of \(J^\pm\). A sketch of this construction, for a specific operator, is shown in \cref{fig:surrounding}. In the next few sections we will make this picture precise.

\subsection{Radial evolution}\label{ssec:radial-evolution}

A crucial feature of the Hilbert space on \(\S^2\times\S^1\), is its Verma module structure under the Kac--Moody algebra. In order to reproduce this structure, via a path integral on \(Y=\B^3\times \S^1\), we need to find an expression of the current modes, \(J^\pm_\sfn\), that can be inserted in the path integral. Key to doing so is solving the self-duality and conservation equations, for \(J^\pm\) on \(Y\). We take the metric on \(Y\) to be,
\begin{equation}
	\dd{s_Y^2} = \dd{r^2} + \underset{\blue{\smqty{\rotcoloneqq \\ \dd{s_{\Sigma_r}^2}}}}{\blue{\underbrace{\textcolor{black}{r^2 \dd{s_{\S^2}} + \dd{s_{\S^1}^2}}}}},
\end{equation}
i.e. we fix the \(\S^1\) radius to be one, throughout, and label the spatial slice where the \(\S^2\) is of radius \(r\in\closed{0}{r_0}\) by \(\Sigma_r\), where \(\Sigma_{r_0}\) is the slice on which the states, in \cref{ssec:S2xS1}, are prepared. With these conventions, it is useful to decompose the currents as,
\begin{equation}
	J^\pm = \dd{r}\w J_r^\pm + J^\pm_{\Sigma_r},
\end{equation}
where \(J_r^\pm\in \Omega^1\qty(\Sigma_r)\) and \(J_{\Sigma_r}^\pm\in \Omega^2\qty(\Sigma_r)\). In this decomposition, the self-duality equations, \(\star_Y J^\pm = \pm J^\pm\), are expressed as a relation between \(J^\pm_r\) and \(J^\pm_{\Sigma_r}\):
\begin{equation}\label{eq:sd'}
	J^\pm_r = \pm\star_r J^\pm_{\Sigma_r}.
\end{equation}
Here, and in the following, we denote by \(\star_r\), the Hodge star on \(\Sigma_r\), induced by the metric on the slices. Moreover, from the conservation equations, \(\dd{J^\pm}=0\), we read a Gauss law and a dynamical equation controlling the radial evolution of the current. The Gauss law, written equivalently in terms of \(J_{\Sigma_r}^\pm\) or \(J^\pm_r\), using \cref{eq:sd'}, reads:
\begin{equation}
	\sd{J^\pm_{\Sigma_r}} = 0 \qquad\Leftrightarrow\qquad \adj{\sd} J^\pm_r = 0,
\end{equation}
The bold differential, \(\sd\), denotes the differential on \(\Sigma_r\) and the codifferential is defined with respect to the Hodge-star on \(\Sigma_r\). Similarly, the radial evolution equations both in terms of \(J_{\Sigma_r}^\pm\) and \(J^\pm_r\) read:
\begin{equation}\label{eq:radial}
	\pd_r J^\pm_{\Sigma_r} \mp \sd\star_r J^\pm_{\Sigma_r} = 0 \qquad\Leftrightarrow\qquad \star_r\pd_r\star_r J^\pm_r \mp \star_r \sd J^\pm_r = 0.
\end{equation}
As explained in \cref{ssec:current-Maxwell,app:Laplacian}, \(J^\pm_{\Sigma_r}\), as a closed two-form, can be expanded in the (orthonormalised) basis, \(\sV^2_\parallel\qty(\Sigma_r)\), given by \cref{eq:Vbasis}, of Hodge-duals of eigen-one-forms of the Beltrami operator on \(\Sigma_r\):
\begin{equation}\label{eq:expansion-Sigma-r}
	J^\pm_{\Sigma_r} = J^\pm_0(r) \star_r \phi_0(r) + \sum_{\substack{\sfn\in\sN_\perp^* \\ \sigma=\pm}} J^\pm_{\sfn \sigma}(r) \star_r \phi_{\sfn \sigma}(r),
\end{equation}
where we now explicitly denote the dependence of the coefficients, \(J^\pm_\sfn(r)\), and the basis forms, \(\phi_{\sfn \sigma}(r)\),  on \(r\). Plugging this expansion into \cref{eq:radial}, yields a system of first order ordinary differential equations for the coefficients:
\begin{equation}\label{eq:diffeq-J}
	\begin{aligned}
		\pd_r J_0^\pm(r) + \frac{1}{r} J_0^\pm(r)                                                  & = 0  \\
		\pd_r J_{\sfn \sigma}^\pm(r) + \qty[\bbA^\pm_{\sfn}(r)]_{\sigma \tau} J_{\sfn \tau}^\pm(r) & = 0,
	\end{aligned}
\end{equation}
together with boundary conditions fixing \(J_0^\pm(r_0)=Q_0^\pm\), and \(J^\pm_{\sfn \sigma}(r_0)=Q^\pm_{\sfn \sigma}\). These are precisely the modes that were used to build creation and annihilation operators on \(\Sigma_{r_0}\). Note that this differential equation factorises modes of different energy, i.e. different \(\sfn\), but mixes modes on the \(\sigma\tau\)-plane which we shall call the Beltrami plane.

The differential equation for the zero-mode, can be immediately integrated, yielding
\begin{equation}\label{eq:J0-solution}
	J_0^\pm(r) = \frac{r_0}{r}\, Q_0^\pm.
\end{equation}
For the non-zero-modes, the matrix \(\bbA^\pm_{\sfn}(r)\) is given as
\begin{equation}\label{eq:Amatrix}
	\qty[\bbA^\pm_{\sfn}(r)]_{\sigma \tau} \coloneqq \ip{\phi_{\sfn \sigma}}{\star_r\pd_r\star_r \phi_{\sfn \tau}} \mp \ip{\phi_{\sfn \sigma}}{\star_r\sd\phi_{\sfn \tau}}.
\end{equation}
We provide details about the explicit construction of the basis \(\sV\qty(\Sigma_r)\), as well as all the entries of the matrix \(\bbA^\pm_{\sfn}\) in \cref{app:Laplacian,app:matrices}. The solution of the above equation takes the form of a radially ordered integral:\footnote{That is, simply, a time-ordered integral, with the radius playing the role of time.}
\begin{align}\label{eq:Jn-solution}
	J^\pm_{\sfn \sigma}(r) & = \qty[\bbU^\pm_{\sfn}(r,r_0)]_{\sigma \tau} Q^\pm_{\sfn \tau}, \qq{with} \\
	\bbU^\pm_{\sfn}(r,r_0) & \coloneqq \rexp(\int_r^{r_0} \dd{r'} \bbA^\pm_{\sfn}(r)).
\end{align}
Altogether \(J^\pm_{\Sigma_r}\) reads, then
\begin{equation}
	J^\pm_{\Sigma_r} = Q_0^\pm \frac{r_0}{r} \star_r \phi_0(r) + \sum_{\sfn,\sigma, \tau} Q^\pm_{\sfn \sigma} \qty[\bbU^\pm_{\sfn}(r,r_0)]_{\sigma\tau}\star_r \phi_{\sfn \tau}(r).
\end{equation}

For illustration purposes, let us concentrate temporarily on a subfamily of the basis \(\sV\qty(\Sigma_r)\), returning subsequently to the general case, deferring the details to the appendix. This subfamily concerns the eigenforms with no momentum along the \(\S^1\) of \(\Sigma_r\), i.e. the one-forms
\begin{equation}\label{eq:phi lm}
	\phi_{\ell m \sigma}(r) \coloneqq \frac{1}{\sqrt{2}}\qty(\Psi_{\ell m}\dd{\eta} + \sigma\qty(\frac{\ell(\ell+1)}{r^2})^{-1/2}\star_r\sd{\qty(\Psi_{\ell m}\dd{\eta})}).
\end{equation}
In the above, \(\Psi_{\ell m}\), are the real spherical harmonics, normalised on \(\S^2_r\), with \(\ell\) and \(m\) the usual angular momentum and magnetic quantum numbers, with \(\ell\neq 0\), and \(\eta\) is the coordinate on \(\S^1\). One can check that these are orthonormalised eigenforms of the Beltrami operator on \(\Sigma_r\),  with eigenvalue $\sigma\qty(\frac{\ell(\ell+1)}{r^2})^{1/2}$. For this family the matrix \(\bbA^\pm_{\sfn}(r)\) becomes
\begin{equation}
	\bbA^\pm_{\ell m}(r) = \frac{1}{2r}\mqty(1\mp 2\sqrt{\ell(\ell+1)} & 1 \\ 1 & 1\pm 2\sqrt{\ell(\ell+1)}).
\end{equation}
In this case \(\bbA^\pm_{\ell m}(r)\) commutes with itself at different radii so the radially ordered exponential reduces to a regular exponential, yielding, finally as solution
\begin{equation}\label{eq:Ulm}
	\begin{aligned}
		\bbU_{\ell m}^\pm(r,r_0) = \frac{1}{2+4\ell}\Bigg[\qty(\frac{r}{r_0})^{-\ell-1}\mqty(1+2\ell\mp 2\sqrt{\ell(\ell+1)} & 1  \\ 1 & 1+2\ell\pm 2\sqrt{\ell(\ell+1)}) & \\[0.5em]
		+ \qty(\frac{r}{r_0})^{\ell}\mqty(1+2\ell\pm2\sqrt{\ell(\ell+1)}                                                     & -1 \\ -1 & 1+2\ell\mp2\sqrt{\ell(\ell+1)})& \Bigg].
	\end{aligned}
\end{equation}

Let us return to the general case. With the solutions \cref{eq:J0-solution,eq:Jn-solution}, we can equivalently express the operators acting on \(\Sigma_{r_0}\) in terms of the fields \(J^\pm\), as
\begin{align}
	Q_0^\pm             & = \int_{\Sigma_r} \frac{r}{r_0} \phi_0(r) \w J^\pm_{\Sigma_r}                                    \label{eq:Q0} \\
	Q_{\sfn \sigma}^\pm & = \int_{\Sigma_r} \qty[\bbU^\pm_{\sfn}(r,r_0)]_{\sigma \tau}^{-1} \phi_{\sfn \tau}(r) \w J^\pm_{\Sigma_r}.
\end{align}
Note how, as they should, the modes \(Q_i^\pm\) (where \(i\) stands either for \(0\) or \(\sfn \sigma\)), are actually independent of \(r\). In other words, connecting to the discussion of \cref{ssec:current-Maxwell}, \(\inv{\bbU^\pm_{i}(r,r_0)}\cdot\phi_{i}(r)\) provide a basis of (anti-)chiral one-forms on \(Y\):
\begin{equation}
	\bbP_\pm\dd\qty(\inv{\bbU^\pm_{i}(r,r_0)}\cdot\phi_{i}(r)) = 0,
\end{equation}
subject to a gauge condition. In this case, the natural gauge condition is given by the Coulomb gauge, that they lie in the kernel of \(\sd\star_r i^*_{\Sigma_r}\).

Knowing the expression of the charges in terms of the field content, we can now obtain how such charges act on states, \(\ket{\cL}\), of the form \cref{eq:state L}. We simply perform the path integral inserting \(Q^\pm_i\) on a small \(\Sigma_r\), and perform the OPE with the path integral insertion sitting at the centre, sending subsequently \(r\to 0\):
\begin{equation}\label{eq:Q|L>}
	Q_i^\pm \ket{\cL} = \lim_{r\to 0}\int_{\cC[\,\cdot\,]} \DD{a} \ex{-S[a]}\ \int_{\Sigma_r} \inv{\bbU^\pm_{i}(r,r_0)}\cdot\phi_{i}(r) \w J^\pm_{\Sigma_r} \times \cL\qty(\S^1\times\set{0}),
\end{equation}
where, in \cref{eq:Q|L>}, only smooth configurations of \(J^\pm_{\Sigma_r}\) contribute, for the same reasons as in \cref{sec:local state-op}. This is a direct four-dimensional analogue of \cref{eq:2d J|0>}. The behaviour of this insertion as \(r\to 0\), requires some clarification. Let us illustrate the situation for the empty state:
\begin{equation}
	\ket{\mathbf{1}} \coloneqq \int_{\cC[\,\cdot\,]} \DD{a} \ex{-S[a]}.
\end{equation}

The radial evolution matrix, \(\bbU_{\sfn}^\pm(r,r_0)\), gives rise to a smooth mode, behaving as 
\begin{align}
	&\sim\qty(\frac{r}{r_0})^\ell
\intertext{at the origin, \(r\to 0\), and a divergent one, behaving as}
	&\sim\qty(\frac{r}{r_0})^{-\ell-1}.
\end{align}
In order to single out the configurations of the currents that contribute to the path integral, we must project out the divergent modes. This is implemented by the projector to the kernel of \(\bbU_\sfn^\pm(0,r_0)\), which we will denote as \(\pr_{\sfn}^\pm\).\footnote{Note that \(\bbU_\sfn^\pm(0,r_0)\) from \cref{eq:Ulm} (cf. also \cref{app:matrices}) has rank 1, reflecting precisely the fact that there is a smooth and a divergent mode.} Therefore, the part of \(J^\pm_{\Sigma_r}\) that contributes in the path integral has an expansion:
\begin{equation}\label{eq:J-smooth}
	\eval{J^\pm_{\Sigma_r}}_\t{smooth} = Q_0^\pm \blue{\underset{\sim r \sin\theta\, \dd{\varphi}\w\dd{\theta}}{\underbrace{\textcolor{black}{\frac{r_0}{r} \star_r \phi_0(r)}}}} + \sum_{\sfn,\sigma, \rho,\tau} Q^\pm_{\sfn \sigma}\, \textcolor{green}{\underset{\t{basis of smooth two-forms}}{\underbrace{\textcolor{black}{\pr_{\sfn \sigma \tau}^\pm \qty[\bbU^\pm_{\sfn}(r,r_0)]_{\tau \rho} \star_r \phi_{\sfn \rho}(r)}}}},
\end{equation}

From this it immediately follows that the empty state is uncharged:
\begin{equation}
	Q_0^\pm\ket{\mathbf{1}} = \lim_{r\to 0}\int_{\cC[\,\cdot\,]} \DD{a} \ex{-S[a]}\ \int_{\Sigma_r} \frac{r}{r_0} \phi_0(r) \w J^\pm_{\Sigma_r} = 0,
\end{equation}
as the smooth part of \(J^\pm_{\Sigma_r}\) goes to zero at \(r\to 0\). The action of the non-zero modes is also straighforward:
\begin{equation}
	Q_{\sfn \sigma}^\pm\ket{\mathbf{1}} =\lim_{r\to 0}\int_{\cC[\,\cdot\,]} \DD{a} \ex{-S[a]}\ \int_{\Sigma_r} \qty[\bbU^\pm_{\sfn}(r,r_0)]_{\sigma \tau}^{-1} \phi_{\sfn \tau}(r) \w J^\pm_{\Sigma_r} = \int_{\cC[\,\cdot\,]} \DD{a} \ex{-S[a]}\ \pr^\pm_{\sfn \sigma \tau} Q^\pm_{\sfn \tau},
\end{equation}
where here, and henceforth, repeated indices \(\set{\sigma,\tau,\ldots}\) on the Beltrami plane are contracted. From this it follows that, if \(v^\pm_{\sfn}\) is any vector in the image of \(\bbU_\sfn^\pm(0,r_0)\), i.e. orthogonal to \(\pr_\sfn^\pm\), the combination \(v_{\sfn \sigma}^\pm Q_{\sfn \sigma}^\pm\) annihilates the empty state:
\begin{equation}
	v_{\sfn \sigma}^\pm Q_{\sfn \sigma}^\pm \ket{\mathbf{1}} = 0.
\end{equation}
This behaviour is in fact general. This linear combination of charges will take the role of annihilation operators, for path integral states, with its Hermitian conjugate becoming the new creation operators.

\subsection{Squeeze, squeeze, squeeze}\label{ssec:squeeze}

Let us make the above intuition precise. As before, we take \(v^\pm_\sfn\) to be a vector orthogonal to \(\pr_\sfn^\pm\),
\begin{equation}
	\pr_{\sfn}^\pm \cdot v^\pm_{\sfn} = 0.
\end{equation}
Of course, since \(\pr_\sfn^\pm\) is a rank-1 projector on a two-dimensional plane, \(v_\sfn^\pm\) is unique, up to rescaling and a global phase. We will take \(v_\sfn^\pm\) to be such that
\begin{equation}\label{eq:bogoliubov-normalisation}
	\qty(v_\sfn^\pm)^\dagger\cdot \vec{\sigma}_z \cdot v_\sfn^\pm = \pm 1,
\end{equation}
where \(\vec{\sigma}_z\) is the Pauli \(z\)-matrix. With this choice of \(v^\pm_\sfm\), it is straightforward to check that the operators
\begin{equation}\label{eq:bogoliubov1}
	\begin{array}{ccccc}
		\cB_{\sfn\pm}         & \coloneqq & v_{\sfn \sigma}^\pm Q_{\sfn \sigma}^\pm        & = & v_{\sfn\pm}^\pm \cA_{\sfn\pm} + v_{\sfn\mp}^\pm \cA_{\sfn\mp}^\dagger,               \\[0.3em]
		\cB^\dagger_{\sfn\pm} & \coloneqq & \conj{v}_{\sfn \sigma}^\pm Q_{\sfn \sigma}^\mp & = & \conj{v}_{\sfn\pm}^\pm \cA^\dagger_{\sfn\pm} + \conj{v}_{\sfn\mp}^\pm \cA_{\sfn\mp}, \\
	\end{array}
\end{equation}
where the overline denotes complex conjugation, satisfy the correct Kac--Moody algebra of ladder operators:
\begin{equation}\label{eq:KM-B}
	\comm{\cB_{\sfn \sigma}}{\cB_{\sfm \tau}^\dagger} = \k\sqrt{\lambda_{\sfn}(r_0)} \delta_{\sfn\sfm}\delta_{\sigma \tau},
\end{equation}
with all other commutators vanishing.

Notice, however, that \cref{eq:bogoliubov1} is a Bogoliubov transformation. The new set of oscillators can be obtained by a unitary transformation of the old oscillators:
\begin{equation}\label{eq:bogoliubov-squeeze}
	\begin{aligned}
		\fS_\sfn\qty(v_\sfn) \cA_{\sfn\pm} \fS^\dagger_\sfn\qty(v_\sfn)                       & = \cB_{\sfn\pm} \qq{and} \\
		\fS_\sfn\qty(\conj{v}_\sfn) \cA_{\sfn\pm}^\dagger \fS^\dagger_\sfn\qty(\conj{v}_\sfn) & = \cB^\dagger_{\sfn\pm}.
	\end{aligned}
\end{equation}
The operator \(\fS_\sfn(v_\sfn)\) is known as  a squeezing operator, and in particular as a as \emph{non-separable two-mode} squeezing operator \cite{Caves:1985zz,bishopGeneralTwomodeSqueezed1988,mankoPhotonStatisticsGeneric1997,Gazeau:2022pzz}. Its explicit form, in terms of the original set of oscillators, follows by an application of the Baker--Campbell--Hausdorff formula, and is given as:
\begin{equation}\label{eq:squeeze}
	\fS_\sfn\qty(v_\sfn) = \exp(-\rho_\sfn\qty(\ex{\ii \xi_\sfn} \cA^\dagger_{\sfn+}\cA^\dagger_{\sfn-}-\ex{-\ii \xi_\sfn} \cA_{\sfn+}\cA_{\sfn-}))\,\exp(\ii \psi_{\sfn+} \cA^\dagger_{\sfn+}\cA_{\sfn+} +\ii \psi_{\sfn-} \cA^\dagger_{\sfn-}\cA_{\sfn-}),
\end{equation}
where \(\rho_\sfn,\xi_\sfn,\psi_{\sfn\pm}\in \R\) are related to \(v^\pm_\sfn\) as
\begin{equation}\label{eq:bogoliubov-coefficients-squeezing}
	\begin{aligned}
		v^\pm_{\sfn\pm} & = \ex{\ii\psi_{\sfn\pm}}\cosh(\rho_\sfn)                 \\
		v^\pm_{\sfn\mp} & = \ex{\ii\qty(\psi_{\sfn\pm}+\xi_\sfn)}\sinh(\rho_\sfn).
	\end{aligned}
\end{equation}
Apart from the phases \(\psi_{\sfn\pm}\) --- which are nevertheless necessary to capture the correct vectors \(v_\sfn^\pm\) --- this corresponds to an \(\su(1,1)\) squeezing transformation.

A salient feature of squeezing transformations is that applying the squeezing operator to the a ground state of the original set of operators,
\begin{equation}
	\cA_{\sfn\pm}\ket{\vec{0}}_{\cA_{\sfn}} = 0,
\end{equation}
yields a ground state of the new set of operators. Namely, the state
\begin{equation}
	\ket{\vec{0}}_{\cB_\sfn} \coloneqq \fS_\sfn(v_\sfn)\ket{\vec{0}}_{\cA_\sfn},
\end{equation}
satisfies
\begin{equation}
	\cB_{\sfn\pm}\ket{\vec{0}}_{\cB_\sfn} = 0.
\end{equation}
These states are known as \emph{squeezed vacua}. Excited states can be built in the standard way, by applying creation operators of the new set of modes on the squeezed vacuum. They are related to the excited states of the original set of modes by a squeezing transformation:
\begin{equation}
	\cB_{\sfn\pm}^\dagger\ket{\vec{0}}_{\cB_\sfn} = \fS_\sfn(v_\sfn)\cA_{\sfn\pm}^\dagger\ket{\vec{0}}_{\cA_\sfn}
\end{equation}
In other words, they remain squeezed.

Finally, recall that our setup involves infinitely many modes. Demanding that the new ground states are annihillated by \(\cB_{\sfn\pm}\) for all \(\sfn\in\sN_\perp^*\), identifies it with an all-mode squeezed state that is only pairwise non-separable. In other words there exists a squeezing operator acting on all modes defined as:
\begin{equation}
	\fS \underset{\t{\tiny for}}{\overset{\t{\tiny shorthand}}{\coloneqq\joinrel=\joinrel=\joinrel=\joinrel=}} \fS\qty(\set{v_\sfn}_{\sfn\in\sN_\perp^*}) \coloneqq \prod_{\sfn\in\sN_\perp^*}\fS_\sfn\qty(v_\sfn),
\end{equation}
Note here, that there is no ordering amgiguity in defining \(\fS\), as modes at different levels commute. It is for the same reason that it also acts on each energy level separately, i.e.
\begin{equation}
	\fS \cA_{\sfn\pm} \adj{\fS} = \cB_{\sfn\pm}, \qquad \fall \sfn\in\sN_\perp^*,
\end{equation}
without mixing ladder operators at different levels. As a consequence, there are all-mode squuezed vacua
\begin{equation}
	\ket{\vec{0}}_\cB \coloneqq \fS \ket{\vec{0}}_\cA,
\end{equation}
annihillated by all \(\cB\) modes,
\begin{equation}
	\cB_{\sfn\pm}\ket{\vec{0}}_{\cB} = 0, \qquad \fall\sfn\in\sN_\perp^*,
\end{equation}
where \(\ket{\vec{0}}_\cA\) is the ground state annihilated by all \(\cA\) modes,
\begin{equation}
	\cA_{\sfn\pm}\ket{\vec{0}}_{\cA} = 0, \qquad \fall\sfn\in\sN_\perp^*.
\end{equation}

As an explicit illustration of this construction, let us focus again on the family without momentum on the \(\S^1\), corresponding to \(\sfn=(\ell,m,0)\), and governed by the evolution matrix \cref{eq:Ulm}. The magnetic quantum number, \(m\), does not enter any of the formulas, so we will suppress it, denoting the modes just by the angular momentum number, \(\ell\). The Bogoliubov coefficients are given by
\begin{align}
	v^+_\ell & = \frac{-1}{2\qty(\ell(\ell+1))^{1/4}}\mqty(\sqrt{\ell}+\sqrt{\ell+1}   \\[0.5em] \sqrt{\ell}-\sqrt{\ell+1}), \\[0.5em]
	v^-_\ell & = \frac{-\ii}{2\qty(\ell(\ell+1))^{1/4}}\mqty(\sqrt{\ell}-\sqrt{\ell+1} \\[0.5em] \sqrt{\ell}+\sqrt{\ell+1}),
\end{align}
corresponding to the squeezing parameters
\begin{equation}
	\rho_\ell = \frac{1}{4}\log(\frac{\ell+1}{\ell}),\quad \xi_\ell = \pi, \quad \psi_{\ell +} = \pi, \qq{and} \psi_{\ell -} = \frac{3\pi}{2}.
\end{equation}
This yields a two-mode squeezing operator as:
\begin{equation}
	\fS_\ell(v_\ell) = \exp(\frac{1}{4}\log(\frac{\ell+1}{\ell})\qty(\cA_{\ell +}^\dagger\cA_{\ell -}^\dagger-\cA_{\ell -}^\dagger\cA_{\ell -}))\exp(\ii \pi\qty(\cA_{\ell +}^\dagger\cA_{\ell +} + \frac{3}{2}\cA_{\ell -}^\dagger\cA_{\ell -})).
\end{equation}

\subsection{The correspondence}

After this interlude on squeezed states, we return to line operators, and the states they prepare, via the path integral. First we consider the states of the Wilson--'t Hooft operators, \cref{eq:W-tH},
\begin{equation}
	\ket{\W_{n,m}} \coloneqq \int_{\cC[\,\cdot\,]} \DD{a} \ex{-S[a]}\ \W_{n,m}\qty(\S^1\times\set{0}).
\end{equation}
Let us see the action of the charges on these states.

Starting wit the zero-mode, we have:
\begin{equation}
	Q_0^\pm \ket{\W_{n,m}} = \lim_{r\to 0}\int_{\cC[\,\cdot\,]} \DD{a} \ex{-S[a]}\ \int_{\Sigma_r} \frac{r}{r_0} \phi_0(r)\w J^\pm_{\Sigma_r} \times \W_{n,m}\qty(\S^1\times\set{0}).
\end{equation}
We will need the OPE between \(J^\pm_{\Sigma_r}\) and the Wilson--'t Hooft loop. In fact, here we will only need the Wick contraction, as all the regular terms are bound to vanish, by regularity of the rest of the zero-mode integrand. The desired OPE is:
\begin{equation}\label{eq:JW OPE}
	J^\pm_{\Sigma_r} \times \W_{n,m}(\S^1\times\set{0}) \sim \qty(\ii\, n \ev{J^\pm_{\Sigma_r} \int_{\S^1} a} + \ii\, m \ev{J^\pm_{\Sigma_r} \int_{\S^1} \check{a}})  \W_{n,m}(\S^1\times\set{0}),
\end{equation}
where the twiddle indicates that it is considered up to regular terms. To continue on we will need the two-point function of the gauge-fields on \(\R^3\times \S^1\).\footnote{Remember, we are taking this OPE in the limit \(r\to 0\), so the three-ball, \(\B^3_{r_0}\) is identical to the whole \(\R^3\).} Parametrising \(\R^3\) by coordinates \(x\) and the circle by an angle, \(\eta\), the two-point function reads:
\begin{equation}
	\ev{a_\mu(x,\eta)a_\nu(0,0)} = \frac{4}{\Im\tt}\; \frac{g_{\mu \nu}\; \sinh\norm{x}}{\norm{x}\qty(\cosh\norm{x}-\cos\eta)}+\t{gauge-dependent terms}.
\end{equation}
Integrating one of the gauge fields on the circle, to get the holonomy, and differentiating the other one to get the current yields immediately:
\begin{equation}
	\ev{\qty(\int_{\Sigma_r} \frac{r}{r_0} \phi_0(r)\w J^\pm_{\Sigma_r})\qty(\int_{\S^1} a)} = \pm \frac{\pi}{\sqrt{2}\,r_0}\frac{1}{\Im\tt}.
\end{equation}
From here, electric--magnetic duality implies
\begin{equation}
	\ev{\qty(\int_{\Sigma_r} \frac{r}{r_0} \phi_0(r)\w J^\pm_{\Sigma_r})\qty(\int_{\S^1} \check{a})} = \pm \frac{\pi}{\sqrt{2}\,r_0}\frac{\tt^\mp}{\Im\tt},
\end{equation}
and therefore, finally:
\begin{equation}
	Q_0^\pm \ket{\W_{n,m}} = \pm \pi\ii\; \frac{1}{\sqrt{2}r_0} \frac{n+\tt^\mp m}{\Im\tt} \ket{\W_{n,m}}.
\end{equation}
Compare this to \cref{eq:Jnm S2xS1}. The state \(\ket{\W_{n,m}}\) has exactly the same zero-mode charges, as the Kac--Moody primaries \(\ket{n,m}\). It is tempting, therefore, to identify the Wilson--'t Hooft state with \(\ket{n,m}\).

This is almost the right answer. To iron out that almost we need the action of the non-zero modes. First, note that the OPE \cref{eq:JW OPE} cannot give any singular contribution upon integrated against any of the higher-modes. The reason is, simply, that \(\ev{J^\pm_{\Sigma_r} \int_{\S^1} a}\) and \(\ev{J^\pm_{\Sigma_r} \int_{\S^1} a}\) are proportional to the volume form on the sphere. But the one-forms \(\phi_\sfn(r)\), that it will be integrated against, are periodic on the circle. Hence the net answer is zero. Therefore, as far as non-zero modes are concerned, the OPE produces only regular contributions. We are therefore in the same territory as for the empty state that we discussed previously. Hence again we have that
\begin{equation}
	\cB_{\sfn\pm}\ket{\W_{n,m}} = 0, \qquad \fall \sfn\in\sN_\perp^*.
\end{equation}
This is the precise sense, in which Wilson--'t Hooft lines are \emph{primary operators}. They have definite one-form charge, and the states they generate are primary states of the Kac--Moody algebra \cref{eq:KM-B}, sitting at the bottom of the Verma module labelled by \(n\) and \(m\).

Reconciling the above fact with our digression on squeezed states, the Wilson--'t Hooft lines correspond to \emph{squeezed primary states}:
\begin{equation}
	\ket{\W_{n,m}} = \fS \ket{n,m}.
\end{equation}
Inverting this relation, we have, equivalently that the vacua \(\ket{n,m}\) correspond to:
\begin{empheq}[box=\obox]{equation}
	\ket{n,m} = \adj{\fS}\ket{\W_{n,m}} = \int_{\cC[\,\cdot\,]} \DD{a}\ex{-S[a]}\ \adj{\fS}\W_{n,m}\qty(\S^1\times\set{0}),
\end{empheq}
where \(\adj{\fS}\W_{n,m}\) denotes the line operator obtained by shrinking \(\adj{\fS}\) --- expressed in terms of the currents --- onto the Wilson--'t Hooft loop, à la \cref{fig:surrounding}, or equivalently via the OPE:
\begin{equation}\label{eq:unsqueezed}
	\adj{\fS}\W_{n,m}\qty(\S^1\times\set{0}) \coloneqq \lim_{r\to 0} \adj{\fS}[J^\pm_{\Sigma_r}] \times \W_{n,m}\qty(\S^1\times\set{0}).
\end{equation}

Let us pause here to comment on the ground state, i.e. the state \(\ket{0,0}\), of zero energy. By the above discussion, it corresponds to the operator
\begin{equation}
	\ket{0,0} \quad\leftrightsquigarrow\quad \adj{\fS}\qty(\S^1\times\set{0}).
\end{equation}
This is clearly not the identity operator. In one sense, it is almost the identity operator, as this is the only primary line operator it sees. It is completely transparent to the one-form charges. However, it is also as far as one can get from the identity operator as it contains photon excitations of arbitrary frequency --- indeed, of all frequencies. This was, to some extent, anticipated in \cite{Belin:2018jtf}, where it was shown that, in a generic CFT, the identity operator, cannot prepare the vacuum state on the torus (or generally, on any spatial slice other than the sphere). Our result is consistent with that statement, while still tractable in this example.

Continuing on, and moving up in the Verma module, all the other states can be reached by acting with \(\cB_{\sfn\pm}^\dagger\) on the squeezed primaries, i.e.
\begin{align}
	\ket{\W_{n,m};\set{N_{\sfn \sigma}}_{\sfn\in\sN_\perp^*}^{\sigma=\pm}} & \coloneqq \prod_{\substack{\sfn\in\sN_\perp^*                                  \\ \sigma=\pm}}\qty(\cB^\dagger_{\sfn \sigma})^{N_{\sfn \sigma}} \ket{\W_{n,m}} \nn
	                                                                   & = \int_{\cC[\,\cdot\,]} \DD{a} \ex{-S[a]}\ \prod_{\substack{\sfn\in\sN_\perp^* \\ \sigma=\pm}}\qty(\cB^\dagger_{\sfn \sigma})^{N_{\sfn \sigma}}\W_{n,m}\qty(\S^1\times\set{0}).
\end{align}
The modes \(\cB_{\sfn \pm}\) and \(\cB_{\sfn \pm}^\dagger\) can be written, in terms of the local operators \(J^\pm\) as
\begin{align}\label{eq:Bn-lowering}
	\cB_{\sfn \pm}\qty(\S^1\times\set{0})         & = \lim_{r\to 0} \int_{\Sigma_r} v_{\sfn \sigma}^\pm\qty[\bbU^\pm_{\sfn}(r,r_0)]_{\sigma \tau}^{-1} \phi_{\sfn \tau}(r) \w J^\pm_{\Sigma_r},         \\
	\label{eq:Bn-raising}
	\cB_{\sfn \pm}^\dagger\qty(\S^1\times\set{0}) & = \lim_{r\to 0} \int_{\Sigma_r} \conj{v}_{\sfn \sigma}^\pm \qty[\bbU^\mp_{\sfn}(r,r_0)]_{\sigma \tau}^{-1} \phi_{\sfn \tau}(r) \w J^\pm_{\Sigma_r},
\end{align}
respectively. Acting them on Wilson--'t Hooft operators gives precisely the \emph{descendant operators} that we discussed in \cref{ssec:line-ops}. They are related to the descendant states \cref{eq:generic state}, by a squeezing transformation. For example, a single excited mode is
\begin{empheq}[box=\obox]{equation}
	\cA^\dagger_{\sfn\pm}\ket{n, m} = \adj{\fS}\cB^\dagger_{\sfn\pm}\ket{\W_{n,m}} = \int_{\cC[\,\cdot\,]} \DD{a}\ex{-S[a]}\ \adj{\fS} \cB_{\sfn\pm}^\dagger\W_{n,m}\qty(\S^1\times\set{0}),
\end{empheq}
and similarly for the higher-excited states.

\begin{figure}[!htbp]
	\begin{align*}
		\ket{0,0} \quad                      & \leftrightsquigarrow\quad \vcenter{\hbox{\def\svgwidth{0.4\textwidth}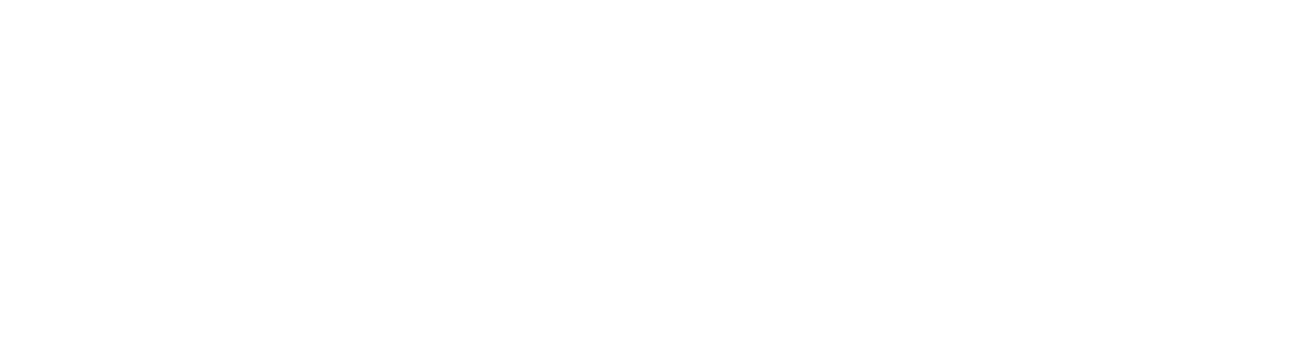}} \\[1em]
		\ket{n,m} \quad                      & \leftrightsquigarrow\quad \vcenter{\hbox{\def\svgwidth{0.4\textwidth}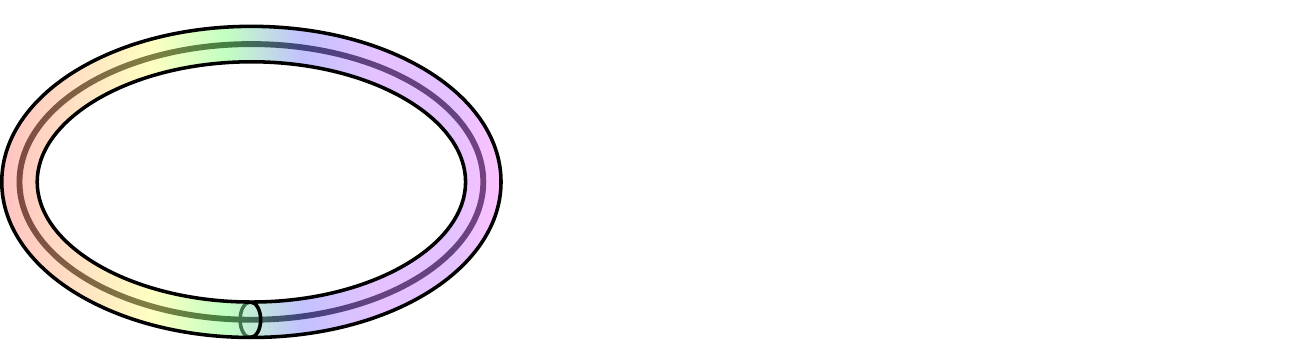}}      \\[1em]
		\cA_{\sfn\pm}^\dagger\ket{n,m} \quad & \leftrightsquigarrow\quad \vcenter{\hbox{\def\svgwidth{0.4\textwidth}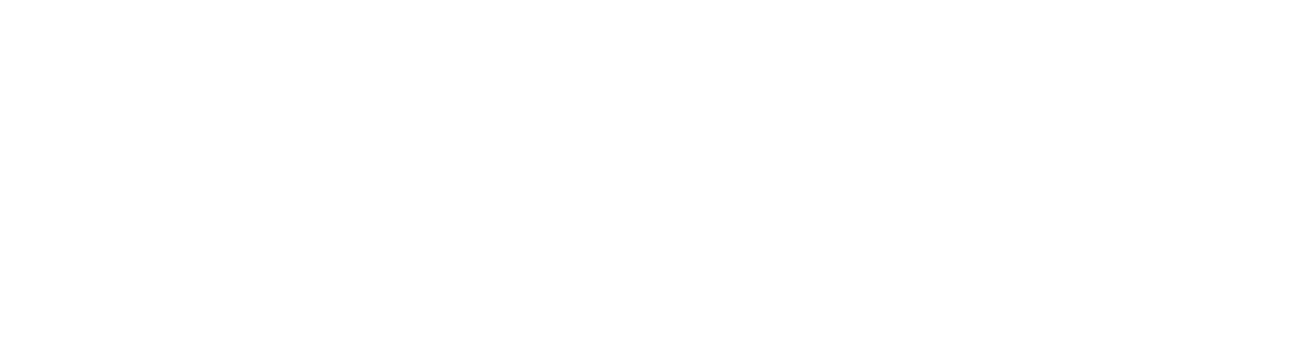}}   \\[0.5em]
		                                     & \hspace{1em}\vdots
	\end{align*}
	\caption{The states and their corresponding operators. (Top) The squeezing operator, \(\adj{\fS}\qty(\S^1\times\set{0})\), containing photons of all frequencies, represented by a rainbow-coloured line prepares the vacuum state. (Middle) The squeezing operator, surrounding a Wilson--'t Hooft line of charges \(n, m\) prepares the primary state \(\ket{n,m}\). (Bottom)  The squeezing operator, on top of the mode \(\cB^\dagger_{\sfn\pm}\) on top of a Wilson--'t Hooft line prepares the descendant \(\cA^\dagger_{\sfn \pm}\ket{n,m}\).}
	\label{fig:state-operator explicit}
\end{figure}

To recap, we have just constructed a one-to-one map between line operators on \(\R^3\times \S^1\) and states on \(\S^2\times \S^1\). The line operators that we have in our disposal are the Wilson--'t Hooft lines and modulated versions thereof, i.e. smeared with modes of the basic gauge-invariant operator, \(f\), or equivalently \(J^\pm\). Each different allowed smearing, that is, each different smooth configuration of \(J^\pm\) gives a different state. The Wilson--'t Hooft lines are the primary operators of the Kac--Moody algebra. Modulated operators are their descendants. They prepare states on \(\S^2\times\S^1\), that are orthonormal and span the entire Hilbert space. These states are related to the standard energy eigenstates on the Hilbert space by a squeezing transformation. This map is displayed in \cref{fig:state-operator explicit}.

\subsubsection*{Energies and overlaps}

In this last paragraph, we will, briefly, compare the squeezed states we have arrived at to the energy eigenstates. We begin by calculating the average energy of the squeezed primary states \(\ket{\W_{n,m}}\). A quick computation shows that their average energy is
\begin{equation}
	\mel{\W_{n,m}}{H_{\Sigma_{r_0}}}{\W_{n,m}} = \Delta_{n,m} + E_{0,\fS}(r_0),
\end{equation}
where
\begin{equation}
	E_{0,\fS}(r_0) = \sum_{\sfn, \sigma} \sinh[2](\rho_\sfn) \sqrt{\lambda_\sfn(r_0)},
\end{equation}
is essentially measuring a zero-point energy, which has to do with the fact that the Hamiltonian \(H_{\Sigma_{r_0}}\) is not normal-ordered with respect to the new set of ladder operators. More importantly,
\begin{equation}
	\Delta_{n,m} = \frac{\pi}{2 r_0^2}\frac{\abs{n+\tt\,m}^2}{\Im\tt}
\end{equation}
is the same as the energy of the (unsqueezed) primaries \cref{eq:Dnm S2xS1}, as well as the instanton weights in the path integral \cref{eq:Dnm PI}. This property reflects and refines the observation of Verlinde \cite{Verlinde:1995mz} and the argument of Kapustin \cite{Kapustin:2005py}, that the Wilson--'t Hooft lines have a quantum number \(\Delta_{n,m}\) akin to a  scaling weight.\footnote{Do notice, however, that these are not the eigenvalues of the dilation operator of the theory under consideration.}

For the first excited states, an elementary computation reveals their average energy
\begin{equation}
	\mel{\W_{n,m}}{\cB_{\sfn \sigma} H_{\Sigma_{r_0}} \cB^\dagger_{\sfn \sigma}}{\W_{n,m}} = \Delta_{n,m} +  \qty(\cosh[2](\rho_\sfn)+\sinh[2](\rho_\sfn))\sqrt{\lambda_\sfn(r_0)} + E_{0,\fS}(r_0).
\end{equation}
In the no-squeezing limit, \(\rho_\sfn\to 0\), this becomes precisely the energy of the first excited energy eigenstates.

Finally we can also compute overlaps between the squeezed and the unsqueezed primaries. This is very much facilitated by the disentangling formula of \(\fs\fu(1,1)\) squeezing operators \cite{loNormalOrderingSU1995,dasguptaDisentanglementFormulasAlternative1996}:
\begin{align}
	\exp(\alpha_\sfn \cX_\sfn^+ -\conj{\alpha}_\sfn \cX_\sfn^-) & =\exp(\ex{\ii\arg(\alpha_\sfn)}\tanh\abs{\alpha_\sfn} \cX_\sfn^+) \nn
	                                                            & \phantom{=~}\gray{\times} \exp(-2\log\cosh\abs{\alpha_\sfn} \cX_\sfn^0)\exp(-\ex{-\ii\arg(\alpha_\sfn)}\tanh\abs{\alpha_\sfn} \cX_\sfn^-),
\end{align}
where
\begin{align}
	\cX_\sfn^+ & \coloneqq \frac{1}{\sqrt{2}}\frac{\cA^\dagger_{\sfn+}\cA^\dagger_{\sfn-}}{\k\sqrt{\lambda_\sfn}},                        \\
	\cX_\sfn^- & \coloneqq \frac{1}{\sqrt{2}}\frac{\cA_{\sfn+}\cA_{\sfn-}}{\k\sqrt{\lambda_\sfn}},                                        \\
	\cX_\sfn^0 & \coloneqq \frac{1}{2}\qty(1+\frac{\cA^\dagger_{\sfn+}\cA_{\sfn+}+\cA^\dagger_{\sfn-}\cA_{\sfn-}}{\k\sqrt{\lambda_\sfn}})
\end{align}
are the generators of \(\fs\fu(1,1)\):
\begin{equation}
	\comm{\cX_\sfn^+}{\cX_\sfn^-} = -2\cX_\sfn^0, \qquad \comm{\cX_\sfn^0}{\cX_\sfn^\pm} = \pm\cX_\sfn^\pm.
\end{equation}
In this case, the phases, \(\psi_{\sfn\pm}\), of the squeezing operator, \cref{eq:squeeze}, do not matter as the phase-shift operator they furnish is normal ordered, and thus, acts as the identity on \(\ket{n,m}\). In total we find:
\begin{equation}
	\braket{n,m}{\W_{n,m}} = \prod_{\sfn\in\sN_\perp^*} \qty(\cosh(\rho_\sfn))^{-2}.
\end{equation}

\section{Discussion}\label{sec:discussion}

In this paper, we studied CFTs with continuous generalised global symmetries. We showed that an invertible continuous \((p+1)\)-form symmetry in a \((2p+2)\)-dimensional unitary CFT automatically enhances to an infinite collection of codimension-one, i.e. zero-form, conserved charges, labelled by  (anti-)chiral \(p\)-forms. The algebra of these charges is spectrum-generating (up to decoupled neutral factors) and characterises completely the CFT. The dynamics of the CFT are those of free \(p\)-forms. Along a similar vein, we showed that a non-invertible continuous \((p+1)\)-form symmetry, leads to a non-invertible current algebra, which we describe in terms of the fusion rules of the symmetry generators. As before, it characterises completely the dynamics, leading, in this case, to an \(\O(2)\) \(p\)-form gauge theory. In the invertible case, and focussing on \(p=1\), hence in four-dimensional CFTs, we constructed the representation theory of this algebra (for the invertible case), which we showed reproduces the path integral calculation, as a non-trivial check. This allowed us to describe explicitly the Hilbert space of Maxwell theory on arbitrary closed spatial manifolds. On non-trivial topologies, the full Hilbert space consists of the usual photon Hilbert space, as well as of Verma modules, built on top of states with one-form symmetry charge. Additionally, we showed that the full spectrum of states on \(\S^2\times\S^1\) can be obtained by a path integral on \(\B^3\times\S^1\), with various operator insertions. The radial evolution on the ball acts as a squeezing transformation between the path integral states and the energy eigenstates, requiring that line operators are subsequently dressed with a squeezing operator, containing all photon frequencies, to reach the energy eigenstates. Notably, the vacuum state of Maxwell theory on \(\S^2\times\S^1\) is not produced by the path integral with no insertions, but by the path integral with a squeezing operator inserted. Nevertheless, this construction leads to a one-to-one correspondence between line operators on \(\R^3\times\S^1\) and states on \(\S^2\times\S^1\) and a classification of line operators in terms of the current algebra. In short, Wilson--'t Hooft lines,
\begin{equation}
	\W_{n,m}\qty(\S^1) = \exp(\ii n \int_{\S^1} a + \ii m \int_{\S^1} \check{a}),
\end{equation}
are charged under the one-form symmetries, \(Q_0^\pm\), \cref{eq:Q0}:
\begin{equation}
	Q_0^\pm \times \W_{n,m}\qty(\S^1) \sim \pm\; \frac{\pi\ii}{\sqrt{2} r_0} \frac{n+\tt^\mp m}{\Im\tt} \W_{n,m}\qty(\S^1),
\end{equation}
and annihilated by the lowering operators \(\cB_{\sfn\pm}\), \cref{eq:Bn-lowering}, of the Kac--Moody algebra:
\begin{equation}
	\cB_{\sfn\pm} \times \W_{n,m}\qty(\S^1) \sim 0.
\end{equation}
This defines them as primary operators. They have definite scaling weight, as defined in the body of the paper, given by
\begin{align}
	\frac{1}{\k}Q_0^+ Q_0^-\times \W_{n,m}\qty(\S^1) & \sim \Delta_{n,m} \W_{n,m}\qty(\S^1)           \\
	\Delta_{n,m}                                     & =\frac{\pi}{2 r_0^2}\frac{\abs{n+\tt\,m}^2}{\Im\tt}.
\end{align}
The path integral on \(\B^3\times\S^1\) with an insertion of \(\W_{n,m}\qty(\S^1)\), dressed by a squeezing operator, \cref{eq:unsqueezed}, prepares the primary energy eigenstates, \(\ket{n,m}\) \cref{eq:primary1}. Descentant line operators are given by acting with raising operators, \(\cB_{\sfn\pm}^\dagger\), \cref{eq:Bn-raising}, on the primaries. Again, the path-inegral on \(\B^3\times\S^1\), with a descendant, dressed with a squeezing operator, produces the descendant states \cref{eq:generic state}.

We comment on several questions we have left unanswered and possible generalisations.

\paragraph{Non-invertible symmetries and orbifold branches}

In \cref{ssec:current-Maxwell} we described a non-invertible current algebra. This was described in terms of the fusion rules of non-invertible topological operators. The free theory that realises the algebra is, in that case, that of an \(\O(2)\) \(p\)-form gauge theory. One very interesting question is to construct the representation theory of these non-invertible current algebras. At \(p=0\), or equivalently in \(d=2\), an important role is played by tube algebras \cite{ocneanuChiralityOperatorAlgebras,Lin:2022dhv} and lasso actions \cite{Chang:2018iay} associated to the fusion of the topological operators. A careful study of this, should land exactly on the orbifold branch of a compact scalar, which has a much simpler algebraic description \cite{Schoutens:2015uia}. Similarly, exploiting and generalising representation theory of higher-dimensional non-invertible symmeties \cite{Bartsch:2022mpm,Bartsch:2022ytj,Lin:2022xod,Bhardwaj:2022lsg} should give an algebraic point of view to orbifold branches of gauge-theories, operning a window to new BKT-like phase transitions at points of enhanced symmetry.

\paragraph{Non-abelian current algebras}

Despite the absence of non-abelian higher-form symmeries, one could still consider the physics of a four-dimensional CFT with a non-abelian version of our higher-dimensional current algebra, schematically of the form:
\begin{equation}
	\comm{Q_{\sfn}^\fa}{Q_{\sfm}^\fb} = \sff^{\fa\fb}_{~~\fc} \sfT_{\sfn\sfm}^{\sfl} Q_{\sfl}^{\fc}\; +\; \k \sqrt{\lambda_{\sfn}}\delta^{\fa\fb}\delta_{\sfn\sfm},
\end{equation}
where the \(\fu(1)\times\fu(1)\) indices, \(\pm\), got replaced by some Lie algebra, \(\fg\), indices, \(\set{\fa,\fb,\fc,\cdots}\). There are a few indicators that such an algebra might be hiding behind four-dimensional supercofnormal field theories. For instance, the form of the Vafa--Witten partition functions \cite{Vafa:1994tf}, for \(\cN=4\) super-Yang--Mills (SYM) theory and their relation to two-dimensional RCFTs is reminiscent of our exact formulas for the Maxwell partition fuction and their two-dimensional analogues. A more concrete indicator is Kapustin's definition of scaling weight for \(\half\)BPS operators in \(\cN=4\) SYM \cite{Kapustin:2005py}, which is arrived at in a similar way as the one for Wilson--'t Hooft operators, which as we saw, is intimately linked to the current algebra.

\paragraph{Modularity and factorisation}

The form, \cref{eq:parti-S1-Sigma-pathint} of the partition function begs for an investigation of the modularity properties of the partition function, as well as a possible ``holomorphic factorisation'' of the partition function. Both of these directions deserve separate attention. On the modularity side, if higher-dimensional CFTs behave like two-dimensional, then swapping the thermal circle for a spatial one-cycles should leave the partition function unaffected. Harnessing that statement leads to a generalised Cardy formula \cite{Shaghoulian:2015kta} and a universal form of the Casimir energy \cite{Luo:2022tqy}. Our exact form of the partition function of Maxwell theory, can serve as a testing ground for modularity in higher-dimensional CFTs. Relatedly, but with a different goal in mind, one could imagine making \(q\) complex and study whether the Maxwell partition function factorises in a holomorphic and anti-holomorphic piece:
\begin{equation}
	\parti_\t{Maxwell}\qty(q,\bar{q}) \overset{?}{=} \parti(q)\,\parti\qty(\bar{q}).
\end{equation}
This would open a new window towards the physics of chiral one-forms (or more generally chiral \(p\)-forms), whose importance in six-dimensional superconformal field theories and in M-theory makes them subject of constant study \cite{Perry:1996mk,Pasti:1996vs,Bekaert:1998yp,Henningson:2001wh,Sen:2015nph,Sen:2019qit,Andriolo:2021gen,Evnin:2022kqn,Fliss:2023uiv}.

\paragraph{Other topologies}

The choice of \(\S^2\times\S^1\) for our state-operator correspondence is a natural one for that the unique one-cycle allows line operators wrapping it and the unique two-cycle allows a single flux. It is not clear that this is the only choice, though. For example, states on a \(\T^3\) topology are prepared via a four-dimensional path integral with some two-cycle filled in. It is likely that to get the complete set of states we have to consider a mixture of states prepared by filling in every two-cycle. Therefore, the state-operator map, would, in that case, include a sum over topologies, with a given boundary, in the path integral. This is not the case here, because of the requirement that the \(\S^1\) remains non-trivial.

\paragraph{Surgery and  overlaps}

Given our state-operator correspondence, we now have a resolution of the identity on \(\S^2\times\S^1\), in terms of line operators:
\begin{equation}
	\id = \sum_{n,m\in\Z}\sum_{\set{N_{\sfn \sigma}}} \ketbra{\W_{n,m};\set{N_{\sfn \sigma}}}.
\end{equation}
We can use this to simplify correlation functions on manifolds that can be cut along an \(\S^2\times\S^1\). Relatedly, states can be glued to produce partition functions or correlation functions on manifolds surgered along \(\S^2\times\S^1\). This fact opens up a new perspective on partition functions. Take, for example, partition functions in $d$ Euclidean dimensions with the topology of the sphere. These are known to be related to counting problems through F-theorem arguments \cite{Jafferis:2011zi}, at least in $d=3$. On the other hand, by the logic above, this partition function should be related to the overlap between states prepared by computing path integrals on some half spaces which do not necessarily produce the vacuum of the theory. An example would amount to considering the Hopf fibration of the sphere and consider states defined on the $\T^2$ gluing surface. Now we know those states, while having a simple path integral construction, are described by complicated superpositions of high-energy eigenstates. Therefore, these partition functions could have a more natural counting interpretation shedding light on the exact nature of the F-theorem. At this point this is just a plausibility argument, as we have only focused on theories with higher-form symmetries in even dimensions. It would be interesting to study this in more depth.

\paragraph*{Acknowledgements} We thank Chen Yang for discussions and an initial collaboration. It is a pleasure to acknowledge helpful discussions with Jeremias Aguilera Damia, Dionysios Anninos, Jackson Fliss, Keivan Namjou, Ryan Thorngren, Luigi Tizzano, David Tong, and Yifan Wang. SV is supported by the NWO Spinoza prize awarded to Erik Verlinde. SV thanks ULB Brussels and International Solvay Institutes for hospitality during the workshop ``Symmetries, Anomalies and Dynamics
of Quantum Field theory,'' as well as CERN, for hospitality during the ``GenHET meeting in String Theory,'' where some of the results in this paper were communicated. Both authors would like to credit Physics Sessions Initiative, and acknowledge the hospitality of KITP at UCSB. Research at KITP was supported in part by the National Science Foundation under Grant No. NSF PHY-1748958.

\appendix

\section{The transversal Laplacian}\label{app:Laplacian}
\newcommand{\trig}{T}

In this appendix we will give some outline some spectral properties of the transversal Laplacian on three dimensional manifolds, which we denote generically by \(\Sigma\), and, subsequently, explicitly construct its spectrum on \(\Sigma_r = \S^2_r\times\S^1\).

\subsection{Spectral properties}

First, by transversal Laplacian, on \(p\)-forms, on a \(d\)-dimensional, closed manifold \(M\), we mean:
\begin{equation}
	\tlapl_p^{M} \coloneqq \eval{\lapl}_{\Omega^p(M)\cap \ker\cdd},
\end{equation}
where \(\lapl\) denotes the Hodge Laplacian, \(\lapl=\qty(\dd+\cdd)^2\). On non-zero-modes, the transversal Laplacian is equivalent to \(\cdd\dd\) on \(p\)-forms. However these operators have, potentially, different number of zero-modes. We care about \(\tlapl_p^{M}\) which has
\begin{equation}
	\dim\ker\tlapl_p^{M} = \dim\ker\lapl_p = \b_p(M),
\end{equation}
where \(\b_p(M)\) is the \(p\)-th Betti number of \(M\). The zero-modes are given, simply, by the harmonic \(p\)-forms on \(M\). Finally, the transversal Laplacian is a self-adjoint operator, and hence, its eigenforms provide a basis for coclosed (also known as transversal) \(p\)-forms. By Poincaré duality, the spectrum of the Hodge Laplacian for \(0\leq p\leq d\) is determined by the spectrum of the transversal Laplacian for \(0\leq p\leq \left\lfloor \frac{d-1}{2}\right\rfloor\).

Let us now focus on three dimensional manifolds, \(\Sigma\), one-forms (i.e. \(p=1\)), and non-zero-modes.\footnote{The story can be adapted, more generally, to \(p\)-forms on \((2p+1)\)-dimensional manifolds, mutatis mutandis \cite{gierMultiplicityEigenvaluesHodge2015}.} There is another seld-adjoint operator acting on one-forms, the Beltrami operator:
\begin{equation}
	\star\dd : \Omega^1(\Sigma)\to \Omega^1(\Sigma).
\end{equation}
The transversal Laplacian is simply the square of the Beltrami operator, \(\tlapl_1^\Sigma = \qty(\star\dd)^2\). As such, they commute, so they can be diagonalised simultaneously, and, moreover, the non-zero spectrum of the transversal Lalpacian consists of the squares of the eigenvalues of the Beltrami operator. Since \(\star\dd\) is self-adjoint its eigen-one-forms provide an (orthonormalisable) basis of coclosed one-forms.

On a generic closed, oriented three-dimensional manifold, \(\Sigma\), the Beltrami operator has simple spectrum \cite{encisoNondegeneracyEigenvaluesHodge2010}. The same holds for the Hodge Laplacian \cite{encisoNondegeneracyEigenvaluesHodge2010} with an appropriate clarification on the word generic.\footnote{More precisely, it is simple on the complement of a codimension-1 set of metrics, but not on the complement of a codimension-2 set of metrics \cite{kepplingerDimensionBeltramiOperator2022}.} For most of our applications, and importantly, for the state-operator correspondence, we are interested in very non-generic manifolds, with high-degree of symmetry, and therefore degeneracy, such as products of spheres. There, the spectrum of the Hodge Laplacian is actually (at least) twofold degenerate. On \(\S^3\) it is a classic result, see e.g. \cite{Rubin:1984tc,Elizalde:1996nb,achourExplicitVectorSpherical2016} for an account of it. On \(\S^2\times\S^1\), we construct explicitly the spectrum below, in \cref{ssec:spectrum S2xS1}. Finally, on \(\T^3\), it is straightforward to construct the spectrum and see that it is spanned by
\begin{equation*}
	\exp(\ii \vec{k}\cdot \vec{\theta})\,\omega(\vec{k}).
\end{equation*}
In the above, \(\vec{\theta}=(\theta_1,\theta_2,\theta_3)\) are the three angles of the torus, \(\vec{k} = (k_1,k_2,k_3)\), with \(k_i=\frac{n_i}{L_i}\), where \(n_i\in\Z\) and \(L_i\) are the radii of each circle, and \(\omega(\vec{k})=\omega(\vec{k})_i \dd{x^i}\) is an eigenform --- or equivalently \(\omega(\vec{k})_i\) is an eigenvector --- of the matrix \(\ii\, \varepsilon_{ij\ell}k^\ell\). It is then easy to verify that there are two such eigenvectors, both corresponding to the eigenvalues \(\norm{\vec{k}}^2\) for the transversal Laplacian, verifying its twofold degeneracy.

\subsection{\texorpdfstring{Eigenforms and eigenvalues on \(\S^2\times\S^1\)}{Eigenforms and eigenvalues on S²×S¹}}\label{ssec:spectrum S2xS1}

WNow we move on to constructing the spectrum on \(\Sigma_r = \S^2_r\times\S^1\). We use the following metric and coordinates:
\begin{equation}
	\dd{s}^2_{\Sigma_r} = r^2\qty(\dd{\theta^2} + \sin\theta \dd{\varphi^2}) + \dd{\eta^2},
\end{equation}
where \(\varphi\in\ropen{0}{2\pi}\) and \(\theta\in\ropen{0}{\pi}\) are the angles on the sphere and \(\eta\in\ropen{0}{2\pi}\) is the angle on the circle. We will denote the eigen-one-forms of \(\tlapl_1^{\Sigma_r}\)  as \(\Phi\), subscripted by their various quanutm numbers.

\subsubsection*{Zero-mode} The easiest to construct is the zero-mode. This is given by the unique harmonic form on \(\Sigma_r\), \(\dd{\eta}\). Therefore, the normalised zero-mode is given by
\begin{equation}\label{eq:F0}
	\Phi_0 \coloneqq \frac{\dd{\eta}}{\sqrt{\vol(\Sigma_r)}} = \frac{\dd{\eta}}{2\sqrt{2} \pi\, r}.
\end{equation}

\subsubsection*{Non-zero modes}

Let us introduce some convenient notation. First, we write
\begin{equation}
	\trig_k(\eta) \coloneqq
	\begin{cases}
		\frac{1}{\sqrt{\pi}}\cos(k \eta),       & k>0  \\[1em]
		\frac{1}{\sqrt{2\pi}},                  & k=0  \\[1em]
		\frac{1}{\sqrt{\pi}}\sin(\abs{k} \eta), & k<0,
	\end{cases}
\end{equation}
with \(\eta\in\ropen{0}{2\pi}\). These are the real, orthonormal eigenfunctions of the Laplacian on \(\S^1\), with eigenvalue \(k^2\):
\begin{equation}
	\tlapl_0^{\S^1} \trig_k(\eta) = -\nabla^2 \trig_k(\eta) = k^2\; \trig_k(\eta).
\end{equation}
Similarly, we denote by \(\Upsilon_{\ell m}(\theta,\varphi)\) the real spherical harmonics on \(\S^2_r\):
\begin{equation}
	\Upsilon_{\ell m}(\theta,\varphi) \coloneqq
	\frac{\cN_{\ell m}}{r}\mathrm{P}_\ell^{\abs{m}}(\cos\theta)
	\begin{cases}
		\cos(m\varphi),       & m>0  \\[1em]
		1,                    & m=0  \\[1em]
		\sin(\abs{m}\varphi), & m<0,
	\end{cases}
\end{equation}
where,
\begin{equation}
	\cN_{\ell m} = (-1)^m \sqrt{2} \sqrt{\frac{2\ell +1}{4\pi} \frac{(\ell-\abs{m})!}{(\ell+\abs{m})!}}\,,
\end{equation}
and \(\mathrm{P}_\ell^m(x)\) are the associated Legendre polynomials. In the above, \(\ell\!\in\!\Z_{\geq 0}\) and \(m\) are integers, with \(-\ell\leq m \leq \ell\). These are the real, orthonormalised eigenfunctions of the Laplacian on \(\S^2_r\):
\begin{align}
	\tlapl_0^{\S^2_r} \Upsilon_{\ell m}(\theta,\varphi) = -\nabla^2 \Upsilon_{\ell m}(\theta,\varphi) = \underset{\mqty{\rotcoloneqq \\ \rho_{\ell m}(r)}}{{\frac{\ell(\ell+1)}{r^2}}} \Upsilon_{\ell m}(\theta,\varphi).
\end{align}
The degeneracy of the eigenvalue \(\rho_{\ell m}(r)\) is
\begin{equation}
	D_{\ell m} = 2\ell +1.
\end{equation}
Using these two building blocks, we can build all the non-zero eigen-one-forms.

\paragraph{Momentumless.}

A first family is given by eigenforms with no momentum along the \(\S^1\), i.e. \(k=0\). These take the form
\begin{equation}
	\Phi^{(1)}_{\ell m} \coloneqq \Upsilon_{\ell m}(\theta,\varphi) \frac{\dd{\eta}}{\sqrt{2\pi}}, \label{eq:F1lm}
\end{equation}
with \(\ell >0\) and \(-\ell \leq m \leq \ell\), and have eigenvalue \(\rho_{\ell m}(r)\):
\begin{equation}
	\tlapl_1^{\Sigma_r} \Phi^{(i)}_{\ell m} = \frac{\ell(\ell+1)}{r^2} \Phi^{(i)}_{\ell m}.
\end{equation}
One can explicitly check that
\begin{equation}
	\begin{aligned}
		\Phi^{(2)}_{\ell m} & \coloneqq \frac{1}{\sqrt{\rho_{\ell m}(r)}}\star_r\sd \Phi^{(1)}_{\ell m}                                                                                                                                                           \\
		                    & = \frac{1}{\sqrt{2\pi}}\frac{r}{\sqrt{\ell(\ell+1)}} \qty(\frac{1}{\sin\theta}\,\pd_\varphi \Upsilon_{\ell m}(\theta,\varphi) \dd{\theta}- \sin\theta\, \pd_\theta \Upsilon_{\ell m}(\theta,\varphi) \dd{\varphi}), \label{eq:F2lm}
	\end{aligned}
\end{equation}
is also an eigen-one-form of \(\tlapl_1^{\Sigma_r}\), with the same eigenvalue, \(\rho_{\ell m}(r)\). All of these modes are orthonormalised:
\begin{equation}
	\ip{\Phi_{\ell m}^{(i)}}{\Phi_{\ell' m'}^{(i')}} = \delta_{\ell\ell'}\delta_{mm'}\delta_{ii'}, \qquad i,i'\in\set{1,2}.
\end{equation}

\paragraph{Momentumful.}

The rest of the modes, are modes with momentum along the \(\S^1\), i.e. \(k\neq 0\). They are given by
\begin{equation}
	\begin{aligned}
		\Phi^{(1)}_{\ell m k} & \coloneqq \frac{1}{\sqrt{\rho_{\ell m}(r)+k^2}}\qty(\sqrt{\frac{\rho_{\ell m}(r)}{k^2}}\Upsilon_{\ell m}(\theta,\varphi)\dd\trig_k(\eta)- \sqrt{\frac{k^2}{\rho_{\ell m}(r)}}\dd\Upsilon_{\ell m}(\theta,\varphi)\,\trig_k(\eta)) \\
		                      & = \qty(\frac{\ell(\ell+1)}{r^2}+k^2)^{-\frac{1}{2}}\left[\frac{\sqrt{\ell(\ell+1)}}{r} \frac{\pd_\eta\trig_k(\eta)}{k} \Upsilon_{\ell m}(\theta,\varphi) \dd{\eta} \right. \\
		                      & \hspace{5em} \left. - k \trig_k(\eta)\,\frac{r}{\sqrt{\ell(\ell+1)}}\qty(\pd_\varphi \Upsilon_{\ell m}(\theta,\varphi)\, \dd{\varphi} + \pd_\theta \Upsilon_{\ell m}(\theta,\varphi) \dd{\theta}) \right], \label{eq:F1lmk}
	\end{aligned}
\end{equation}
and
\begin{equation}
	\begin{aligned}
		\Phi^{(2)}_{\ell m k} & \coloneqq \frac{1}{\sqrt{\rho_{\ell m}(r)+k^2}} \star_r\sd\Phi^{(2)}_{\ell m k} \\
		                      & = \frac{1}{\sqrt{\rho_{\ell m}(r)\,k^2}} \star_r \qty[\dd \Upsilon_{\ell m}(\theta,\varphi) \w \dd\trig_k(\eta)] \\
		                      & = \frac{r}{\sqrt{\ell(\ell+1)}} \frac{\pd_\eta\trig_k(\eta)}{k} \qty(\frac{1}{\sin\theta}\,\pd_\varphi \Upsilon_{\ell m}(\theta,\varphi)\,\dd{\theta}- \sin\theta\, \pd_\theta \Upsilon_{\ell m}(\theta,\varphi) \dd{\varphi}). \label{eq:F2lmk}
	\end{aligned}
\end{equation}
Their eigenvalue is \(\lambda_{\ell m k}(r) \coloneqq \rho_{\ell m}(r)+k^2\):
\begin{equation}
	\tlapl_1^{\Sigma_r} \Phi^{(i)}_{\ell m k} = \qty(\frac{\ell(\ell+1)}{r^2}+k^2)\Phi^{(i)}_{\ell m k} \eqqcolon. \lambda_{\ell  m k}(r) \Phi^{(i)}_{\ell m k}
\end{equation}
and are again orthonormalised:
\begin{equation}
	\ip{\Phi_{\ell m k}^{(i)}}{\Phi_{\ell' m' k}^{(i')}} = \delta_{\ell\ell'}\delta_{mm'} \delta_{k k'}\delta_{ii'}.
\end{equation}

We write collectively, both for the case with momentum and without momentum, the modes as \(\Phi^{(i)}_{\ell m k}\), by defining \(\Phi^{(i)}_{\ell m 0}\coloneqq \Phi^{(i)}_{\ell m}\). Therefore, a complete orthonormal basis of the space of coclosed one-forms on \(\Sigma_r\) is given by
\begin{equation}
	\sB^1_\perp(\Sigma_r)  \coloneqq \Bigg\{\Phi_0, \Big\{\Phi_{\ell m k}^{(i)}\Big\}_{\ell \in \Z_{>0},\; m\in\closed{-\ell}{\ell},\; k\in\Z}^{i=1,2}\Bigg\}.
\end{equation}
The degeneracy of the eigenvalue \(\lambda_{\ell m k}(r)\) on one-forms is
\begin{equation}
	D_{\ell m k}^{(1)} = 2(2\ell +1).
\end{equation}

Finally, the Hodge duals of the above forms provide the spectrum of the longitudinal Laplacian on two-forms, \(\eval{\lapl}_{\Omega^2(\Sigma_r)\cap \ker\sd}\). Relatedly, an orthonormal basis for those is:
\begin{equation}
	\sB^2_\parallel(\Sigma_r)  \coloneqq \Bigg\{\star_r\Phi_0,\Big\{\star_r\Phi_{\ell m k}^{(i)}\Big\}_{\ell \in \Z_{>0},\; m\in\closed{-\ell}{\ell},\; k\in\Z}^{i=1,2}\Bigg\}.
\end{equation}

\subsection{A convenient basis}

As we discussed above, the transversal Laplacian commutes with the operator \(\star_r\sd\). Therefore they can be diagonalised simultaneously. The one-forms that diagonalise the operator \(\star_r\sd\) are related to the above basis as follows:
\begin{equation}
	\phi_{\ell m k \sigma} \coloneqq \frac{1}{\sqrt{2}}\qty(\Phi^{(1)}_{\ell m k}+\sigma\,\Phi^{(2)}_{\ell m k}), \qquad \sigma=\pm.
\end{equation}
They are again orthonormal and statisfy
\begin{equation}
	\star_r\sd \phi_{\ell m k \sigma} = \sigma\sqrt{\lambda_{\ell m k}(r)} \ \phi_{\ell m k \sigma}.
\end{equation}
Therefore, we can also write the following two bases for coclosed one-forms and closed two-forms respectively:
\begin{equation}\label{eq:Vbasis}
	\begin{aligned}
		\sV^{1}_{\perp}(\Sigma_r)   & \coloneqq \Bigg\{\phi_0(r),\Big\{\phi_{\sfn\sigma}(r)\Big\}_{\sfn=(\ell, m, k)}^{\sigma=\pm}\Bigg\},                  \\
		\sV^2_{\parallel}(\Sigma_r) & \coloneqq \Bigg\{\star_r \phi_0(r),\Big\{\star_r \phi_{\sfn \sigma}(r)\Big\}_{\sfn=(\ell, m, k)}^{\sigma=\pm}\Bigg\},
	\end{aligned}
\end{equation}
where we also renamed the zero-mode to \(\phi_0(r)\equiv \Phi_0(r)\). When we need to expand both one- and two- forms in these bases, we will refer to them collectively as \(\sV(\Sigma_r)\). This basis has the advantage that it makes the Kac--Moody structure clear.

\section{Current algebra in the generic case}\label{app:KM-simple-spectrum}

In this appendix we explain how the Kac--Moody algebra of \cref{ssec:current-Maxwell} gets modified in the generic --- in the sense of \cite{encisoNondegeneracyEigenvaluesHodge2010,kepplingerDimensionBeltramiOperator2022} --- case of a closed three-dimensional manifold, where the spectrum of the transversal Laplacian is simple, i.e. the eigenvalues are non-degenerate (cf. \cref{app:Laplacian}). We denote the three-dimensional Riemannian manifold we are treating, by a slight abuse of notation, as \(\Sigma\), where it should be understood that we are actually considering the pair \((\Sigma,g)\) where \(g\) is a Riemannian metric on \(\Sigma\) used to define the Hodge-star. We are not interested in varying the metric to obtain genericity results, like in \cite{encisoNondegeneracyEigenvaluesHodge2010,kepplingerDimensionBeltramiOperator2022}, so we suppress it.

The Beltrami operator, \(\star\dd\), will still be used to diagonalise the current algebra \cref{eq:1-KM}/\cref{eq:comm-QL}. However, in the generic case, if \(\sqrt{\lambda_\sfn}\) appears in the spectrum of \(\star\dd\), \(-\sqrt{\lambda_\sfn}\) doesn't, and vice versa. So we will label the orthonormalised eigen-one-forms of \(\star\dd\) as
\begin{equation}\label{eq:phin-generic}
	\set{\phi_{\sfn_{\pm}}}_{\sfn_\pm \in \sN_\pm},
\end{equation}
(where \(\sN_\pm\) is the index-set containing the labels \(\sfn_\pm\)) satisfying
\begin{equation}\label{eq:beltrami-generic}
	\star\dd{\phi_{\sfn_\pm}} = \pm \sqrt{\lambda_{\sfn_\pm}} \phi_{\sfn_\pm},
\end{equation}
where \(\lambda_{\sfn_\pm}>0\) is the corresponding (non-zero) eigenvalue of the transversal Laplacian on \(\phi_{\sfn_\pm}\). The labelling above, is such that eigenforms with label \(\sfn_-\) contain the negative spectrum of the Beltrami operator, and those with \(\sfn_+\), the positive spectrum. The genericity result mentioned above, implies that typically, \(\lambda_{\sfn_+}\neq \lambda_{\sfn_-}\) for all \(\sfn_+,\sfn_-\). This is unlike the case we have treated in the main text --- valid in highly symmetric, and fine-tuned, manifolds, such as products of spheres. Moreover, on a three-dimensional manifold, the spectrum of \(\star\dd\) accumulates at \(+\infty\) and at \(-\infty\) \cite{weylUeberSpektrumHohlraumstrahlung1912,safarovAsymptoticBehaviorSpectrum1984,baerCurlOperatorOdddimensional2019}, which guarantees that both signs appear in \cref{eq:beltrami-generic}. As before, we exclude the kernel of \(\star\dd\), as this is treated, unambiguously, via the harmonic forms, \(\set{\phi_{0\sfi}}_{\sfi=1}^{\b_1(\Sigma)}\), of \(\Sigma\). The one-forms \(\phi_{\sfn_\pm}\) are such that
\begin{equation}
	\ip{\phi_{\sfn_\pm}}{\phi_{\sfm_\pm}}_\Sigma = \delta_{\sfn_\pm \sfm_\pm}, \qq{and} \ip{\phi_{\sfn_+}}{\phi_{\sfn_-}}_\Sigma = 0.
\end{equation}

As explained in \cref{app:Laplacian}, \(\set{\phi_{0\sfi}, \phi_{\sfn_{\pm}}}\) provide a complete basis of coclosed one-forms on \(\Sigma\). We use this basis to expand \(\Lambda^\pm\) in \cref{eq:1-KM}/\cref{eq:comm-QL}, and the dual, two-form basis, given by the Hodge stars of the above, to expand \(J^\pm\):
\begin{align}
	i^*_\Sigma \Lambda^\mp & = \sum_{\sfi=1}^{\b_1(\Sigma)} \Lambda^\mp_{0\sfi} \phi_{0\sfi} + \sum_{\sfs=\pm}\;\sum_{\sfn_\sfs \in\sN_\sfs} \Lambda^\mp_{\sfn_\sfs} \star_\Sigma \phi_{\sfn_\sfs},   \\
	i^*_\Sigma J^\pm       & = \sum_{\sfi=1}^{\b_2(\Sigma)} Q^\pm_{0\sfi} \star_{\Sigma}\phi_{0\sfi} + \sum_{\sfs=\pm}\;\sum_{\sfn_\sfs \in\sN_\sfs} Q^\pm_{\sfn_\sfs} \star_\Sigma \phi_{\sfn_\sfs}.
\end{align}
The difference with \cref{eq:J-modeexp} is subtle, but important. This expansion leads to the mode algebra:
\begin{equation}
	\comm{Q^\pm_{\sfn_\sfs}}{Q^\mp_{\sfm_\sft}} = \pm \k\sfs\sqrt{\lambda_{\sfn_\sfs}} \delta_{\sfn_\sfs \sfm_\sft}.
\end{equation}
Defining ladder operators
\begin{equation}
	\cA_{\sfn_\sfs} \coloneqq
	\begin{cases}
		Q^+_{\sfn_\sfs} & \sfs=+ \\
		Q^-_{\sfn_\sfs} & \sfs=- \\
	\end{cases}, \qq{and} 	\cA^\dagger_{\sfn_\sfs} \coloneqq
	\begin{cases}
		Q^-_{\sfn_\sfs} & \sfs=+  \\
		Q^+_{\sfn_\sfs} & \sfs=-, \\
	\end{cases},
\end{equation}
leads to the algebra
\begin{equation}\label{eq:osc-comm-app}
	\comm{\cA_{\sfn_\sfs}}{\cA^\dagger_{\sfm_\sft}} = \k \sqrt{\lambda_{\sfn_\sfs}} \delta_{\sfn_\sfs \sfm_\sft}.
\end{equation}
Note that we still have two decoupled algebras, one for each sign of \(\sfs\), i.e. one for each side of the spectrum of \(\star\dd\). In terms of these modes, the Hamiltonian takes the form
\begin{equation}
	H_\Sigma = \frac{1}{\k}\sum_{\sfi=1}^{\b_2(\Sigma)} Q_{0\sfi}^+ Q_{0\sfi}^- + \frac{1}{\k}\sum_{\sfs=\pm}\;\sum_{\sfn_\sfs\in\sN_\sfs} \cA^\dagger_{\sfn_\sfs}\cA_{\sfn_\sfs} + E_0,
\end{equation}
with
\begin{equation}
	E_0 = \frac{1}{2}\sum_{\sfs}\sum_{\sfn_\sfs} \sqrt{\lambda_{\sfn_\sfs}}.
\end{equation}
As can be seen from \cref{eq:osc-comm-app}, the operators \(\cA^\dagger_{\sfn_\sfs}\) raise the energy by \(\sqrt{\lambda_{\sfn_\sfs}}\), and  \(\cA_{\sfn_\sfs}\) lower it by the same amount.

From here on, it is an easy exercise to repeat the steps explained in the main text and arrive at the conclusion that the Hilbert space is spanned by
\begin{equation}
	\ket{\vec{n},\vec{m};\set{N_{\sfn_\sfs}}_{\sfn_\sfs\in\sN_\sfs}^{\sfs=\pm}} \coloneqq \prod_{\sfs=\pm}\prod_{\sfn\in\sN_\sfs} \qty(\cA_{\sfn_\sfs}^\dagger)^{N_{\sfn_\sfs}}\ket{\vec{n},\vec{m}},
\end{equation}
with \(\ket{\vec{n},\vec{m}}\) the primary states, given by \cref{eq:primary1,eq:primary2}, with the appropriate tweak that they are annihilated by all \(\cA_{\sfn_\sfs}\). The rest of the discussion, including \cref{eq:xch 4d} and its implications follow immediately.

\section{Details on the radial evolution}\label{app:matrices}

Here we collect some details about the radial evolution on \(\B^3\times\S^1\). For the reader's convenience we repeat the radial evolution equation, \cref{eq:diffeq-J} here:
\begin{equation}\label{eq:app-diffeq-J}
	\pd_r J_{\sfn \sigma}^\pm(r) + \qty[\bbA^\pm_{\sfn}(r)]_{\sigma \tau} J_{\sfn \tau}^\pm(r) = 0.
\end{equation}
The matrix \(\bbA_{\sfn}^\pm(r)\) is given, in the basis \(\sV(\Sigma_r)\) \cref{eq:Vbasis}, by
\begin{align}
	\bbA_{\sfn}^\pm(r) & = \ip{\phi_{\sfn \sigma}}{\star_r\pd_r\star_r \phi_{\sfn \tau}} \mp \ip{\phi_{\sfn \sigma}}{\star_r\sd\phi_{\sfn \tau}} = \nn
	                   & =  \frac{1}{2r} \frac{\frac{\ell(\ell+1)}{r^2}}{\frac{\ell(\ell+1)}{r^2}+k^2}\mqty(1                                          & 1 \\ 1 & 1) \mp \sqrt{\frac{\ell(\ell+1)}{r^2}+k^2}\mqty(\dmat{1,-1}),
\end{align}
where \(\sfn=(\ell, m, k)\). Writing
\begin{equation}
	J^\pm_{\sfn \sigma}(r) = \qty[\bbU_{\sfn}^\pm(r,r_0)]_{\sigma \tau} J^\pm_{\sfn \sigma}(r_0),
\end{equation}
the initial value problem consisting of \cref{eq:app-diffeq-J} and its boundary conditions at \(r=r_0\) gets mapped to the matrix ordinary differential equation:
\begin{equation}\label{eq:matrix ODE}
	\dot{\bbU}_{\sfn}^\pm(r,r_0) + \bbA_{\sfn}^\pm(r) \bbU_{\sfn}^\pm(r,r0) = 0,
\end{equation}
where the dot indicates derivative with respect to \(r\), together with boundary conditions
\begin{equation}
	\bbU_{\sfn}^\pm(r_0,r_0) = \id,
\end{equation}
This has the unique solution
\begin{equation}
	\bbU_{\sfn}^\pm(r,r_0) = \rexp\qty(\int_r^{r_0} \dd{r'}\bbA_{\sfn}^\pm\qty(r')),
\end{equation}
where \(\rexp\) denotes the radially ordered exponential:
\begin{equation}
	\rexp\qty(\int_r^{r_0} \dd{r'}\bbO\qty(r')) \coloneqq \sum_{N=0}^\infty \frac{1}{N!} \int_r^{r_0}\! \dd{r_1} \int_r^{r_0}\! \dd{r_2} \cdots \int_r^{r_0}\! \dd{r_N} \mathrm{R}\qty\big(\bbO(r_1)\bbO(r_2)\cdots \bbO(r_N)),
\end{equation}
with
\begin{equation}
	\mathrm{R}\qty(\bbO_1(r_1) \bbO_2(r_2)) \coloneqq
	\begin{cases}
		\bbO_1(r_1) \bbO_2(r_2) & \t{if}\ r_1<r_2, \\
		\bbO_2(r_2) \bbO_1(r_1) & \t{if}\ r_2<r_1.
	\end{cases}
\end{equation}

When \(k=0\), it holds that \(\comm{\bbA_{\sfn}^\pm(r_1)}{\bbA_{\sfn}^\pm(r_2)}=0\) for all radii, and hence the ordered exponential reduces to a regular one. The solution is given in this case by \cref{eq:Ulm}:
\begin{equation}
	\begin{aligned}
		\bbU_{\ell m}^\pm(r,r_0) = \frac{1}{2+4\ell}\Bigg[\qty(\frac{r}{r_0})^{-\ell-1}\mqty(1+2\ell\mp 2\sqrt{\ell(\ell+1)} & 1  \\ 1 & 1+2\ell\pm 2\sqrt{\ell(\ell+1)}) & \\[0.5em]
		+ \qty(\frac{r}{r_0})^{\ell}\mqty(1+2\ell\pm2\sqrt{\ell(\ell+1)}                                                     & -1 \\ -1 & 1+2\ell\mp2\sqrt{\ell(\ell+1)})& \Bigg].
	\end{aligned}
\end{equation}
Here and onwards we suppress the magnetic quantum number \(m\) in the labelling of the radial evolution data, as they don't depend on it. The eigenvalues of \(\bbU_\ell^\pm(r,r_0)\) are
\begin{equation}
	\delta_{(1)\ell}^\pm = \qty(\frac{r}{r_0})^{-\ell -1} \qq{and} \delta_{(2)\ell}^\pm= \qty(\frac{r}{r_0})^{\ell},
\end{equation}
with eigenvectors
\begin{equation}
	u_{(1)\ell}^\pm = \mqty(1+2\ell \mp 2\sqrt{\ell(\ell+1)} \\[0.5em] 1) \qq{and} u_{(2)\ell}^\pm =\mqty(-1-2\ell \mp 2\sqrt{\ell(\ell+1)} \\[0.5em] 1),
\end{equation}
respectively. It is evident that at \(r\to 0\) one of the eigenvalues vanishes and, thus, the evolution matrix becomes rank one. Its kernel is spanned by \(v_{(2)\ell}^\pm\). The projector \(\mathbf{\Pi}_{\ell}^\pm\), appearing from \cref{eq:J-smooth} onwards, is, in this case:
\begin{equation}
	\mathbf{\Pi}_{\ell}^\pm = \frac{u_{(2)\ell}^\pm\otimes u_{(2)\ell}^\pm}{\norm{u_{(2)\ell}^\pm}^2} = \frac{1}{2(1+2\ell)}\mqty(1+2\ell \pm 2\sqrt{\ell(\ell+1)} & -1 \\[0.5em] -1 & 1+2\ell \mp 2\sqrt{\ell(\ell+1)}).
\end{equation}

For \(k\neq 0\), \(\bbA_\sfn^\pm(r)\) does not commute with itself at different radii and thus we have to use the radially ordered exponential. While this we cannot solve exactly for arbitrary radius, the crucial feature our the state-operator correspondence is its behaviour as \(r\to 0\). In that limit, the differential equation \cref{eq:matrix ODE} reduces to the differential equation for \(k=0\), and reveals that \(\bbU_{\ell m k}^\pm(r,r_0)\), behaves identically to its \(k=0\) eigenpart. Namely it has a singular and a regular part, scaling as 
\begin{equation}
	\sim\qty(\frac{r}{r_0})^{-\ell -1} \qq{and} \sim \qty(\frac{r}{r_0})^{\ell},
\end{equation}     
respectively. Correspondingly, \(\bbU_{\ell m k}^\pm(0,r_0)\) becomes rank-1.

\printbibliography

\end{document}

%% file: preamble.tex
\usepackage{lmodern}
\usepackage{amsfonts,amssymb,amsthm}
\usepackage{upgreek,mathrsfs}
\usepackage{bold-extra}
\usepackage{microtype}

\makeatletter
\g@addto@macro\bfseries{\boldmath}
\makeatother

\PassOptionsToPackage{dvipsnames}{xcolor}
\definecolor{maroon}{RGB}{167,74,74}
\definecolor{magreen}{RGB}{59,125,37}
\definecolor{mablue}{RGB}{59,136,195}
\definecolor{mayellow}{RGB}{242,147,24}
\colorlet{mabrown}{maroon!50!magreen}
\colorlet{maorange}{maroon!50!mayellow}
\colorlet{macyan}{magreen!50!mablue}

\colorlet{red}{maroon}
\colorlet{green}{magreen}
\colorlet{blue}{mablue}
\colorlet{yellow}{mayellow}
\colorlet{brown}{mabrown}
\colorlet{orange}{maorange}
\colorlet{cyan}{macyan}

\newcommand{\blue}[1]{\textcolor{mablue}{#1}}

\newcommand{\green}[1]{\textcolor{magreen}{#1}}
\newcommand{\gray}[1]{\textcolor{black!40}{#1}}

\RequirePackage{hyperref}
\hypersetup{colorlinks,
	linkcolor=mablue,
	citecolor=magreen,
	urlcolor=maorange,
	anchorcolor=mablue,
	filecolor=mablue,
	menucolor=mablue}

\makeatletter
\renewcommand\section{\@startsection{section}{1}{\z@}%
  {-3.5ex \@plus -1.3ex \@minus -.7ex}%
  {2.3ex \@plus.4ex \@minus .4ex}%
  {\large\bfseries}}
\renewcommand\subsection{\@startsection{subsection}{2}{\z@}%
	{-2.3ex\@plus -1ex \@minus -.5ex}%
	{1.2ex \@plus .3ex \@minus .3ex}%
	{\normalsize\bfseries}}
\renewcommand\subsubsection{\@startsection{subsubsection}{3}{\z@}%
	{-2.3ex\@plus -1ex \@minus -.5ex}%
	{1ex \@plus .2ex \@minus .2ex}%
	{\normalsize\bfseries}}
\renewcommand\paragraph{\@startsection{paragraph}{4}{\z@}%
	{1.75ex \@plus1ex \@minus.2ex}%
	{-1em}%
	{\normalsize\bfseries}}
\renewcommand\subparagraph{\@startsection{subparagraph}{5}{\parindent}%
	{1.75ex \@plus1ex \@minus .2ex}%
	{-1em}%
	{\normalsize\bfseries}}
\makeatother

\usepackage[style=numeric-comp, eprint=true, natbib, backend=biber, maxbibnames=10, giveninits=true, sorting=none, useprefix=true]{biblatex}

\DeclareFieldFormat*{title}{\emph{#1}}

\DeclareFieldFormat*{journaltitle}{#1}

\DeclareFieldFormat{pages}{#1}

\DeclareFieldFormat{labelalpha}{\textsc{#1}}

\renewbibmacro{in:}{}

\AtEveryBibitem{\clearfield{urldate}}
\AtEveryBibitem{\clearfield{month}}
\AtEveryBibitem{\clearfield{day}}
\AtEveryBibitem{\clearfield{issn}}
\AtEveryBibitem{\clearfield{version}}

\usepackage[inline]{enumitem}
\RequirePackage{mdframed}

\usepackage{empheq}
\newcommand*\obox[1]{%
	\fcolorbox{white}{blue!17}{\hspace{0.5em}#1\hspace{0.5em}}}

\RequirePackage[thicklines]{cancel}

\newcommand{\nn}{\nonumber \\}

\let\coloneqq\relax
\let\eqqcolon\relax
\makeatletter
\newcommand*{\coloneqq}{\mathrel{\rlap{\raisebox{0.3ex}{$\m@th\cdot$}}\raisebox{-0.3ex}{$\m@th\cdot$}}=}
\newcommand*{\eqqcolon}{\mathrel{=\llap{\raisebox{0.3ex}{$\m@th\cdot$}}\llap{\raisebox{-0.3ex}{$\m@th\cdot$}}}}
\makeatother

\newcommand{\rotcoloneqq}{\text{\rotatebox{90}{$\coloneqq$}}}

\newcommand{\demeqq}{\overset{!}{=}}
\renewcommand{\geq}{\geqslant}
\renewcommand{\leq}{\leqslant}

\usepackage{physics}
\newcommand{\parti}{\mathcal{Z}}

\let\f\relax
\newcommand{\f}[2]{{#1}_{\qty[#2]}} 
\newcommand{\gf}[2]{{#1}^{\qty[#2]}} 

\newcommand{\sd}[1]{\qty[#1]} 
\usepackage{stackengine} 

\newcommand{\suchthat}{\;\middle|\;}
\newcommand{\pd}{\partial} 

\newcommand{\w}{\mathbin{\scalebox{0.8}{$\wedge$}}}

\newcommand{\ctb}{\mathrm{T}^*}

\newcommand{\inv}[1]{{{#1}^{-1}}} 
\newcommand{\adj}[1]{{#1^{\dagger}}} 
\newcommand{\trans}[1]{{{#1}^{\top}}} 
\newcommand{\conj}[1]{{\mkern 1.3mu\overline{\mkern-1.3mu#1\mkern-1.3mu}\mkern 1.3mu}} 
\newcommand{\half}{\tfrac{1}{2}} 
\newcommand{\Half}{\frac{1}{2}}

\renewcommand{\vec}{\boldsymbol}
\newcommand{\ex}[1]{\mathrm{e}^{#1}}
\newcommand{\ii}{\mathrm{i}}

\newcommand{\set}[1]{\qty{#1}}

\let\C\undefined
\let\U\undefined
\newcommand{\R}{\mathbb{R}} 
\newcommand{\C}{\mathbb{C}} 
\newcommand{\Z}{\mathbb{Z}} 
\renewcommand{\S}{\mathbb{S}} 
\newcommand{\T}{\mathbb{T}} 
\newcommand{\D}{\mathbb{D}} 
\newcommand{\B}{\mathbb{B}} 

\newcommand{\U}{\mathrm{U}}

\renewcommand{\O}{\mathrm{O}}

\newcommand{\SL}{\mathrm{SL}}

\newcommand{\alg}[1]{\mathfrak{#1}}

\newcommand{\su}{\alg{su}}

\newcommand*{\id}{\text{\usefont{U}{bbold}{m}{n}1}}

\newcommand{\dual}[1]{\check{#1}}

\renewcommand{\H}{\mathrm{H}} 

\renewcommand{\t}[1]{{\text{#1}}}

\newcommand{\blob}{\bullet}

\usepackage{amsthm}
\theoremstyle{definition}

\def \cA {\mathcal{A}}
\def \cB {\mathcal{B}}
\def \cC {\mathcal{C}}

\def \cG {\mathcal{G}}
\def \cH {\mathcal{H}}

\def \cJ {\mathcal{J}}

\def \cL {\mathcal{L}}

\def \cN {\mathcal{N}}
\def \cO {\mathcal{O}}

\def \cV {\mathcal{V}}

\def \cX {\mathcal{X}}

\def \sB {\mathscr{B}}

\def \sN {\mathscr{N}}

\def \sV {\mathscr{V}}

\def \fS {\mathfrak{S}}

\def \fa {\mathfrak{a}}
\def \fb {\mathfrak{b}}
\def \fc {\mathfrak{c}}

\def \fg {\mathfrak{g}}

\def \fs {\mathfrak{s}}

\def \fu {\mathfrak{u}}

\def \bbA {\mathbb{A}}
\def \bbB {\mathbb{B}}

\def \bbE {\mathbb{E}}

\def \bbG {\mathbb{G}}
\def \bbH {\mathbb{H}}

\def \bbJ {\mathbb{J}}

\def \bbO {\mathbb{O}}
\def \bbP {\mathbb{P}}
\def \bbQ {\mathbb{Q}}

\def \bbU {\mathbb{U}}

\def \bb1 {{\mathbb{1}}}

\usepackage[outline]{contour}

\def \sfT {\mathsf{T}}

\def \sff {\mathsf{f}}

\def \sfi {\mathsf{i}}
\def \sfj {\mathsf{j}}

\def \sfl {\mathsf{l}}
\def \sfm {\mathsf{m}}
\def \sfn {\mathsf{n}}

\def \sfs {\mathsf{s}}
\def \sft {\mathsf{t}}

\newcommand{\inst}{\t{\scshape inst}} 
\newcommand{\osc}{\t{\scshape osc}} 

\newcommand{\cdd}{\dd^{\dagger}}

\newcommand{\W}{\mathrm{W}}

\newcommand{\g}{\mathrm{g}} 
\renewcommand{\k}{\mathrm{k}} 
\renewcommand{\tt}{\mathrm{t}} 
\newcommand{\pr}{\vec{\Pi}}

\usepackage{csquotes}
\usepackage{import}

\DeclareMathOperator{\spec}{spec}

\newcommand{\lapl}{\triangle}
\newcommand{\tlapl}{\Box}
\renewcommand{\b}{\mathrm{b}}

\newcommand{\inp}[2]{#1\mathbin{\lrcorner}#2} 
\newcommand{\fall}{{{}^\forall\!}}

\newcommand{\cpos}{\mathbf{q}}
\newcommand{\cmom}{\mathbf{p}} 

\def\fdiffd{\mathrm{D}}
\DeclareDocumentCommand\fdifferential{ o g d() }{ 
	\IfNoValueTF{#2}{
		\IfNoValueTF{#3}
		{\fdiffd\IfNoValueTF{#1}{}{^{#1}}}
		{\mathinner{\fdiffd\IfNoValueTF{#1}{}{^{#1}}\argopen(#3\argclose)}}
	}
	{\mathinner{\fdiffd\IfNoValueTF{#1}{}{^{#1}}#2} \IfNoValueTF{#3}{}{(#3)}}
}
\DeclareDocumentCommand\DD{}{\fdifferential}

\DeclareDocumentCommand\variation{ o g d() }{ 
	\IfNoValueTF{#2}{
		\IfNoValueTF{#3}
		{\updelta \IfNoValueTF{#1}{}{^{#1}}}
		{\mathinner{\updelta \IfNoValueTF{#1}{}{^{#1}}\argopen(#3\argclose)}}
	}
	{\mathinner{\updelta \IfNoValueTF{#1}{}{^{#1}}#2} \IfNoValueTF{#3}{}{(#3)}}
}
\DeclareDocumentCommand\var{}{\variation} 

\def\sdiffd{\mathbf{d}}
\DeclareDocumentCommand\sdifferential{ o g d() }{ 
	\IfNoValueTF{#2}{
		\IfNoValueTF{#3}
		{\sdiffd\IfNoValueTF{#1}{}{^{#1}}}
		{\mathinner{\sdiffd\IfNoValueTF{#1}{}{^{#1}}\argopen(#3\argclose)}}
	}
	{\mathinner{\sdiffd\IfNoValueTF{#1}{}{^{#1}}#2} \IfNoValueTF{#3}{}{(#3)}}
}
\DeclareDocumentCommand\sd{}{\sdifferential}

\DeclareDocumentCommand\detp{}{\opbraces{\determinant{}'}}

\DeclareMathOperator{\volume}{vol}
\DeclareDocumentCommand\vol{}{\opbraces{\volume}}

\DeclareMathOperator{\kernel}{ker}
\DeclareDocumentCommand\ker{}{\opbraces{\kernel}}

\DeclareMathOperator{\character}{ch}
\DeclareDocumentCommand\ch{}{\opbraces{\character}}

\DeclareDocumentCommand\xch{}{\opbraces{\boldsymbol{\character}}}

\DeclareMathOperator{\harm}{Harm}

\DeclareMathOperator{\radialexp}{Rexp}
\DeclareDocumentCommand\rexp{}{\opbraces{\radialexp}}

\renewcommand{\ip}[2]{\left\langle#1,#2\right\rangle}

\RequirePackage{interval}
\intervalconfig{soft open fences}
\newcommand{\ropen}[2]{\interval[open right]{#1}{#2}}

\newcommand{\open}[2]{\interval[open]{#1}{#2}}
\newcommand{\closed}[2]{\interval{#1}{#2}}

\usepackage[noabbrev]{cleveref}

\crefname{subsection}{subsection}{subsections}
\crefname{equation}{}{}

\numberwithin{equation}{section}

%% file: figures/state-operator-local.pdf_tex
\begingroup%
  \makeatletter%
  \providecommand\color[2][]{%
    \errmessage{(Inkscape) Color is used for the text in Inkscape, but the package 'color.sty' is not loaded}%
    \renewcommand\color[2][]{}%
  }%
  \providecommand\transparent[1]{%
    \errmessage{(Inkscape) Transparency is used (non-zero) for the text in Inkscape, but the package 'transparent.sty' is not loaded}%
    \renewcommand\transparent[1]{}%
  }%
  \providecommand\rotatebox[2]{#2}%
  \newcommand*\fsize{\dimexpr\f@size pt\relax}%
  \newcommand*\lineheight[1]{\fontsize{\fsize}{#1\fsize}\selectfont}%
  \ifx\svgwidth\undefined%
    \setlength{\unitlength}{492.17517282bp}%
    \ifx\svgscale\undefined%
      \relax%
    \else%
      \setlength{\unitlength}{\unitlength * \real{\svgscale}}%
    \fi%
  \else%
    \setlength{\unitlength}{\svgwidth}%
  \fi%
  \global\let\svgwidth\undefined%
  \global\let\svgscale\undefined%
  \makeatother%
  \begin{picture}(1,0.72139163)%
    \lineheight{1}%
    \setlength\tabcolsep{0pt}%
    \put(0,0){\includegraphics[width=\unitlength,page=1]{state-operator-local.pdf}}%
    \put(0.80435004,0.10124014){\color[rgb]{0,0,0}\rotatebox{5.8038777}{\makebox(0,0)[lt]{\lineheight{1.25}\smash{\begin{tabular}[t]{l}$\scriptstyle\R^{d}$\end{tabular}}}}}%
    \put(0,0){\includegraphics[width=\unitlength,page=2]{state-operator-local.pdf}}%
    \put(4.1133933,-0.293079){\color[rgb]{0,0,0}\rotatebox{5.8038777}{\makebox(0,0)[lt]{\lineheight{1.25}\smash{\begin{tabular}[t]{l}$\R^{d}$\end{tabular}}}}}%
    \put(0,0){\includegraphics[width=\unitlength,page=3]{state-operator-local.pdf}}%
    \put(0.48929253,0.24123744){\color[rgb]{0,0,0}\makebox(0,0)[lt]{\lineheight{1.25}\smash{\begin{tabular}[t]{l}$\cO(0)$\end{tabular}}}}%
    \put(0,0){\includegraphics[width=\unitlength,page=4]{state-operator-local.pdf}}%
    \put(0.11796724,0.05310369){\color[rgb]{0,0,0}\makebox(0,0)[lt]{\lineheight{1.25}\smash{\begin{tabular}[t]{l}$\cH_{\S^{d-1}}\ni\ket{\cO}$\end{tabular}}}}%
    \put(0.66401798,0.69887){\color[rgb]{0,0,0}\makebox(0,0)[lt]{\lineheight{1.25}\smash{\begin{tabular}[t]{l}$\scriptstyle\R\times\S^{d-1}$\end{tabular}}}}%
  \end{picture}%
\endgroup%

%% file: figures/state-operator-nonlocal.pdf_tex
\begingroup%
  \makeatletter%
  \providecommand\color[2][]{%
    \errmessage{(Inkscape) Color is used for the text in Inkscape, but the package 'color.sty' is not loaded}%
    \renewcommand\color[2][]{}%
  }%
  \providecommand\transparent[1]{%
    \errmessage{(Inkscape) Transparency is used (non-zero) for the text in Inkscape, but the package 'transparent.sty' is not loaded}%
    \renewcommand\transparent[1]{}%
  }%
  \providecommand\rotatebox[2]{#2}%
  \newcommand*\fsize{\dimexpr\f@size pt\relax}%
  \newcommand*\lineheight[1]{\fontsize{\fsize}{#1\fsize}\selectfont}%
  \ifx\svgwidth\undefined%
    \setlength{\unitlength}{461.41734686bp}%
    \ifx\svgscale\undefined%
      \relax%
    \else%
      \setlength{\unitlength}{\unitlength * \real{\svgscale}}%
    \fi%
  \else%
    \setlength{\unitlength}{\svgwidth}%
  \fi%
  \global\let\svgwidth\undefined%
  \global\let\svgscale\undefined%
  \makeatother%
  \begin{picture}(1,0.78646798)%
    \lineheight{1}%
    \setlength\tabcolsep{0pt}%
    \put(0,0){\includegraphics[width=\unitlength,page=1]{state-operator-nonlocal.pdf}}%
    \put(-2.29158718,0.27014753){\color[rgb]{0,0,0}\rotatebox{5.8038777}{\makebox(0,0)[lt]{\lineheight{1.25}\smash{\begin{tabular}[t]{l}$\scriptstyle\R^{d}$\end{tabular}}}}}%
    \put(0,0){\includegraphics[width=\unitlength,page=2]{state-operator-nonlocal.pdf}}%
    \put(-2.62764625,0.41947697){\color[rgb]{0,0,0}\makebox(0,0)[lt]{\lineheight{1.25}\smash{\begin{tabular}[t]{l}$\cO(0)$\end{tabular}}}}%
    \put(0,0){\includegraphics[width=\unitlength,page=3]{state-operator-nonlocal.pdf}}%
    \put(-3.02372388,0.21880233){\color[rgb]{0,0,0}\makebox(0,0)[lt]{\lineheight{1.25}\smash{\begin{tabular}[t]{l}$\cH_{\S^{d-1}}\ni\ket{\cO}$\end{tabular}}}}%
    \put(-2.4412737,0.90761507){\color[rgb]{0,0,0}\makebox(0,0)[lt]{\lineheight{1.25}\smash{\begin{tabular}[t]{l}$\scriptstyle\R\times\S^{d-1}$\end{tabular}}}}%
    \put(0,0){\includegraphics[width=\unitlength,page=4]{state-operator-nonlocal.pdf}}%
    \put(0.6978478,0.7631518){\color[rgb]{0,0,0}\makebox(0,0)[lt]{\lineheight{1.25}\smash{\begin{tabular}[t]{l}$\scriptstyle\R\times\S^{2}\times\S^{1}$\end{tabular}}}}%
    \put(0,0){\includegraphics[width=\unitlength,page=5]{state-operator-nonlocal.pdf}}%
    \put(0.76377964,0.25214264){\color[rgb]{0,0,0}\makebox(0,0)[lt]{\lineheight{1.25}\smash{\begin{tabular}[t]{l}$\scriptstyle\B^{3}\times\S^{1}$\end{tabular}}}}%
    \put(0,0){\includegraphics[width=\unitlength,page=6]{state-operator-nonlocal.pdf}}%
    \put(0.76377964,0.42224587){\color[rgb]{0,0,0}\makebox(0,0)[lt]{\lineheight{1.25}\smash{\begin{tabular}[t]{l}$\scriptstyle\S^{2}\times\S^{1}$\end{tabular}}}}%
    \put(0.64723458,0.01921563){\color[rgb]{0,0,0}\makebox(0,0)[lt]{\lineheight{1.25}\smash{\begin{tabular}[t]{l}$\cL\qty(\S^{1}\times\set{0})$\end{tabular}}}}%
  \end{picture}%
\endgroup%

%% file: figures/surrounding.pdf_tex
\begingroup%
  \makeatletter%
  \providecommand\color[2][]{%
    \errmessage{(Inkscape) Color is used for the text in Inkscape, but the package 'color.sty' is not loaded}%
    \renewcommand\color[2][]{}%
  }%
  \providecommand\transparent[1]{%
    \errmessage{(Inkscape) Transparency is used (non-zero) for the text in Inkscape, but the package 'transparent.sty' is not loaded}%
    \renewcommand\transparent[1]{}%
  }%
  \providecommand\rotatebox[2]{#2}%
  \newcommand*\fsize{\dimexpr\f@size pt\relax}%
  \newcommand*\lineheight[1]{\fontsize{\fsize}{#1\fsize}\selectfont}%
  \ifx\svgwidth\undefined%
    \setlength{\unitlength}{512.60922409bp}%
    \ifx\svgscale\undefined%
      \relax%
    \else%
      \setlength{\unitlength}{\unitlength * \real{\svgscale}}%
    \fi%
  \else%
    \setlength{\unitlength}{\svgwidth}%
  \fi%
  \global\let\svgwidth\undefined%
  \global\let\svgscale\undefined%
  \makeatother%
  \begin{picture}(1,0.39041436)%
    \lineheight{1}%
    \setlength\tabcolsep{0pt}%
    \put(0,0){\includegraphics[width=\unitlength,page=1]{surrounding.pdf}}%
    \put(-1.29597119,0.85287448){\color[rgb]{0,0,0}\rotatebox{25.851489}{\makebox(0,0)[lt]{\lineheight{1.25}\smash{\begin{tabular}[t]{l}$\displaystyle\int_{\S^2_\varepsilon\times\gamma} \alpha \wedge J^\pm$\end{tabular}}}}}%
    \put(0,0){\includegraphics[width=\unitlength,page=2]{surrounding.pdf}}%
    \put(-1.24696187,0.39848449){\color[rgb]{0,0,0}\makebox(0,0)[lt]{\lineheight{1.25}\smash{\begin{tabular}[t]{l}$\t{W}_{n,m}(\gamma)$\end{tabular}}}}%
    \put(-0.7976425,0.39848449){\color[rgb]{0,0,0}\makebox(0,0)[lt]{\lineheight{1.25}\smash{\begin{tabular}[t]{l}$J^\pm_\sfn\t{W}_{n,m}(\gamma)$\end{tabular}}}}%
    \put(0,0){\includegraphics[width=\unitlength,page=3]{surrounding.pdf}}%
    \put(0.43503417,0.19783753){\color[rgb]{0,0,0}\makebox(0,0)[lt]{\lineheight{1.25}\smash{\begin{tabular}[t]{l}{\Large $\displaystyle\ =\  \sum_{\sfn} \alpha_\sfn$}\end{tabular}}}}%
    \put(0.07721936,0.10011392){\color[rgb]{0,0,0}\makebox(0,0)[lt]{\lineheight{1.25}\smash{\begin{tabular}[t]{l}$\t{W}_{n,m}(\gamma)$\end{tabular}}}}%
    \put(0.6865595,0.10011392){\color[rgb]{0,0,0}\makebox(0,0)[lt]{\lineheight{1.25}\smash{\begin{tabular}[t]{l}$J^\pm_\sfn\t{W}_{n,m}(\gamma)$\end{tabular}}}}%
    \put(0.24958936,0.34827044){\color[rgb]{0,0,0}\makebox(0,0)[lt]{\lineheight{1.25}\smash{\begin{tabular}[t]{l}$\displaystyle\int_{\S^2_\varepsilon\times\gamma} \alpha \wedge J^\pm$\end{tabular}}}}%
    \put(0,0){\includegraphics[width=\unitlength,page=4]{surrounding.pdf}}%
  \end{picture}%
\endgroup%

%% file: figures/vacuum-operator.pdf_tex
\begingroup%
  \makeatletter%
  \providecommand\color[2][]{%
    \errmessage{(Inkscape) Color is used for the text in Inkscape, but the package 'color.sty' is not loaded}%
    \renewcommand\color[2][]{}%
  }%
  \providecommand\transparent[1]{%
    \errmessage{(Inkscape) Transparency is used (non-zero) for the text in Inkscape, but the package 'transparent.sty' is not loaded}%
    \renewcommand\transparent[1]{}%
  }%
  \providecommand\rotatebox[2]{#2}%
  \newcommand*\fsize{\dimexpr\f@size pt\relax}%
  \newcommand*\lineheight[1]{\fontsize{\fsize}{#1\fsize}\selectfont}%
  \ifx\svgwidth\undefined%
    \setlength{\unitlength}{623.25374573bp}%
    \ifx\svgscale\undefined%
      \relax%
    \else%
      \setlength{\unitlength}{\unitlength * \real{\svgscale}}%
    \fi%
  \else%
    \setlength{\unitlength}{\svgwidth}%
  \fi%
  \global\let\svgwidth\undefined%
  \global\let\svgscale\undefined%
  \makeatother%
  \begin{picture}(1,0.28020014)%
    \lineheight{1}%
    \setlength\tabcolsep{0pt}%
    \put(0,0){\includegraphics[width=\unitlength,page=1]{vacuum-operator.pdf}}%
    \put(4.29797871,0.12749072){\color[rgb]{0,0,0}\makebox(0,0)[lt]{\lineheight{1.25}\smash{\begin{tabular}[t]{l}$\adj{\fS}\t{W}_{n,m}\qty(\S^1\times\set{0})$\end{tabular}}}}%
    \put(0,0){\includegraphics[width=\unitlength,page=2]{vacuum-operator.pdf}}%
    \put(0.42311641,0.12682072){\color[rgb]{0,0,0}\makebox(0,0)[lt]{\lineheight{1.25}\smash{\begin{tabular}[t]{l}$\adj{\fS}\qty(\S^1\times\set{0})$\end{tabular}}}}%
    \put(0,0){\includegraphics[width=\unitlength,page=3]{vacuum-operator.pdf}}%
    \put(8.3532586,0.12749056){\color[rgb]{0,0,0}\makebox(0,0)[lt]{\lineheight{1.25}\smash{\begin{tabular}[t]{l}$\adj{\fS}\cB_{\sfn\pm}^\dagger\t{W}_{n,m}\qty(\S^1\times\set{0})$\end{tabular}}}}%
    \put(0,0){\includegraphics[width=\unitlength,page=4]{vacuum-operator.pdf}}%
  \end{picture}%
\endgroup%

%% file: figures/nm-primary.pdf_tex
\begingroup%
  \makeatletter%
  \providecommand\color[2][]{%
    \errmessage{(Inkscape) Color is used for the text in Inkscape, but the package 'color.sty' is not loaded}%
    \renewcommand\color[2][]{}%
  }%
  \providecommand\transparent[1]{%
    \errmessage{(Inkscape) Transparency is used (non-zero) for the text in Inkscape, but the package 'transparent.sty' is not loaded}%
    \renewcommand\transparent[1]{}%
  }%
  \providecommand\rotatebox[2]{#2}%
  \newcommand*\fsize{\dimexpr\f@size pt\relax}%
  \newcommand*\lineheight[1]{\fontsize{\fsize}{#1\fsize}\selectfont}%
  \ifx\svgwidth\undefined%
    \setlength{\unitlength}{623.25374573bp}%
    \ifx\svgscale\undefined%
      \relax%
    \else%
      \setlength{\unitlength}{\unitlength * \real{\svgscale}}%
    \fi%
  \else%
    \setlength{\unitlength}{\svgwidth}%
  \fi%
  \global\let\svgwidth\undefined%
  \global\let\svgscale\undefined%
  \makeatother%
  \begin{picture}(1,0.28020014)%
    \lineheight{1}%
    \setlength\tabcolsep{0pt}%
    \put(0,0){\includegraphics[width=\unitlength,page=1]{nm-primary.pdf}}%
    \put(0.42311643,0.12749084){\color[rgb]{0,0,0}\makebox(0,0)[lt]{\lineheight{1.25}\smash{\begin{tabular}[t]{l}$\adj{\fS}\t{W}_{n,m}\qty(\S^1\times\set{0})$\end{tabular}}}}%
    \put(0,0){\includegraphics[width=\unitlength,page=2]{nm-primary.pdf}}%
    \put(-3.45174586,0.12682072){\color[rgb]{0,0,0}\makebox(0,0)[lt]{\lineheight{1.25}\smash{\begin{tabular}[t]{l}$\adj{\fS}\qty(\S^1\times\set{0})$\end{tabular}}}}%
    \put(0,0){\includegraphics[width=\unitlength,page=3]{nm-primary.pdf}}%
    \put(4.47839633,0.12749056){\color[rgb]{0,0,0}\makebox(0,0)[lt]{\lineheight{1.25}\smash{\begin{tabular}[t]{l}$\adj{\fS}\cB_{\sfn\pm}^\dagger\t{W}_{n,m}\qty(\S^1\times\set{0})$\end{tabular}}}}%
    \put(0,0){\includegraphics[width=\unitlength,page=4]{nm-primary.pdf}}%
  \end{picture}%
\endgroup%

%% file: figures/nm-descendant.pdf_tex
\begingroup%
  \makeatletter%
  \providecommand\color[2][]{%
    \errmessage{(Inkscape) Color is used for the text in Inkscape, but the package 'color.sty' is not loaded}%
    \renewcommand\color[2][]{}%
  }%
  \providecommand\transparent[1]{%
    \errmessage{(Inkscape) Transparency is used (non-zero) for the text in Inkscape, but the package 'transparent.sty' is not loaded}%
    \renewcommand\transparent[1]{}%
  }%
  \providecommand\rotatebox[2]{#2}%
  \newcommand*\fsize{\dimexpr\f@size pt\relax}%
  \newcommand*\lineheight[1]{\fontsize{\fsize}{#1\fsize}\selectfont}%
  \ifx\svgwidth\undefined%
    \setlength{\unitlength}{623.25374573bp}%
    \ifx\svgscale\undefined%
      \relax%
    \else%
      \setlength{\unitlength}{\unitlength * \real{\svgscale}}%
    \fi%
  \else%
    \setlength{\unitlength}{\svgwidth}%
  \fi%
  \global\let\svgwidth\undefined%
  \global\let\svgscale\undefined%
  \makeatother%
  \begin{picture}(1,0.28020014)%
    \lineheight{1}%
    \setlength\tabcolsep{0pt}%
    \put(0,0){\includegraphics[width=\unitlength,page=1]{nm-descendant.pdf}}%
    \put(-3.63216358,0.12749084){\color[rgb]{0,0,0}\makebox(0,0)[lt]{\lineheight{1.25}\smash{\begin{tabular}[t]{l}$\adj{\fS}\t{W}_{n,m}\qty(\S^1\times\set{0})$\end{tabular}}}}%
    \put(0,0){\includegraphics[width=\unitlength,page=2]{nm-descendant.pdf}}%
    \put(-7.50702587,0.12682072){\color[rgb]{0,0,0}\makebox(0,0)[lt]{\lineheight{1.25}\smash{\begin{tabular}[t]{l}$\adj{\fS}\qty(\S^1\times\set{0})$\end{tabular}}}}%
    \put(0,0){\includegraphics[width=\unitlength,page=3]{nm-descendant.pdf}}%
    \put(0.42311642,0.12749056){\color[rgb]{0,0,0}\makebox(0,0)[lt]{\lineheight{1.25}\smash{\begin{tabular}[t]{l}$\adj{\fS}\cB_{\sfn\pm}^\dagger\t{W}_{n,m}\qty(\S^1\times\set{0})$\end{tabular}}}}%
    \put(0,0){\includegraphics[width=\unitlength,page=4]{nm-descendant.pdf}}%
  \end{picture}%
\endgroup%